\documentclass[aps,prx,reprint,floatfix,showkeys,longbibliography]{revtex4-1}
\usepackage[utf8]{inputenc}
\usepackage[english]{babel}

\usepackage{lipsum}
\usepackage[hidelinks,colorlinks=true]{hyperref}
\usepackage[usenames,dvipsnames,svgnames,table]{xcolor}
\usepackage{graphicx}

\definecolor{myred}{RGB}{228,26,28}
\definecolor{myblue}{RGB}{55,126,184}
\definecolor{myorange}{RGB}{225,127,0}
\definecolor{mygreen}{RGB}{77,175,74}
\definecolor{mylila}{RGB}{152,78,163}
\definecolor{mybrown}{RGB}{153,76,0}
\definecolor{mygray}{RGB}{153,153,153}
\definecolor{darkred}{rgb}{0.8,0,0}
\definecolor{mydarkgreen}{RGB}{0,102,0}
\definecolor{mydarkbrown}{RGB}{102,52,0}
\definecolor{Orange}{RGB}{235,129,27}
\definecolor{Green}{RGB}{35,55,59}

\usepackage{bm}
\usepackage{bbm}
\usepackage{amsmath,amsfonts,amssymb}

\usepackage{tikz} 
\usetikzlibrary{decorations,backgrounds,patterns} 
\usepackage{pgfplots} 
\usepackage{pgfplotstable}

\usepgfplotslibrary{external}
\pgfplotsset{compat=newest}
\usetikzlibrary{pgfplots.groupplots}
\usetikzlibrary{calc}
\usetikzlibrary{3d}
\usetikzlibrary{positioning}
\usepgfplotslibrary{patchplots}
\usepgfplotslibrary{fillbetween}
\tikzset{>=stealth}

\newcommand{\idest}{\emph{i.e. }}

\renewcommand{\d}{\mathrm{d}}
\newcommand{\D}{\partial}

\renewcommand{\i}{\mathrm{i}}

\newcommand{\A}{\bm{A}}

\newcommand{\tree}{\mathcal{T}}
\newcommand{\graph}{G}

\newcommand{\eigenvalue}{\lambda}

\newcommand{\rest}{\mathbb{O}}

\newcommand{\trajectory}{\bm{m}}
\newcommand{\trajectorymeso}{\bm{M}}

\newcommand{\pathintegralmeasure}{\mathcal{D}}

\newcommand{\trajectoryenergymicro}{e}
\newcommand{\trajectoryheatmicro}{q}
\newcommand{\trajectoryworkmicro}{w}
\newcommand{\trajectoryentropymicro}{s}
\newcommand{\trajectoryepmicro}{\sigma}
\newcommand{\trajectoryentropyflowmicro}{s_e}
\newcommand{\trajectoryobservablemicro}{o}

\newcommand{\trajectoryenergymeso}{E}
\newcommand{\trajectoryheatmeso}{Q}
\newcommand{\trajectoryworkmeso}{W}
\newcommand{\trajectoryentropymeso}{S}
\newcommand{\trajectoryepmeso}{\Sigma}
\newcommand{\trajectoryentropyflowmeso}{S_e}
\newcommand{\trajectoryobservablemeso}{O}
\newcommand{\countingfield}{\gamma}
\newcommand{\microgeneratingfunction}{g}
\newcommand{\mesogeneratingfunction}{G}

\newcommand{\mesoobservable}{O}
\newcommand{\meanfieldobservable}{\mathcal{\mesoobservable}}
\newcommand{\size}{M}
\newcommand{\reservoirdimension}{L}

\newcommand{\arrheniusprefactor}{\Gamma}
\newcommand{\stateenergy}{\epsilon}
\newcommand{\dimension}{N}
\newcommand{\microstate}{\alpha}
\newcommand{\mesostate}{\bm{\dimension}}
\newcommand{\density}{n}
\newcommand{\invtemperature}{\beta}

\newcommand{\force}{f}

\newcommand{\microrates}{w}
\newcommand{\mesorates}{W}

\newcommand{\microprobability}{p}
\newcommand{\mesoprobability}{P}

\newcommand{\microenergy}{e}
\newcommand{\microheat}{q}
\newcommand{\microwork}{w}
\newcommand{\microentropy}{s}

\newcommand{\microep}{\sigma}
\newcommand{\mesoenergy}{E}
\newcommand{\mesoheat}{Q}
\newcommand{\mesowork}{W}
\newcommand{\mesoentropy}{S}

\newcommand{\mesoep}{\Sigma}
\newcommand{\microfreeenergy}{a}
\newcommand{\mesofreeenergy}{A}
\newcommand{\meanfieldenergy}{\mathcal{\mesoenergy}}
\newcommand{\meanfieldheat}{\mathcal{\mesoheat}}
\newcommand{\meanfieldwork}{\mathcal{\mesowork}}
\newcommand{\meanfieldentropy}{\mathcal{\mesoentropy}}
\newcommand{\meanfieldfreeenergy}{\mathcal{\mesofreeenergy}}

\newcommand{\potential}{u}
\newcommand{\multiplicity}{\Omega}
\newcommand{\meanfieldprobability}{\overline{n}}
\newcommand{\meanfieldrates}{k}
\newcommand{\statenumber}{q}
\newcommand{\protocol}{\lambda}
\newcommand{\reservoir}{\nu}
\newcommand{\microcurrent}{j}
\newcommand{\mesocurrent}{J}
\newcommand{\meanfieldcurrent}{\mathcal{\mesocurrent}}

\newcommand{\field}{\pi}
\newcommand{\hamiltonian}{H}
\newcommand{\action}{\mathcal{L}}
\newcommand{\scgf}{\mathcal{G}}
\newcommand{\lagrangemultiplier}{\lambda}
\newcommand{\ratefct}{\Phi}

\begin{document}
\title{Stochastic thermodynamics of all-to-all interacting many-body systems}
\author{Tim Herpich}
\email{Electronic Mail: tim.herpich@uni.lu}
\author{Tommaso Cossetto}
\author{Gianmaria Falasco}
\author{Massimiliano Esposito}
\email{Electronic Mail: massimiliano.esposito@uni.lu}
\affiliation{Complex Systems and Statistical Mechanics, Department of Physics and Materials Science, University of Luxembourg, L-1511 Luxembourg, Luxembourg}
\date{\today}

\begin{abstract}
We provide a stochastic thermodynamic description across scales for $\dimension$ identical units with all-to-all interactions that are driven away from equilibrium by different reservoirs and external forces. We start at the microscopic level with Poisson rates describing transitions between many-body states. We then identify an exact coarse graining leading to a mesoscopic description in terms of Poisson transitions between system occupations. We proceed studying macroscopic fluctuations using the Martin-Siggia-Rose formalism and large deviation theory. In the macroscopic limit ($\dimension ~\to ~ \infty$), we derive the exact nonlinear (mean-field) rate equation describing the deterministic dynamics of the most likely occupations. We identify the scaling of the energetics and kinetics ensuring thermodynamic consistency (including the detailed fluctuation theorem) across microscopic, mesoscopic and macroscopic scales. The conceptually different nature of the ``Shannon entropy'' (and of the ensuing stochastic thermodynamics) at different scales is also outlined. Macroscopic fluctuations are calculated semi-analytically in an out-of-equilibrium Ising model. Our work provides a powerful framework to study thermodynamics of nonequilibrium phase transitions.
\end{abstract}
\maketitle

\onecolumngrid
\section{Introduction}

Interacting many body systems can give rise to a very rich variety of emergent behaviors such as phase transitions.
At equilibrium, their thermodynamic properties have been the object of intensive studies and are nowadays well understood \cite{stanley1971, goldenfeld1992, yeomans1992oxford, landau1994pergamon}, see also \cite{kadanoff2009jsp} for a more philosophical perspective. When driven out-of-equilibrium, these systems are known to give rise to complex dynamical behaviors \cite{schulman1980jpa, gaveau1987jpa, zia1995, bouchaud1997, odor2008, aceborn2005rmp, vandenbroeck2001epl, garrido1987jps, blote1990jpa, tjon1973physica}.
While most of the works are focused on their ensemble averaged description, in recent years progress was also made in characterizing their fluctuations \cite{garrahan2009jpa, gingrich2014pre, garrahanprl2014, bertini2015rpm, jack2010ptp}.
However, little is known about their thermodynamic description.
For instance, thermodynamics of nonequilibrium phase transitions started to be explored only recently \cite{herpich2018prx, imparato2019prl, herpich2019pre, verley2017epl, imparato2013pre, imparato2012prl, imparato2015njp, noa2019pre, sasa2015njp, noa2019pre, tomte2012prl, tomte2005pre, zhang2016jsm, crosato2018pre}.
There is a pressing need to develop methodologies to study thermodynamic quantities such as heat work and dissipation, not only at the average but also at the fluctuation level. 
To do so one has to start from stochastic thermodynamics that has proven instrumental to systematically infer the thermodynamics of small systems that can be driven arbitrarily far from equilibrium \cite{seifert2012rpp, broeck2015physica, sekimoto2010, zhang2012pr, ge2012pr, jarzynski2010ar}. 
This theory consistently builds thermodynamics on top of a Markov dynamics (e.g. master equations \cite{vandenbroeck2010pre} or Fokker-Planck equations \cite{vandenbroeck2010pre2}) describing open systems interacting with their surrounding.
Its predictions have been experimentally validated in a broad range of fields ranging from electronics to single molecules and Brownian particles \cite{pekkola2013rmp, ciliberto2017prx}.
It has been particularly successful in studying the performance of small energy converters operating far-from-equilibrium and their power-efficiency trade-off \cite{seifert2012rpp, esposito2009prl, verley2014nc, polettini2017prl, shiraishi2016prl, pietzonka2018prl}.
Until now, most of the focus has been on systems with finite phase space or few particle systems.
However there are exceptions.
Interacting systems have started to be considered in the context of energy conversion to asses whether they can trigger synergies in large ensembles of interacting energy converters. Beside few works such as \cite{herpich2018prx, imparato2019prl}, most other studies are restricted to mean-field treatments \cite{herpich2019pre, verley2017epl, imparato2012prl, imparato2013pre, imparato2015njp, esposito2013pre, verley2019}.
Another exception are chemical reaction networks which provide an interesting class of interacting systems. Indeed, while molecules in ideal solution are by definition noninteracting from an energetic standpoint, the stoichiometry of non-unimolecular reactions creates correlations amongst molecular species which generate entropic interactions. In the macroscopic limit, the mean field dynamics is exact and nonlinear \cite{nicolis1977, ge2011rsi, anderson2015} and can give rise to all sorts of complex behaviors \cite{epstein1996jpc}. The thermodynamics of chemical reaction networks has started to raise some attention in recent years \cite{gaspard2004jcp, polettini2015jcp, rao2016prx, rao2018jcp, lazarescu2019jcp}.

The main achievement of this paper is to provide a consistent nonequilibrium thermodynamic description across scales of many body systems with all-to-all interactions. We do so by considering $\dimension$ identical units with all-to-all (or infinite range) interactions. Each unit is composed of $\statenumber$ discrete states and undergoes transitions caused by one or more reservoirs. It may also be driven by an external force. The thermodynamics of this open many-body system is formulated at the ensemble averaged and fluctuating level, for finite $\dimension$ as well as in the macroscopic limit $\dimension \to \infty$.   

At the \emph{microscopic level}, the system is characterized by microstates which correspond to the many-body states (i.e. they define the state of each of the units). Poisson rates describe the transitions between the microstates triggered by the reservoirs. These rates satisfy local detail balance, i.e. their log-ratio is the entropy change in the reservoir caused by the transition \cite{rao2018njp}. It implicitly assumes that the system is weakly coupled to reservoirs which instantaneously relax back to equilibrium after an exchange with the system. By linking the stochastic dynamics with the physics, this crucial property ensures a consistent nonequilibrium thermodynamics description of the system, in particular a detailed fluctuation theorem and an ensuing second law at the ensemble averaged level. The entropy of a state is given by minus the logarithm of the probability to find the system in that state and the ensemble averaged entropy is the corresponding Shannon entropy. 

Because we assume all units to be identical in the way they interact with each other and with the reservoirs, we show that the \emph{microscopic} stochastic dynamics can be exactly coarse grained to a \emph{mesoscopic level}, where each system state specifies the unit occupations (i.e. the exact number of units which are in each of the unit states).
The mesoscopic rates describing transition between occupations satisfy a local detailed balance. At this level, the entropy of a state is given by minus the logarithm of the probability to find the system in that state plus the internal entropy given by the logarithm of the number of microstates inside a mesostate, reflecting the fact that the units are energetically indistinguishable. 
We demonstrate that stochastic thermodynamics is invariant under this exact coarse-graining of the stochastic dynamics, provided one considers initial conditions which are uniform within each mesostate, or for systems in stationary states.

We then consider the \emph{macroscopic limit} ($\dimension \to \infty$). Using a path integral representation of the stochastic dynamics (Martin-Siggia-Rose formalism), we identify the scaling in system size of the rates and of the energy that is necessary to ensure that the macroscopic fluctuations (i.e. the fluctuations that scale exponentially with $\dimension$) satisfy a detailed fluctuation theorem and are thus thermodynamically consistent. We show via the path-integral representation that the stochastic dynamics exactly reduces to a mean-field rate equation with nonlinear rates governing the evolution of the deterministic variables which correspond to the most likely values of the occupation of each unit state. Remarkably, the nonlinear rates still satisfy local detailed balance and the entropy of each deterministic occupation is given by minus their logarithm. The entropy is thus a Shannon entropy for deterministic variables exclusively arising from the entropy inside the mesostates and not from the probability distribution to be on a mesostates. Indeed, this latter narrows down around its single or multiple (in case of phase transition) most likely values and gives rise to a vanishing stochastic entropy.
We finally use our methodology to calculate macroscopic fluctuations in a semi-analytically solvable Ising model in contact with two reservoirs and displaying a nonequilibrium phase transition.

The plan of the paper is as follows.
First, in Sec. \ref{sec:stochasticdynamics}, the many-body model is introduced and the stochastic dynamics is formulated. Moreover, the exact coarse-graining scheme is presented and the asymptotic mean-field equations are derived.
Next, in Sec. \ref{sec:stochasticthermodynamics}, using the formalism of stochastic thermodynamics and Martin-Siggia-Rose, the fluctuating thermodynamic quantities are formulated at different scales and the conditions under which they are preserved across these scales are identified. These theoretical results are illustrated via a semi-analytically solvable Ising model. We conclude with a summary and perspectives in Sec. \ref{sec:conclusion}.

\section{Stochastic Dynamics}
\label{sec:stochasticdynamics}
\subsection{Microscopic Description}

We consider a system that consists of $\dimension$ all-to-all interacting identical and classical units that consist of $\statenumber$ states $i$ with energies $\stateenergy_i( \protocol_{t} )$ that are varying in time according to a known protocol $\protocol_{t}$ of an external driving. The system is coupled with multiple heat reservoirs $\reservoir = 1,2, ~ \ldots ~ , \reservoirdimension$ at inverse temperatures $\invtemperature^{(\reservoir)}$. Each unit is assumed to be fully connected, \idest any state of a given unit can be reached within a finite number of steps from all other states of that unit, so that the global system is irreducible.
Moreover, we suppose that all units are subjected to generic nonconservative forces $\force_{ij}^{(\reservoir)}$. Depending on whether a transition is aligned with or acting against the nonconservative force, the latter fosters or represses the transition from state $j$ to $i$. For generality, the force is assumed to be different depending on which heat reservoir $\reservoir$ the system is exchanging energy with during the transition from $j$ to $i$. Until explicitly states otherwise, we will take $\dimension$ to be finite in the following.

The many-body system is unambiguously characterized by a microstate
\begin{align}
\microstate = (\microstate_1,\ldots,\microstate_i,\ldots,\microstate_{\dimension}),
\quad \microstate_i=1,2,\ldots , \statenumber .
\end{align} 
The system energy consists of the state occupation of the units and the interactions between them. For all-to-all interactions, we readily determine the energy of the system in a microstate $\microstate$ as follows,
\begin{align}
\microenergy_{\microstate}(\bm{\protocol}_t) &= 
\sum\limits_{i=1}^{\statenumber} \Big\lbrace \dimension_i(\microstate) \,  \stateenergy_i(\protocol_{t}) + \frac{\potential_i(\protocol'_{t})}{2\dimension} \, \dimension_i(\microstate) \big[ \dimension_i(\microstate) - 1 \big] 
+ \sum\limits_{j<i} \frac{\potential_{ij}(\protocol'_{t})}{\dimension} \dimension_i (\microstate) \, \dimension_j (\microstate) \Big\rbrace ,  \label{eq:hamiltonian}
\end{align}
where $\potential_i(\protocol'_{t})/ \dimension$ and $\potential_{ij}(\protocol'_{t})/ \dimension$ denote the pair potential of units occupying the same or different single-unit states, respectively. These interactions can be tuned by an external driving according to a known protocol $\protocol'_{t}$, hence $\bm{\protocol}_t ~=~ \big(\protocol_{t},\protocol'_{t}\big)^{\top}$. Moreover, $\dimension_i(\microstate)$ refers to the number of units $\dimension_i$ occupying the single-unit state $i$ for a given microstate $\microstate$.

The stochastic jump process is governed by an irreducible Markovian master equation which describes the time evolution of the microscopic probability $\microprobability_{\microstate}$ for the system to be in the microstate $\microstate$ as follows,
\begin{align} \label{eq:micromasterequation} 
\D_t \microprobability_{\microstate}(t) = \sum\limits_{ \microstate' } \microrates_{\microstate\microstate'}(\bm{\protocol}_t) \, \microprobability_{\microstate'}(t)  \, , 
\end{align} 
with the microscopic rates $\microrates_{\microstate \microstate'}(\bm{\protocol}_t)$ for transitions from $\microstate'$ to $\microstate$ that in general depend on the current value of the driving parameter $\bm{\protocol}_t$. We note that probability conservation is ensured by the stochastic property of the transition rate matrix, $ \sum_{\microstate} \microrates_{\microstate \microstate'}(\bm{\protocol}_t) = 0$. The transition from $\microstate'$ to $\microstate$ is induced by one of the $\reservoirdimension$ heat reservoirs, thus
\begin{align}  \label{eq:microtransitionrates} 
\microrates_{\microstate\microstate'}(\bm{\protocol}_t) = \sum\limits_{\reservoir=1}^{\reservoirdimension}
\microrates^{(\reservoir)}_{\microstate\microstate'}(\bm{\protocol}_t) .
\end{align}
Here, for simplicity we assume that the transition rates are additive in the reservoirs $\reservoir$. A more general treatment can be made following the procedure described in Ref. \cite{rao2018njp}.
The microscopic transition rates that specify the heat reservoir satisfy the microscopic local detailed balance condition separately,
\begin{align} \label{eq:microlocaldetailedbalance}
\frac{\microrates^{(\reservoir)}_{\microstate\microstate'} (\bm{\protocol}_t)}{\microrates^{(\reservoir)}_{\microstate'\microstate} (\bm{\protocol}_t)} = 
\exp \Big\lbrace -\invtemperature^{(\reservoir)} \left[ \microenergy_{\microstate}(\bm{\protocol}_t) - \microenergy_{\microstate'}(\bm{\protocol}_t) - \bm{\force}_{\microstate \microstate'}^{(\reservoir)} \right] \Big\rbrace ,
\end{align}
which in turn ensures the thermodynamic consistency of the system. Here, $ \bm{\force}^{(\reservoir)}_{\microstate\microstate'} $ is the element of the nonconservative force vector $\bm{\force}^{(\reservoir)}$ that is equal to $\force_{ij}^{(\reservoir)}$, if the microscopic transition from $\microstate' \to \microstate$ corresponds to a single-unit transition from $j \to i$. If the transition rates are kept constant, $\bm{\protocol}_t = \bm{\protocol}$, the dynamics will relax into a unique stationary state, $\D_t \microprobability_{\microstate}^s(\bm{\protocol}) = 0$. If furthermore all heat reservoirs have the same inverse temperature, $\invtemperature^{(\reservoir)} = \invtemperature \; \forall \reservoir$, and the nonconservative forces vanish, $ \bm{\force}^{(\reservoir)} = 0 \; \forall \reservoir $, the stationary distribution coincides with the equilibrium one which satisfies the microscopic detailed balance condition,
\begin{align}
\microrates_{\microstate\microstate'}(\bm{\protocol}) \, \microprobability_{\microstate'}^{eq}(\bm{\protocol}) = \microrates_{\microstate' \microstate} (\bm{\protocol}) \, \microprobability_{\microstate}^{eq}(\bm{\protocol}) . 
\end{align}
The local detailed balance \eqref{eq:microlocaldetailedbalance} implies that the microscopic equilibrium distribution assumes the canonical form,
\begin{align}
\microprobability^{eq}_{\microstate}(\bm{\protocol}) &= \exp \big\lbrace -\invtemperature \left[ \microenergy_{\microstate}(\bm{\protocol}) - \microfreeenergy^{eq}(\bm{\protocol}) \right] \big\rbrace \, ,
\intertext{with the microscopic equilibrium free energy}
\microfreeenergy^{eq}(\bm{\protocol}) &= - \frac{1}{\invtemperature} \, \ln \, \sum\limits_{\microstate} \exp \big[ -\invtemperature \microenergy_{\microstate}(\bm{\protocol}) \big] \, .
\end{align}

\subsection{Mesoscopic Description}

The microscopic state space grows exponentially with the number of units, $|| \microstate || = \statenumber^{\dimension} $. Yet, the complexity of the system can be significantly reduced.
First, we note that due to the all-to-all interactions, there are equi-energetic microstates that are characterized by the same values for the occupation numbers $\dimension_i$. Next, we assume that the units are not only indistinguishable energetically [asymmetric part of the microscopic transition rates \eqref{eq:microlocaldetailedbalance}] but also kinetically [symmetric part of the microscopic transition rates \eqref{eq:microtransitionrates}] because they are all coupled in the same way to the reservoirs. As a result, the microscopic transition rates do not depend on the detailed pair of microstates that they connect but only on the pair of mesostate $\mesostate ~\equiv ~ (\dimension_1,\dimension_2, ~\ldots ~ ,\dimension_{\statenumber})$ that they connect.

Consequentially, the microscopic dynamics can be marginalized into a mesoscopic one, where the mesostate $\mesostate$ now identifies the state of the system.
We denote by $\microstate_{_{\mesostate}}$ the equienergetic microstates $\microstate$ inside a mesostate $\mesostate$, that is microstates for which the relation 
\begin{align} \label{eq:equienergeticmicrostates}
\microenergy_{\microstate_{_{\mesostate}}}(\bm{\protocol}_t) = \mesoenergy_{\mesostate}(\bm{\protocol}_t) ,
\end{align}
holds. The number $\multiplicity_{\mesostate}$ of microstates which belong to a mesostate is given by
\begin{align} \label{eq:multiplicityfactorpolynomial}
\multiplicity_{\mesostate} = \binom{\dimension}{\dimension_1} \binom{\dimension - \dimension_1}{\dimension_2} \cdots \binom{\dimension - \dimension_1 - \ldots - \dimension_{\statenumber-1} }{\dimension_{\statenumber}} = \frac{\dimension !}{\prod\limits_{i=1}^{\statenumber} \dimension_i!} \, .
\end{align}
We introduce the mesoscopic probability to observe the mesostate $\mesostate$ 
\begin{align} \label{eq:mesoscopicprobability}
\mesoprobability_{\mesostate} (t) &\equiv  \sum\limits_{\microstate_{_{\mesostate}}} \, \microprobability_{\microstate_{_{\mesostate}}} (t) \, .
\end{align}
The conditional probability to find the system in a microstate $\microstate_{_{\mesostate}}$ that belongs to that mesostate reads
\begin{align} \label{eq:mesoconditionalprobability} 
\mathbb{\mesoprobability}_{\microstate_{_{\mesostate}}}(t) = \frac{\microprobability_{\microstate_{_{\mesostate}}}(t)}{\mesoprobability_{\mesostate}(t)} .
\end{align}
Probability normalization implies that 
\begin{align} \label{eq:conditonalprobabilityconservation}
\sum_{\microstate_{_{\mesostate}}} \mathbb{\mesoprobability}_{\microstate_{_{\mesostate}}}(t) = 1 .
\end{align}
With Eqs. \eqref{eq:equienergeticmicrostates}, \eqref{eq:mesoscopicprobability} and \eqref{eq:conditonalprobabilityconservation} the microscopic master equation \eqref{eq:micromasterequation} can be exactly coarse-grained as follows, 
\begin{align}
\D_t \mesoprobability_{\mesostate}(t) &= \sum\limits_{\mesostate'} \sum\limits_{\microstate_{_{\mesostate}}} \sum\limits_{\microstate'_{_{\mesostate'}}} \microrates_{\microstate_{_{\mesostate}}, \microstate'_{_{\mesostate'}}}(\bm{\protocol}_t) \; \, \mathbb{\mesoprobability}_{\microstate'_{_{\mesostate'}}}\!(t) \; \mesoprobability_{\mesostate'}(t) 
= \sum\limits_{\mesostate'} \mesorates_{\mesostate \mesostate'}(\bm{\protocol}_t) \; \mesoprobability_{\mesostate'} (t) \, , \label{eq:mesoscopicmasterequation}
\end{align}
with the mesoscopic transition rates $\mesorates_{\mesostate \mesostate'}(\bm{\protocol}_t) = \multiplicity_{\mesostate,\mesostate'} \; \microrates_{_{\mesostate \mesostate'}}(\bm{\protocol}_t)$.
The quantity $\multiplicity_{\mesostate,\mesostate'}$ takes into account that only those microstates $\microstate_{_{\mesostate}}$ and $\microstate'_{_{\mesostate'}}$ contribute to the sum in Eq. \eqref{eq:mesoscopicmasterequation} which are connected to each other. This amounts to determine how many microstates $\microstate$ belong to the mesostate $\mesostate$ under the constraint that they are connected to microstates $\microstate'$ belonging to the mesostate $\mesostate'$.
The combinatorial problem is readily solved by noting that the occupation number that is decremented during the transition corresponds to the wanted quantity, \idest
\begin{align}  \label{eq:multiplicityfactorconstrained}
\multiplicity_{\mesostate,\mesostate'}
&= \sum_{i=1}^{\statenumber} \dimension'_i\, \delta_{\dimension'_i,\dimension_i+1 } \, ,
\end{align}
where $\dimension_i+1$ is understood as $ (\dimension_i+1) \text{ mod } \statenumber $. It is easy to verify that the stochastic property of the transition rate matrix is preserved by the coarse-graining, $ \sum_{\mesostate} \mesorates_{\mesostate \mesostate'}(\bm{\protocol}_t) = 0$.
The mesoscopic transition rates are still consisting of multiple contributions due to the different heat reservoirs,
\begin{align}  \label{eq:mesotransitionrates} 
\mesorates_{\mesostate \mesostate'}(\bm{\protocol}_t) = \sum\limits_{\reservoir=1}^{\reservoirdimension}
\mesorates_{\mesostate \mesostate'}^{(\reservoir)}(\bm{\protocol}_t) , 
\end{align}
that separately preserve the microscopic local detailed balance relation \eqref{eq:microlocaldetailedbalance} at the mesoscopic level,
\begin{align} \label{eq:mesolocaldetailedbalance}
\frac{ \mesorates_{\mesostate \mesostate'}^{(\reservoir)}(\bm{\protocol}_t) }{ \mesorates_{\mesostate'\mesostate}^{(\reservoir)}(\bm{\protocol}_t) } = \exp \Big\lbrace - \invtemperature^{(\reservoir)} \left[ \mesofreeenergy^{(\reservoir)}_{\mesostate}(\bm{\protocol}_t) - \mesofreeenergy^{(\reservoir)}_{\mesostate'}(\bm{\protocol}_t) - \bm{\force}^{(\reservoir)}_{\mesostate,\mesostate'}  \, \right] \Big\rbrace , 
\end{align}
with the notation $\bm{\force}^{(\reservoir)}_{\mesostate,\mesostate'}$ that is defined as $\bm{\force}^{(\reservoir)}_{\microstate,\microstate'}$ in Eq. \eqref{eq:microlocaldetailedbalance}.
Here, we introduced the free energy of a mesostate
\begin{align} \label{eq:mesofreeenergy}
\mesofreeenergy^{(\reservoir)}_{\mesostate}(\bm{\protocol}_t) = \mesoenergy_{\mesostate}(\bm{\protocol}_t) - \frac{1}{\invtemperature^{(\reservoir)}} \, \mesoentropy^{int}_{\mesostate} ,
\end{align}
and used the Boltzmann entropy
\begin{align} \label{eq:internalentropy}
\mesoentropy^{int}_{\mesostate} = \ln \multiplicity_{\mesostate} \, ,
\end{align}
along with the relation
\begin{align}
\frac{\multiplicity_{\mesostate}}{\multiplicity_{\mesostate'}} = \frac{\multiplicity_{\mesostate, \mesostate'}}{\multiplicity_{\mesostate',\mesostate}} ,
\end{align}
which can be seen by using Eqs. \eqref{eq:multiplicityfactorpolynomial} and \eqref{eq:multiplicityfactorconstrained}.
We remark that the dynamically exact coarse-graining of the microscopic dynamics towards a mesoscopic one has already been identified for and applied to an all-to-all interacting Ising model in Ref. \cite{tjon1973physica}.

If the transition rates are kept constant, $\bm{\protocol}_t = \bm{\protocol}$, the dynamics will reach a unique stationary state, $\D_t \mesoprobability_{\mesostate}^s(\bm{\protocol}) = 0$.
If furthermore all heat reservoirs have the same inverse temperature, $\invtemperature^{(\reservoir)} = \invtemperature \; \forall \reservoir$, and the nonconservative forces vanish, $ \bm{\force}^{(\reservoir)} = 0 \; \forall \reservoir$, the stationary distribution coincides with the equilibrium one which satisfies the mesoscopic detailed balance condition,
\begin{align}
\mesorates_{\mesostate\mesostate'}(\bm{\protocol}) \, \mesoprobability_{\mesostate'}^{eq}(\bm{\protocol}) = \mesorates_{\mesostate' \mesostate} (\bm{\protocol})\, \mesoprobability_{\mesostate}^{eq}(\bm{\protocol}),
\end{align}
and, because of Eq. \eqref{eq:mesolocaldetailedbalance}, assumes the canonical form,
\begin{align} \label{eq:mesoscopicequilibriumdistribution}
\mesoprobability^{eq}_{\mesostate}(\bm{\protocol}) = \exp \big\lbrace - \invtemperature \left[ \mesofreeenergy_{\mesostate}(\protocol) - \mesofreeenergy^{eq}(\bm{\protocol}) \right] \big\rbrace ,
\end{align}
with the mesoscopic equilibrium free energy 
\begin{align}
\mesofreeenergy^{eq}(\bm{\protocol}) = -\frac{1}{\invtemperature} \ln \sum\limits_{\mesostate} \exp \big[ -\invtemperature \mesofreeenergy_{\mesostate}(\bm{\protocol}) \big] \, .
\end{align}
The marginalization of the equienergetic microstates significantly reduces the complexity of the system since the mesoscopic state space asymptotically grows like a power law,
\begin{align}
|| \mesostate || = \sum_{\dimension_1=0}^{\dimension} \sum_{\dimension_2=0}^{\dimension_1} \cdots \sum_{\dimension_{\statenumber-1}=0}^{\dimension_{\statenumber-2}} 1 \, \overset{\dimension \to \infty}{\sim} \frac{\dimension^{\statenumber-1}}{(\statenumber-1)!} ,
\end{align}
as opposed to the exponential growth of the microscopic state space.

Since it will be useful further below, we now make two important remarks. 
First, a stationary mesoscopic distribution necessarely implies that all microstates that belong to the respective mesostates are equiprobable. This can be seen by first noting that in the stationary state, the microscopic master equation \eqref{eq:micromasterequation} reduces to $0= \sum_j \microrates_{ij} \, \microprobability_j$. Since the microscopic transition rates \eqref{eq:microtransitionrates} do not depend on the individual microstate $\microstate_{_{\mesostate}}$ belonging to a given mesostate $\mesostate$, it follows that the microscopic probability does not either in the stationary state so that
\begin{align} \label{eq:stationarymicroprobabilities} 
\mathbb{\mesoprobability}_{\microstate_{_{\mesostate}}}^{s} = \frac{1}{\multiplicity_{\mesostate}}, \qquad  \microprobability^{s}_{\microstate_{_{\mesostate}}}(\bm{\protocol}) &= \frac{\mesoprobability^{s}_{\mesostate}(\bm{\protocol})}{\multiplicity_{\mesostate}} . 
\end{align}
A more formal proof is deferred to appendix \ref{sec:spanningtreeappendix}.
Second, any microscopic initial condition of the form $\microprobability_{\microstate'}^{sp}(0) = \mesoprobability_{\mesostate'}^{sp}(0) / \multiplicity_{\mesostate}$ is preserved at all times since the Hamiltonian \eqref{eq:hamiltonian} and thus the microscopic transition rates \eqref{eq:microtransitionrates} do not discriminate between the equienergetic microstates inside the mesostate. Hence 
\begin{align} \label{eq:stationarymicroprobabilitiesTdep} 
\microprobability_{\microstate_{_{\mesostate}}}(t) &= \frac{\mesoprobability_{\mesostate}(t)}{\multiplicity_{\mesostate}}. 
\end{align}
An important implication is that if one only has experimental access to physical observable that do not discriminate among the units, there is no way to drive the system away from equipartition inside the mesostates. The system can therfore be driven arbitrary far from its stationary state in terms of its occupations but the equipartition within mesostates remains unaffected, i.e. \eqref{eq:stationarymicroprobabilitiesTdep} holds.


We demonstrated that for thermodynamically consistent and discrete identical systems with all-to-all interactions there is an exact coarse-graining of the microscopic stochastic dynamics characterized by many-body states towards a mesoscopic stochastic dynamics that is fully characterized by the global occupation of the different unit states. It is however \emph{a priori} not obvious that the thermodynamic structures built on top of these Markov process using stochastic thermodynamics are equivalent. This issue is investigated in the following section.

\section{Stochastic Thermodynamics}
\label{sec:stochasticthermodynamics}

\subsection{Trajectory Definitions}

\begin{figure}[h!]
\begin{center} 

\includegraphics[scale=1]{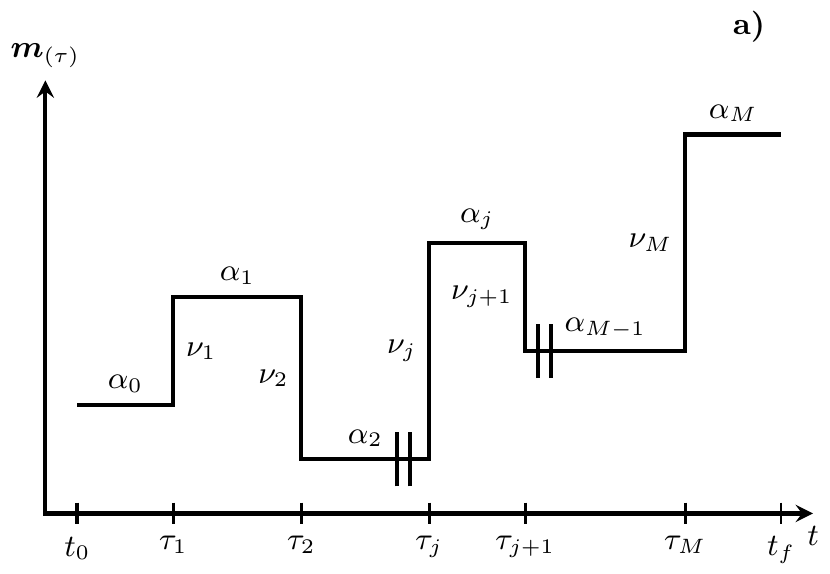}
\includegraphics[scale=1]{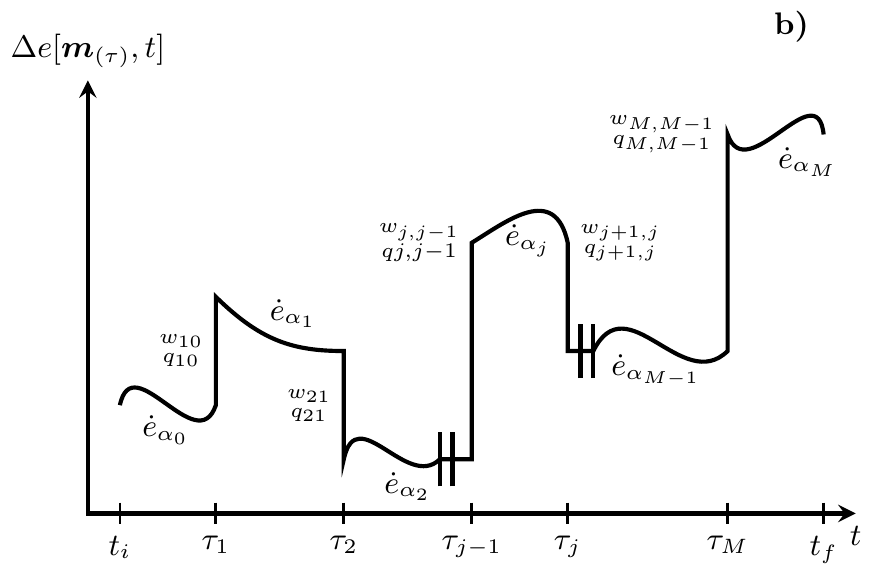}

\caption[Schematic representation of a single trajectory and its energetics]{Schematic representation of a single trajectory $\trajectory_{(\tau)}$ during the time $[t_0,t_f]$ in a) and the corresponding time-integrated first law of thermodynamics for the same trajectory in b). \label{fig:representationtrajectory} }
\end{center}
\end{figure}

After having established the stochastic dynamics at microscopic and mesoscopic scales, the following is devoted to formulating the stochastic thermodynamic quantities across these scales. To this end, we first introduce the fluctuating quantities at the level of a single trajectory. Generically, a trajectory is denoted by $\trajectory_{(\tau)}(t)$. This notation corresponds to the specification of the actual state in the time interval under consideration, $\trajectory_{(\tau)}(t), \,  t ~ \in ~ [t_0,t_f]$. Here, $\tau$ is a parametrization of the trajectory specifying the initial state $ \trajectory_{(\tau)}(t_0) = \microstate_0 $, the subsequent jumps from $\microstate_{j-1}$ to $\microstate_{j}$ as well as the heat reservoir $\reservoir_j$ involved at the instances of time, $ t = \tau_{j}, \; j = 1,\ldots \size$, and the final state, $ \trajectory_{(\tau)}(t_f) = \microstate_M $, where $\size$ is the total number of jumps.
More explicitly, we write
\begin{align}
\trajectory_{(\tau)} = \left\lbrace \trajectory_0 \overset{\reservoir_1, \tau_1}{\longrightarrow} \trajectory_{1} \overset{\reservoir_2, \tau_2}{\longrightarrow} \trajectory_{2} \overset{\reservoir_3, \tau_3}{\longrightarrow} \ldots \overset{\reservoir_j, \tau_j}{\longrightarrow} \trajectory_{j} \overset{\reservoir_{j+1}, \tau_{j+1}}{\longrightarrow} \ldots \trajectory_{\size-1} \overset{ \reservoir_{\size}, \tau_{\size}}{\longrightarrow} \trajectory_{\size} \right\rbrace , 
\end{align}
and refer to Fig. \ref{fig:representationtrajectory} a) for an illustrative example of such a stochastic trajectory.

\noindent In the following, we will use lower scripts to label trajectory-dependent quantities in microscopic representation and write $ \trajectoryobservablemicro[\trajectory_{(\tau)},t]$ for the value the observable $\trajectoryobservablemicro$ takes at time $t$ for the trajectory $\trajectory_{(\tau)}$.
We define the energy associated with the trajectory at time $t$ to be given by the energy of the particular microstate $\microstate$ the system is in for the trajectory under consideration, \idest 
\begin{align} \label{eq:stochasticenergy}
\trajectoryenergymicro[\trajectory_{(\tau)},t] = \sum\limits_{\microstate} \microenergy_{\microstate}(\bm{\protocol}_t)\, \delta_{\microstate,\trajectory_{(\tau)}(t)},
\end{align}
where the Kronecker delta $\delta_{\microstate,\trajectory_{(\tau)}(t)}$ selects the state $\microstate$ in which the trajectory is at the time under consideration.
The stochastic energy is a state function,
\begin{align}
\Delta \trajectoryenergymicro[\trajectory_{(\tau)},t] = \sum\limits_{\microstate} \big[ \microenergy_{\microstate}(\bm{\protocol}_t) \, \delta_{\microstate,\trajectory_{(\tau)}(t)} - \microenergy_{\microstate}(\bm{\protocol}_0) \, \delta_{\microstate,\trajectory_{(\tau)}(0)} \big] , 
\end{align}
as indicated by the notation $\Delta \trajectoryenergymicro$, and its time-derivative \footnote{To determine the time-derivative of the Kronecker delta, we realize that $ \dot{\delta}_{\microstate,\trajectory_{(\tau)}(t)}$ goes from 0 to 1, and from 1 to 0, when $\trajectory_{(\tau)}(t)$ jumps into, or out of the microstate $\microstate$, respectively, at time $t$. Hence the time-derivative consists of a sum of delta functions, with weights 1 and -1, respectively, centered at the times, $\tau_j$, of the jumps.} can be decomposed as follows,
\begin{align} \label{eq:stochasticfirstlaw} 
\d_t \, \trajectoryenergymicro[\trajectory_{(\tau)},t] &= \dot{\trajectoryheatmicro}[\trajectory_{(\tau)},t] + \dot{\trajectoryworkmicro}[\trajectory_{(\tau)},t] , 
\end{align}
with the stochastic heat and work currents
\begin{align} \label{eq:stochasticheat}
\dot{\trajectoryheatmicro}[\trajectory_{(\tau)},t] &= \sum_{\reservoir = 1}^{\reservoirdimension} \sum\limits_{j=1}^{\size} \delta( \reservoir - \reservoir_j) \delta(t-\tau_j) \left[ \trajectoryenergymicro_{\microstate_j}(\bm{\protocol}_{\tau_j}) -  \trajectoryenergymicro_{\microstate_{j-1}}(\bm{\protocol}_{\tau_j}) - \bm{\force}^{(\reservoir_j)}_{\microstate_j,\microstate_{j-1}} \right] \\
&= \sum_{\reservoir = 1}^{\reservoirdimension} \underbrace{ - \sum\limits_{j=1}^{\size} \delta(\reservoir-\reservoir_j) \delta(t-\tau_j) \, \frac{1}{ \invtemperature^{(\reservoir_j)} } \, \ln \frac{\microrates^{(\reservoir_j)}_{\microstate_j, \microstate_{j-1}} (\bm{\protocol}_{\tau_j})}{\microrates^{(\reservoir_j)}_{\microstate_{j-1},\microstate_j} (\bm{\protocol}_{\tau_j})} }_{ \dot{\trajectoryheatmicro}^{(\reservoir)}[\trajectory_{(\tau)},t] } \nonumber \\ 
\dot{\trajectoryworkmicro}[\trajectory_{(\tau)},t] &=
\underbrace{ \sum\limits_{\microstate} \big[ \dot{\bm{\protocol}}_t \cdot \nabla_{\bm{\protocol}_t} \, \microenergy_{\microstate}(\bm{\protocol}_t) \big] \, \delta_{\microstate,\trajectory_{(\tau)}(t)} \Big|_{\trajectory_{(\tau)}(t)} }_{ \dot{\trajectoryworkmicro}_{\bm{\protocol}}[\trajectory_{(\tau)},t] } + \sum_{\reservoir = 1}^{\reservoirdimension} \, \underbrace{ \sum\limits_{j=1}^{\size} \delta(\reservoir-\reservoir_j) \delta(t-\tau_j) \; \bm{\force}^{(\reservoir_j)}_{\microstate_j,\microstate_{j-1}} }_{ \dot{\trajectoryworkmicro}^{(\reservoir)}_{\bm{\force}}[\trajectory_{(\tau)},t] } ,  \label{eq:stochasticwork} 
\end{align}
where we introduced the notation $\nabla_{\bm{\protocol}_t} = \big( \D_{\protocol_{t}},\D_{\protocol'_{t}} \big)^{\top}$ and $\dot{x} |_{\trajectory_{(\tau)}(t)}$ which corresponds to the instantaneous and smooth changes of $x$ along the horizontal segments of the trajectory $\trajectory_{(\tau)(t)}$ in Fig. \ref{fig:representationtrajectory}a).
It will be proven instrumental to split the fluctuating work current into the contribution $\dot{\trajectoryworkmicro}_{\bm{\protocol}}[\trajectory_{(\tau)},t]$ from the nonautonomous driving and the dissipative contribution $ \sum_{\reservoir = 1}^{\reservoirdimension} \dot{\trajectoryworkmicro}^{(\reservoir)}_{\bm{\force}}[\trajectory_{(\tau)},t] $ due to the nonconservative forces.
It is noteworthy that Eq. \eqref{eq:stochasticfirstlaw} is the stochastic first law and ensures energy conservation at the trajectory level \cite{sekimoto1997ptps}. As an illustrative example, Fig. \ref{fig:representationtrajectory} b) shows the time-integrated stochastic first law for the corresponding trajectory in a).

Next, the stochastic system entropy is defined as follows \cite{seifert2005prl,shengentropy2019}
\begin{align} \label{eq:stochasticentropy}
\trajectoryentropymicro[\trajectory_{(\tau)},t] = - \sum\limits_{\microstate} \ln \microprobability_{\microstate}(t) \; \delta_{\microstate,\trajectory_{(\tau)}(t)} , 
\end{align}
and is therefore also a state-function,
\begin{align}
\Delta \trajectoryentropymicro[\trajectory_{(\tau)},t] = - \sum\limits_{\microstate} \big[ \ln \microprobability_{\microstate}(t) \, \delta_{\microstate,\trajectory_{(\tau)}(t)} - \ln \microprobability_{\microstate}(0) \, \delta_{\microstate,\trajectory_{(\tau)}(0)} \big] , 
\end{align}
where we set $k_B \equiv 1$.
Its time-derivative
\begin{align}
\d_t \, \trajectoryentropymicro[\trajectory_{(\tau)},t] &= 
- \sum\limits_{\microstate} \frac{ \D_t \microprobability_{\microstate}(t) }{ \microprobability_{\microstate}(t) } \delta_{\microstate,\trajectory_{(\tau)}(t)} \Big|_{\trajectory_{(\tau)}(t)} + \, \sum\limits_{j=1}^{\size} \delta(t-\tau_j) \ln \frac{\microprobability_{\microstate_{j-1}}(t)}{\microprobability_{\microstate_{j}}(t)} = \dot{\trajectoryentropymicro}_e[\trajectory_{(\tau)},t] + \dot{\sigma}[\trajectory_{(\tau)},t] , \label{eq:stochasticentropybalance}
\end{align}
can be split into the stochastic entropy flow 
\begin{align} 
\label{eq:stochsticentropyflow} 
\dot{\trajectoryentropymicro}_e[\trajectory_{(\tau)},t] &= \sum_{\reservoir = 1}^{\reservoirdimension} - \sum\limits_{j=1}^{\size} \delta( \reservoir - \reservoir_j) \delta(t-\tau_j) \, \ln \frac{\microrates^{(\reservoir_j)}_{\microstate_j, \microstate_{j-1}} (\bm{\protocol}_{\tau_j})}{\microrates^{(\reservoir_j)}_{\microstate_{j-1},\microstate_j} (\bm{\protocol}_{\tau_j})} = \sum_{\reservoir = 1}^{\reservoirdimension} \invtemperature^{(\reservoir)} \, \dot{\trajectoryheatmicro}^{(\reservoir)}[\trajectory_{(\tau)},t] ,
\end{align}
and the stochastic entropy production rate
\begin{align}
\dot{\trajectoryepmicro}[\trajectory_{(\tau)},t] &= - \sum\limits_{\microstate} \frac{ \D_t \microprobability_{\microstate}(t) }{ \microprobability_{\microstate}(t) } \delta_{\microstate,\trajectory_{(\tau)}(t)} \Big|_{\trajectory_{(\tau)}(t)} + \sum_{\reservoir=1}^{\reservoirdimension} \sum\limits_{j=1}^{\size} \delta( \reservoir - \reservoir_j) \delta(t-\tau_j) \, \ln \frac{\microrates^{(\reservoir_j)}_{\microstate_j, \microstate_{j-1}} (\bm{\protocol}_{\tau_j}) \, \microprobability_{\microstate_{j-1}}(t) }{\microrates^{(\reservoir_j)}_{\microstate_{j-1},\microstate_j} (\bm{\protocol}_{\tau_j}) \, \microprobability_{\microstate_{j}}(t) } . 
\label{eq:stochasticentropyproductionrate}
\end{align}
We note that Eq. \eqref{eq:stochasticentropybalance} corresponds to the entropy balance at the trajectory level.

It will prove useful to also consider the time-integrated stochastic first law
\begin{align}
\label{eq:timeintegratedstochasticfirstlaw} 
\Delta \trajectoryenergymicro[\trajectory_{(\tau)},t] &\equiv \sum_{\reservoir = 1}^{\reservoirdimension} \delta \trajectoryenergymicro^{(\reservoir)}[\trajectory_{(\tau)},t] =  \delta \trajectoryheatmicro[\trajectory_{(\tau)},t] + \delta \trajectoryworkmicro[\trajectory_{(\tau)},t] ,
\end{align}
with the time-integrated fluctuating energy current
\begin{align} \label{eq:stochasticenergydecomposition}
\delta \trajectoryenergymicro^{(\reservoir)}[\trajectory_{(\tau)},t] = \int\limits_0^t \d t' \sum\limits_{j=1}^{\size} \delta( \reservoir - \reservoir_j) \delta(t'-\tau_j) [ \trajectoryenergymicro_{\microstate_j} - \trajectoryenergymicro_{\microstate_{j-1}} ] .
\end{align}
and the fluctuating heat and work 
\begin{align} \label{eq:stochasticheatdecomposition}
\delta \trajectoryheatmicro[\trajectory_{(\tau)},t] &= \sum_{\reservoir = 1}^{\reservoirdimension} - \int\limits_0^t \d t' \, \sum\limits_{j=1}^{\size} \delta( \reservoir - \reservoir_j) \delta(t'-\tau_j) \, \frac{1}{ \invtemperature^{(\reservoir_j)} } \, \ln \frac{\microrates^{(\reservoir_j)}_{\microstate_j, \microstate_{j-1}} (\bm{\protocol}_{\tau_j})}{\microrates^{(\reservoir_j)}_{\microstate_{j-1},\microstate_j} (\bm{\protocol}_{\tau_j})}  = \sum_{\reservoir = 1}^{\reservoirdimension} \Big( \underbrace{ \delta \trajectoryenergymicro^{(\reservoir)}[\trajectory_{(\tau)},t] - \delta \trajectoryworkmicro_{\bm{\force}}^{(\reservoir)}[\trajectory_{(\tau)},t] }_{ \delta \trajectoryheatmicro_{\bm{\force}}^{(\reservoir)}[\trajectory_{(\tau)},t] } \Big) \\
\delta \trajectoryworkmicro[\trajectory_{(\tau)},t] &=  \underbrace{ \int\limits_0^t \d t' \sum\limits_{\microstate} \big[ \dot{\bm{\protocol}}_{t'} \cdot \nabla_{\bm{\protocol}_{t'}} \, \microenergy_{\microstate}(\bm{\protocol}_{t'}) \big] \, \delta_{\microstate,\trajectory_{(\tau)}(t')} \Big|_{\trajectory_{(\tau)}(t')} }_{ \delta \trajectoryworkmicro_{\bm{\protocol}}[\trajectory_{(\tau)},t] } + \; \sum_{\reservoir = 1}^{\reservoirdimension} \; \underbrace{ \int\limits_0^t \d t' \sum\limits_{j=1}^{\size} \delta( \reservoir - \reservoir_j) \delta(t'-\tau_j) \bm{\force}^{(\reservoir)}_{\microstate_j,\microstate_{j-1}} }_{ \delta \trajectoryworkmicro_{\bm{\force}}^{(\reservoir)}[\trajectory_{(\tau)},t] } .
\label{eq:stochasticworkdecomposition}
\end{align}
Using Eqs. \eqref{eq:stochasticentropyproductionrate} and \eqref{eq:stochasticheatdecomposition}, the entropy production can be written as follows
\begin{align}  \label{eq:stochasticentropyproduction}
\delta \trajectoryepmicro[\trajectory_{(\tau)},t] = - \ln \frac{\microprobability_{\microstate_{_{\size}}}(t) }{\microprobability_{\microstate_{_0}}(0)} + \sum_{\reservoir = 1}^{\reservoirdimension} \int\limits_0^t \d t' \, \sum\limits_{j=1}^{\size} \delta( \reservoir - \reservoir_j) \delta(t'-\tau_j) \, \ln \frac{\microrates^{(\reservoir_j)}_{\microstate_j, \microstate_{j-1}} (\bm{\protocol}_{\tau_j}) }{\microrates^{(\reservoir_j)}_{\microstate_{j-1},\microstate_j} (\bm{\protocol}_{\tau_j}) } = - \ln \frac{\microprobability_{\microstate_{_{\size}}}(t) }{\microprobability_{\microstate_{_0}}(0)} - \sum_{\reservoir = 1}^{\reservoirdimension} \invtemperature^{(\reservoir)} \delta \trajectoryheatmicro^{(\reservoir)}[\trajectory_{(\tau)},t] .
\end{align}

\subsection{Generating Function Techniques}
\subsubsection{Microscopic Description}

In the preceding section we introduced in detail all the relevant fluctuating thermodynamic quantities. We now present techniques in order to compute the statistics and features of these quantities as they will also prove useful to determine if the thermodynamics is invariant under the dynamically exact coarse-graining in Eq. \eqref{eq:mesoscopicmasterequation}.
To this end, we consider the microscopic generating function related to the change $ \delta \trajectoryobservablemicro[\trajectory_{(\tau)},t]$ of the fluctuating microscopic observable $\trajectoryobservablemicro$ along a trajectory $\trajectory_{(\tau)}$ conditioned to be in a microstate $\microstate$ at time $t$ which is defined as
\begin{align} \label{eq:generatingfunctiondefinition}
\microgeneratingfunction_{\microstate}(\countingfield_{\trajectoryobservablemicro},t) &= \microprobability_{\microstate}(t) \langle \exp \lbrace - \countingfield_{\trajectoryobservablemicro} \, \delta \trajectoryobservablemicro[\trajectory_{(\tau)},t] \rbrace \rangle_{\microstate} ,
\end{align}
where $\langle ~ \cdot ~ \rangle_{\microstate} $ denotes an ensemble average over all trajectories that are in the microstate $\microstate$ at time $t$ and $\countingfield_{\trajectoryobservablemicro}$ is the counting field (also bias). It thus holds that $\microgeneratingfunction(\countingfield_{\trajectoryobservablemicro},t) = \sum_{\microstate} \microgeneratingfunction_{\microstate}(\countingfield_{\trajectoryobservablemicro},t) $. The microscopic generating function can also be expressed as follows
\begin{align}
\label{eq:generatingfunctiondefinitionrewritten}
\microgeneratingfunction(\i \countingfield_{\trajectoryobservablemicro},t) &= \int\limits_{-\infty}^{\infty} \d (\delta \trajectoryobservablemicro) \, \exp [ - \i \, \countingfield_{\trajectoryobservablemicro} \, \delta \trajectoryobservablemicro ] \, \microprobability(\delta \trajectoryobservablemicro,t) , 
\end{align}
where $\microprobability(\delta \trajectoryobservablemicro,t)$ is the probability to observe a change $\delta \trajectoryobservablemicro$ in the microscopic observable $\trajectoryobservablemicro$ until time $t$. The different moments of the microscopic observable $\delta \trajectoryobservablemicro$ are obtained via the associated microscopic generating function as follows,
\begin{align} \label{eq:generatingfunctiondifferentiation} 
\langle \delta \trajectoryobservablemicro^n \rangle =  (-1)^{n} \frac{\D^n }{(\D \countingfield_{\trajectoryobservablemicro})^n} \, \microgeneratingfunction(\countingfield_{\trajectoryobservablemicro},t) \Big|_{\countingfield_{\trajectoryobservablemicro}=0} \; .
\end{align}
The equation of motion for the microscopic generating function has the form of a biased microscopic master equation \cite{esposito2007pre},
\begin{align} 
\D_t \microgeneratingfunction_{\microstate}(\countingfield_{\trajectoryobservablemicro},t) = \sum\limits_{\microstate'} \microrates_{\microstate \microstate'}(\countingfield_{\trajectoryobservablemicro},\bm{\protocol}_t) \, \microgeneratingfunction_{\microstate'}(\countingfield_{\trajectoryobservablemicro},t), \quad  \microrates_{\microstate \microstate'}(\countingfield_{\trajectoryobservablemicro},\bm{\protocol}_t) =  - \countingfield_{\trajectoryobservablemicro} \, \D_t \trajectoryobservablemicro_{\microstate} \, \delta_{\microstate,\microstate'} + \sum_{\reservoir=1}^{\reservoirdimension} \exp \big[ - \countingfield_{\trajectoryobservablemicro} \; \trajectoryobservablemicro^{(\reservoir)}_{\microstate\microstate'}(\bm{\protocol}_t) \big] \, \microrates^{(\reservoir)}_{\microstate \microstate'}(\bm{\protocol}_t) ,  \label{eq:generatingfunctionequationofmotion} 
\end{align}
where $\microrates_{\microstate \microstate'}(\countingfield_{\trajectoryobservablemicro},\bm{\protocol}_t)$ is the microscopic biased generator. The notation $\trajectoryobservablemicro_{\microstate}$ and $\trajectoryobservablemicro_{\microstate \microstate'}$ refers to the value of the stochastic microscopic observable in microstate $\microstate$ and its change during a transition from state $\microstate'$ to $\microstate$ while the system exchanges energy with the reservoir $\reservoir$, respectively.

For state functions, $\delta \trajectoryobservablemicro[\trajectory_{(\tau)},t] = \Delta \trajectoryobservablemicro[\trajectory_{(\tau)},t] = \trajectoryobservablemicro[\trajectory_{(\tau)},t] - \trajectoryobservablemicro[\trajectory_{(\tau)},0]$, the ensemble average over all trajectories in Eq. \eqref{eq:generatingfunctiondefinition} reduces to an ensemble average with respect to the initial microstates of the trajectories only. Consequently, the  microscopic generating function associated with any state function has the simple closed form
\begin{align}
\microgeneratingfunction(\countingfield_{\trajectoryobservablemicro},t) &= \sum\limits_{\microstate,\microstate'} \exp \big\lbrace - \countingfield_{\trajectoryobservablemicro} \, [ \trajectoryobservablemicro_{\microstate}(t) - \trajectoryobservablemicro_{\microstate'}(0) ] \big\rbrace \, \microprobability_{\microstate}(t) \, \microprobability_{\microstate'}(0). \label{eq:generatingfunctionequationofmotionstatefunctions}
\end{align}
Using Eqs. \eqref{eq:stochasticenergy} and \eqref{eq:stochasticentropy}, we have for the microscopic generating functions associated with the stochastic state-like observables energy and entropy,
\begin{align} \label{eq:microscopicgeneratingfunctionenergy} 
\microgeneratingfunction(\countingfield_{\trajectoryenergymicro},t) &= \sum\limits_{\microstate,\microstate'} \exp \big\lbrace - \countingfield_{\trajectoryenergymicro} \, [ \microenergy_{\microstate}(\bm{\protocol}_t) - \microenergy_{\microstate'}(\bm{\protocol}_0) ] \big\rbrace \, \microprobability_{\microstate}(t) \, \microprobability_{\microstate'}(0) \\
\microgeneratingfunction(\countingfield_{\trajectoryentropymicro},t) &= \sum\limits_{\microstate,\microstate'} \exp \big\lbrace \countingfield_{\trajectoryentropymicro} \big[ \ln \microprobability_{\microstate}(t) - \ln \microprobability_{\microstate'}(0) \big] \big\rbrace \, \microprobability_{\microstate}(t) \, \microprobability_{\microstate'}(0) 
. \label{eq:microscopicgeneratingfunctionsystementropy} 
\end{align}
Moreover, substituting Eqs. \eqref{eq:stochasticheat}, \eqref{eq:stochasticwork}, \eqref{eq:stochsticentropyflow} and \eqref{eq:stochasticentropyproductionrate} into Eq. \eqref{eq:generatingfunctionequationofmotion}, we obtain for the microscopic generating functions associated with the currents
\begin{align} \label{eq:microscopicgeneratingfunctionheat}
\D_t \microgeneratingfunction_{\microstate}(\countingfield_{\trajectoryheatmicro},t) &= \sum_{\reservoir=1}^{\reservoirdimension} \sum\limits_{\microstate'} \exp \big\lbrace - \countingfield_{\trajectoryheatmicro} \big[ \microenergy_{\microstate}(\bm{\protocol}_t) - \microenergy_{\microstate'}(\bm{\protocol}_t) - \bm{\force}^{(\reservoir)}_{\microstate \microstate'} \big] \big\rbrace \; \microrates^{(\reservoir)}_{\microstate \microstate'}(\bm{\protocol}_t) \;  \microgeneratingfunction_{\microstate'}(\countingfield_{\trajectoryheatmicro},t) \\ 
\D_t \microgeneratingfunction_{\microstate}(\countingfield_{\trajectoryworkmicro},t) &= - \countingfield_{\trajectoryworkmicro} \, \dot{\bm{\protocol}}_t \cdot \left[ \nabla_{\bm{\protocol}_t} \, \microenergy_{\microstate}(\bm{\protocol}_t) \right] \, \microgeneratingfunction_{\microstate}(\countingfield_{\trajectoryworkmicro},t) + \sum_{\reservoir=1}^{\reservoirdimension} \sum\limits_{\microstate'} \exp \big[ - \countingfield_{\trajectoryworkmicro} \, \bm{\force}^{(\reservoir)}_{\microstate \microstate'} \big] \; \microrates^{(\reservoir)}_{\microstate \microstate'}(\bm{\protocol}_t) \; \microgeneratingfunction_{\microstate'}(\countingfield_{\trajectoryworkmicro},t) \label{eq:microscopicgeneratingfunctionwork} \\ 
\D_t \microgeneratingfunction_{\microstate}(\countingfield_{\trajectoryentropyflowmicro},t) &= \sum_{\reservoir=1}^{\reservoirdimension} \sum\limits_{\microstate'} \exp \big\lbrace  \countingfield_{\trajectoryentropyflowmicro} \big[ \ln \microrates^{(\reservoir)}_{\microstate \microstate'}(\bm{\protocol}_t) - \ln \microrates^{(\reservoir)}_{\microstate' \microstate}(\bm{\protocol}_t)\big] \big\rbrace  \; \microrates^{(\reservoir)}_{\microstate \microstate'}(\bm{\protocol}_t) \; \microgeneratingfunction_{\microstate'}(\countingfield_{\trajectoryentropyflowmicro},t)  \label{eq:microscopicgeneratingfunctionentropyflow} \\ 
\D_t \microgeneratingfunction_{\microstate}(\countingfield_{\trajectoryepmicro},t) &= \countingfield_{\trajectoryepmicro} \, \frac{\D_t \microprobability_{\microstate}(t)}{\microprobability_{\microstate}(t)} \, \microgeneratingfunction_{\microstate}(\countingfield_{\trajectoryepmicro},t) + \sum_{\reservoir=1}^{\reservoirdimension} \sum\limits_{\microstate'} \exp \big\lbrace - \countingfield_{\trajectoryepmicro} \big[ \ln \microrates^{(\reservoir)}_{\microstate \microstate'}(\bm{\protocol}_t) \, \microprobability_{\microstate'}(t) - \ln \microrates^{(\reservoir)}_{\microstate' \microstate}(\bm{\protocol}_t) \microprobability_{\microstate}(t) \big] \big\rbrace \; \microrates^{(\reservoir)}_{\microstate \microstate'}(\bm{\protocol}_t) \; \microgeneratingfunction_{\microstate'}(\countingfield_{\trajectoryepmicro},t) .  \label{eq:microscopicgeneratingfunctionentropyproduction}
\end{align}

\subsubsection{Mesoscopic Description}
\label{sec:meogeneratingfunctions}

We rewrite the microscopic generating function \eqref{eq:generatingfunctiondefinition} as follows
\begin{align} \label{eq:mesoscopicgeneratingfunctiondefinition}\microgeneratingfunction_{\microstate_{_{\mesostate}}}(\countingfield_{\trajectoryobservablemicro},t) = \mesoprobability_{\mesostate}(t) \; \mathbb{\mesoprobability}_{\microstate_{_{\mesostate}}}(t) \, \langle \exp \big\lbrace - \countingfield_{\trajectoryobservablemicro} \, \delta \trajectoryobservablemicro[\trajectory_{(\tau)},t] \big\rbrace \rangle_{\microstate_{_{\mesostate}}} ,  
\end{align}
and define the mesososcopic generating function
\begin{align}
\mesogeneratingfunction_{\mesostate}(\countingfield_{_{\trajectoryobservablemeso}},t) \equiv \sum\limits_{\microstate_{\mesostate}} \microgeneratingfunction_{\microstate_{_{\mesostate}}}(\countingfield_{_{\trajectoryobservablemicro}},t) = \mesoprobability_{\mesostate}(t) \, \sum\limits_{\microstate_{\mesostate}} \mathbb{\mesoprobability}_{\microstate_{_{\mesostate}}}(t) \, \langle \exp \big\lbrace - \countingfield_{_{\trajectoryobservablemicro}} \, \delta o[\trajectory_{(\tau)},t] \big\rbrace \rangle_{\microstate_{_{\mesostate}}}  ,
\label{eq:mesoscopicgeneratingfunction}
\end{align}
where $\langle ~ \cdot ~ \rangle_{\microstate_{_{\mesostate}}} $ and $\langle ~ \cdot ~ \rangle_{\mesostate} $ denote ensemble averages over all trajectories that are in the microstate $\microstate$ belonging to a given mesostate $\mesostate$ and over all those that are in mesostate $\mesostate$ at time $t$, respectively. Moreover, $\trajectoryobservablemeso$ denotes a mesoscopic observable defined along a trajectory propagating in the mesoscopic state space, $\trajectorymeso_{(\tau)}$, which we write as $ \trajectoryobservablemeso[\trajectorymeso_{(\tau)},t]$ in the following.
Since the trajectory observables $\trajectoryobservablemicro = \trajectoryenergymicro,\trajectoryheatmicro,\trajectoryworkmicro,\trajectoryentropyflowmicro$ do not depend on microscopic information [cf. Eqs. \eqref{eq:stochasticenergy}--\eqref{eq:stochasticwork}], we have for those observables that $ \trajectoryobservablemicro[\trajectory_{(\tau)},t] =\trajectoryobservablemeso[\trajectorymeso_{(\tau)},t] $ and Eq. \eqref{eq:mesoscopicgeneratingfunction} closes as follows
\begin{align} 
\mesogeneratingfunction_{\mesostate}(\countingfield_{_{\trajectoryobservablemeso}},t) =  \mesoprobability_{\mesostate}(t) \, \sum\limits_{\microstate_{\mesostate}} \mathbb{\mesoprobability}_{\microstate_{_{\mesostate}}}(t) \, \langle \exp \big\lbrace - \countingfield_{_{\trajectoryobservablemeso}} \, \delta \trajectoryobservablemeso[\trajectorymeso_{(\tau)},t] \big\rbrace \rangle_{\mesostate} = \mesoprobability_{\mesostate}(t) \, \langle \exp \big\lbrace - \countingfield_{_{\trajectoryobservablemeso}} \, \delta \trajectoryobservablemeso[\trajectorymeso_{(\tau)},t] \big\rbrace \rangle_{\mesostate} , \quad \trajectoryobservablemeso = \trajectoryenergymeso,\trajectoryheatmeso,\trajectoryworkmeso,\trajectoryentropyflowmeso . 
\label{eq:mesoscopicgeneratingfunctionclosed}
\end{align}
Thus, the microscopic generating function for the energy \eqref{eq:microscopicgeneratingfunctionenergy} in mesoscopic representation reads
\begin{align}  \label{eq:mesoscopicgeneratingfunctionenergy}
\mesogeneratingfunction(\countingfield_{_{\trajectoryenergymeso}},t) &= \sum\limits_{\mesostate, \mesostate'} \exp \big\lbrace - \countingfield_{_{\trajectoryenergymeso}} \, [ \mesoenergy_{_{\mesostate}}(\bm{\protocol}_t) - \mesoenergy_{_{\mesostate'}}(\bm{\protocol}_0)] \big\rbrace \, \mesoprobability_{\mesostate}(t) \, \mesoprobability_{\mesostate'}(0)=\microgeneratingfunction(\countingfield_{\trajectoryenergymicro},t) ,
\end{align}
and from the microscopic equation of motion for the generating function \eqref{eq:generatingfunctionequationofmotion} we get
\begin{align}
\D_t \mesogeneratingfunction_{\mesostate}(\countingfield_{_{\trajectoryobservablemeso}},t) &= \sum\limits_{\mesostate'} \mesorates_{\mesostate \mesostate'}(\countingfield_{_{\trajectoryobservablemeso}},\bm{\protocol}_t) \; \mesogeneratingfunction_{\mesostate'}(\countingfield_{_{\trajectoryobservablemeso}},t) , \label{eq:mesoscopicgeneratingfunctionequationofmotion} 
\end{align}
with the mesoscopic biased generator
\begin{align}
\mesorates_{\mesostate \mesostate'}(\countingfield_{_{\trajectoryobservablemeso}},\bm{\protocol}_t) &=  - \countingfield_{_{\trajectoryobservablemeso}} \dot{ \trajectoryobservablemeso}_{\mesostate}(\bm{\protocol}_t) \, \delta_{\mesostate,\mesostate'} + \sum_{\reservoir=1}^{\reservoirdimension} \exp \big[ - \countingfield_{_{\trajectoryobservablemeso}} \, \trajectoryobservablemeso^{(\reservoir)}_{\mesostate,\mesostate'}(\bm{\protocol}_t) \big] \; \mesorates^{(\reservoir)}_{\mesostate \mesostate'}(\bm{\protocol}_t) ,  \label{eq:mesoscopicgeneratingfunctionbiasedgenerator} 
\end{align}
for $\trajectoryobservablemeso = \trajectoryenergymeso,\trajectoryheatmeso,\trajectoryworkmeso,\trajectoryentropyflowmeso$. More explicitly, Eqs. \eqref{eq:microscopicgeneratingfunctionheat}, \eqref{eq:microscopicgeneratingfunctionwork} and \eqref{eq:microscopicgeneratingfunctionentropyflow} can be rewritten in mesoscopic representation as follows
\begin{align} 
\D_t \mesogeneratingfunction_{\mesostate}(\countingfield_{_{\trajectoryheatmeso}},t) &= \sum_{\reservoir=1}^{\reservoirdimension} \sum\limits_{\mesostate'} \exp \big\lbrace - \countingfield_{_{\trajectoryheatmeso}} \, \big[ \mesoenergy_{\mesostate}(\bm{\protocol}_t) - \mesoenergy_{\mesostate'}(\bm{\protocol}_t) - \bm{\force}^{(\reservoir)}_{\mesostate \mesostate'} \big] \big\rbrace \; \mesorates^{(\reservoir)}_{\mesostate \mesostate'}(\bm{\protocol}_t) \; \mesogeneratingfunction_{\mesostate'}(\countingfield_{_{\trajectoryheatmeso}},t)\label{eq:mesoscopicgeneratingfunctionheat} \\
\D_t \mesogeneratingfunction_{\mesostate}(\countingfield_{_{\trajectoryworkmeso}},t) &= - \countingfield_{_{\trajectoryworkmeso}} \, \dot{\bm{\protocol}}_t \cdot \left[ \nabla_{\bm{\protocol}_t} \, \mesoenergy_{\mesostate}(\bm{\protocol}_t) \right] \mesogeneratingfunction_{\mesostate}(\countingfield_{_{\trajectoryworkmeso}},t) + \sum_{\reservoir=1}^{\reservoirdimension} \sum\limits_{\mesostate'} \exp \big[ - \countingfield_{_{\trajectoryworkmeso}} \, \bm{\force}^{(\reservoir)}_{\mesostate \mesostate'} \big] \; \mesorates^{(\reservoir)}_{\mesostate \mesostate'}(\bm{\protocol}_t) \; \mesogeneratingfunction_{\mesostate'}(\countingfield_{_{\trajectoryworkmeso}},t)  \label{eq:mesoscopicgeneratingfunctionwork} \\ 
\D_t \mesogeneratingfunction_{\mesostate}(\countingfield_{_{\trajectoryentropyflowmeso}},t) &= \sum_{\reservoir=1}^{\reservoirdimension} \sum\limits_{\mesostate'} \exp \Big\lbrace \countingfield_{_{\trajectoryentropyflowmeso}} \Big[ \ln \mesorates^{(\reservoir)}_{\mesostate \mesostate'}(\bm{\protocol}_t) - \ln \mesorates^{(\reservoir)}_{\mesostate' \mesostate}(\bm{\protocol}_t) - \big( \mesoentropy^{int}_{\mesostate} - \mesoentropy^{int}_{\mesostate'} \big) \Big] \Big\rbrace \;  \mesorates^{(\reservoir)}_{\mesostate \mesostate'}(\bm{\protocol}_t) \; \mesogeneratingfunction_{\mesostate'}(\countingfield_{_{\trajectoryentropyflowmeso}},t) . \label{eq:mesoscopicgeneratingfunctionentropyflow} 
\end{align}
It is easy to verify that $\sum_{\mesostate} \D_t \mesogeneratingfunction_{\mesostate}(\countingfield_{_{\trajectoryobservablemeso}},t) = \sum_{\microstate} \D_t \microgeneratingfunction_{\microstate}(\countingfield_{\trajectoryobservablemicro},t)$ for $\trajectoryobservablemeso = \trajectoryenergymeso,\trajectoryheatmeso,\trajectoryworkmeso,\trajectoryentropyflowmeso$ and $\trajectoryobservablemicro = \trajectoryenergymicro,\trajectoryheatmicro,\trajectoryworkmicro,\trajectoryentropyflowmicro $. Thus, we find that the statistics of the stochastic first law in microscopic representation \eqref{eq:stochasticfirstlaw} is invariant under coarse-graining.

Conversely, the stochastic system entropy \eqref{eq:stochasticentropy} and stochastic entropy production rate \eqref{eq:stochasticentropyproductionrate} are functions of the microscopic ensemble probability. The corresponding equation for the mesoscopic generating function \eqref{eq:mesoscopicgeneratingfunction} would, in general, not be closed and the stochastic entropy balance in microscopic representation \eqref{eq:stochasticentropybalance} is, in general, not invariant under the coarse-graining.
However, an exact coarse-graining is possible whenever the microscopic probabilities are uniform within each mesostate, i.e. \eqref{eq:stationarymicroprobabilitiesTdep} holds. In this case the mesoscopic generating functions associated with the system entropy and entropy production rate read, respectively
\begin{align} 
\mesogeneratingfunction(\countingfield_{_{\trajectoryentropymeso}},t) &= \sum\limits_{\mesostate,\mesostate'} \exp \big\lbrace \countingfield_{_{\trajectoryentropymeso}} \big[ \ln \mesoprobability_{\mesostate}(t) - \ln \mesoprobability^{sp}_{\mesostate'}(0) - \big( \mesoentropy^{int}_{\mesostate}  - \mesoentropy^{int}_{\mesostate'} \big) \big] \big\rbrace \, \mesoprobability_{\mesostate}(t) \mesoprobability_{\mesostate'}^{sp}(0) =  \microgeneratingfunction(\countingfield_{\trajectoryentropymicro},t) \label{eq:mesogeneratingfunctionentropy}  \\ 
\D_t \mesogeneratingfunction_{\mesostate}(\countingfield_{_{\trajectoryepmeso}},t) &=  \countingfield_{_{\trajectoryepmeso}} \frac{\D_t \mesoprobability_{\mesostate}(t)}{\mesoprobability_{\mesostate}(t)} \, \mesogeneratingfunction_{\mesostate}(\countingfield_{_{\trajectoryepmeso}},t) + \sum_{\reservoir=1}^{\reservoirdimension} \sum\limits_{\mesostate'}  \exp \Big\lbrace - \countingfield_{_{\trajectoryepmeso}} \ln \frac{ \mesorates^{(\reservoir)}_{\mesostate \mesostate'}(\bm{\protocol}_t) \, \mesoprobability_{\mesostate'}(t) }{ \mesorates^{(\reservoir)}_{\mesostate' \mesostate}(\bm{\protocol}_t) \, \mesoprobability_{\mesostate}(t) } \Big\rbrace \, \mesorates^{(\reservoir)}_{\mesostate \mesostate'}(\bm{\protocol}_t) \; \mesogeneratingfunction_{\mesostate'}(\countingfield_{_{\trajectoryepmeso}},t) ,  \label{eq:mesogeneratingfunctionentropyproduction} 
\end{align}
with $\sum_{\mesostate} \D_t \mesogeneratingfunction_{\mesostate}(\countingfield_{_{\trajectoryepmeso}},t) = \sum_{\microstate} \D_t \microgeneratingfunction_{\microstate}(\countingfield_{\trajectoryepmicro},t) $.
Hence we conclude that the statistics of the stochastic entropy balance \eqref{eq:stochasticentropybalance} is invariant under the coarse-graining, if one considers initial conditions which are uniform within each mesostate, or for systems in stationary states.

Comparing Eqs. \eqref{eq:microscopicgeneratingfunctionenergy}, \eqref{eq:microscopicgeneratingfunctionheat} and \eqref{eq:microscopicgeneratingfunctionwork} with Eqs. \eqref{eq:mesoscopicgeneratingfunctionenergy}, \eqref{eq:mesoscopicgeneratingfunctionheat} and \eqref{eq:mesoscopicgeneratingfunctionwork}, we note that the evolution of the generating functions associated with the first-law observables, that is energy, heat and work, have the same form in microscopic and mesoscopic representation. In contrast, the mesoscopic generating functions associated with the entropies do not have the same form as the microscopic ones but also contain the internal entropy $\mesoentropy^{int}$. This is due to the coarse-grained degrees of freedom that give rise to Boltzmann entropies \eqref{eq:internalentropy} assigned to the mesostates.
Physically, the conditions for the invariance of the stochastic entropy balance [Eqs. \eqref{eq:mesogeneratingfunctionentropy} and \eqref{eq:mesogeneratingfunctionentropyproduction}] can be understood as follows. If the microscopic degrees of freedom inside the mesostates are not equiprobable, there are transient microscopic currents that can not be captured at the mesoscopic level and which only vanish identically once the uniform probability distributions inside the mesostates are achieved.

So far, we have established two descriptions of the stochastic thermodynamics at the microscopic and mesoscopic level. These two formulations are equivalent for the stochastic first law. In case of the stochastic entropy balance, the microscopic and mesoscopic thermodynamics coincide under the condition that the microstates inside each mesostate are equiprobable. The thermodynamics consistency at each level is ensured by the respective local detailed balance conditions in Eqs. \eqref{eq:microlocaldetailedbalance} and \eqref{eq:mesolocaldetailedbalance}. Alternatively, the thermodynamic consistency is also encoded by the so-called detailed fluctuation theorem for the stochastic entropy production. In the following, we will discuss this symmetry of the fluctuations of the entropy production as it will be of importance further below.

\subsection{Detailed Fluctuation Theorems Across Scales}

Let us consider a forward process that starts from a state that is at equilibrium with respect to the reference reservoir $\reservoir=1$,
\begin{align} \label{eq:initialequilibriummicro}
\microprobability_{\microstate_{_0}}^{eq}(\bm{\protocol}_0) = \exp \big\lbrace - \invtemperature^{(1)} [ \microenergy_{\microstate_{_0}}(\bm{\protocol}_0)-\microfreeenergy^{eq}(\bm{\protocol}_0)] \big\rbrace \, . 
\end{align}
The system then evolves under the driven microscopic Markov process according to the forward protocol $\bm{\protocol}_{t'}, \, t' ~\in ~ [0,t]$.
For the backward process, indicated by the notation ``$\tilde{\phantom{w}}$'', the system is initially prepared in the final equilibrium state of the forward process
\begin{align} \label{eq:finalequilibriummicro}
\microprobability_{\microstate_{_{\size}}}^{eq}(\bm{\protocol}_t) = \exp \big\lbrace - \invtemperature^{(1)} [ \microenergy_{\microstate_{_{\size}}}(\bm{\protocol}_t) - \microfreeenergy^{eq}(\bm{\protocol}_t)] \big\rbrace \, ,
\end{align}
and subsequently evolves under the time-reversed driven microscopic Markov process according to the backward protocol $\tilde{\bm{\protocol}}_{t'} = \bm{\protocol}_{t-t'}, \, t' ~\in ~ [0,t]$.
Then, the following microscopic finite-time detailed fluctuation theorem ensues \cite{cuetara2014pre,rao2018njp}
\begin{align} \label{eq:microdft} 
\ln \frac{ \microprobability \Big( \invtemperature^{(1)} \delta \trajectoryworkmicro_{\bm{\protocol}} + \sum\limits_{\reservoir = 1}^{\reservoirdimension} \big[ \invtemperature^{(\reservoir)} \delta \trajectoryworkmicro_{\bm{\force}}^{(\reservoir)} + \big[ \invtemperature^{(1)} - \invtemperature^{(\reservoir)} \big] \delta \trajectoryenergymicro^{(\reservoir)}  \big] \Big) }{
\tilde{\microprobability} \Big( -\invtemperature^{(1)} \delta \trajectoryworkmicro_{\bm{\protocol}} - \sum\limits_{\reservoir = 1}^{\reservoirdimension} \big[ \invtemperature^{(\reservoir)} \delta \trajectoryworkmicro_{\bm{\force}}^{(\reservoir)} + \big[ \invtemperature^{(1)} - \invtemperature^{(\reservoir)} \big] \delta \trajectoryenergymicro^{(\reservoir)}  \big] \Big) } = \invtemperature^{(1)} \big[ \delta \trajectoryworkmicro_{\bm{\protocol}} -  \Delta \microfreeenergy^{eq}_1 \big] + \sum\limits_{\reservoir = 1}^{\reservoirdimension} \big[ \invtemperature^{(\reservoir)} \delta \trajectoryworkmicro_{\bm{\force}}^{(\reservoir)} + \big[ \invtemperature^{(1)} - \invtemperature^{(\reservoir)} \big] \delta \trajectoryenergymicro^{(\reservoir)}  \big] , 
\end{align}
where $\Delta \microfreeenergy^{eq}_1 = \microfreeenergy^{eq}_1(\bm{\protocol}_t) - \microfreeenergy^{eq}_1(\bm{\protocol}_0) $ denotes the change in global microscopic equilibrium free energy with respect to the reservoir $\reservoir = 1$ along the forward process that only depends on the initial and final value of the driving protocol and thus does not fluctuate.

In fact, this microscopic finite-time detailed fluctuation theorem also holds for the joint probability distribution,
\begin{align} \label{eq:microdftjoint}
\ln \frac{ \microprobability \Big( \invtemperature^{(1)} \delta \trajectoryworkmicro_{\bm{\protocol}} \, , \, \lbrace \delta \microcurrent^{(\reservoir)}_{\bm{\force}} \rbrace \, , \, \lbrace \delta \microcurrent^{(\reservoir)}_{\trajectoryenergymicro} \rbrace \Big) }{
\tilde{\microprobability} \Big( - \invtemperature^{(1)} \delta \trajectoryworkmicro_{\bm{\protocol}} \, , \, \lbrace - \delta \microcurrent^{(\reservoir)}_{\bm{\force}} \rbrace \, , \, \lbrace - \delta \microcurrent^{(\reservoir)}_{\trajectoryenergymicro} \rbrace \Big) }  = \invtemperature^{(1)} \big[ \delta \trajectoryworkmicro_{\bm{\protocol}} - \Delta \microfreeenergy^{eq}_1 \big] + \sum\limits_{\reservoir = 1}^{\reservoirdimension} \big[ \invtemperature^{(\reservoir)} \delta \trajectoryworkmicro_{\bm{\force}}^{(\reservoir)} + \big[ \invtemperature^{(1)} - \invtemperature^{(\reservoir)} \big]  \delta \trajectoryenergymicro^{(\reservoir)}  \big] , 
\end{align}
where we write the time-integrated microscopic autonomous work currents as $ \lbrace \delta \microcurrent^{(\reservoir)}_{\bm{\force}} \rbrace \equiv \big( \, \invtemperature^{(1)} \delta \trajectoryworkmicro_{\bm{\force}}^{(1)}, \ldots ,  \invtemperature^{(\reservoirdimension)} \delta \trajectoryworkmicro_{\bm{\force}}^{(\reservoirdimension)} )$ and the time-integrated microscopic energy currents as $ \lbrace \delta \microcurrent^{(\reservoir)}_{\trajectoryenergymicro} \rbrace \equiv ( [ \invtemperature^{(1)} - \invtemperature^{(2)} ] \delta \trajectoryenergymicro^{(2)} , \ldots, [\invtemperature^{(1)} - \invtemperature^{(\reservoirdimension)} ] \delta \trajectoryenergymicro^{(\reservoirdimension)} \, \big)$. Here, $\microprobability( \invtemperature^{(1)} \delta \trajectoryworkmicro_{\bm{\protocol}} , \lbrace \delta \microcurrent^{(\reservoir)}_{\bm{\force}} \rbrace , \lbrace \delta \microcurrent^{(\reservoir)}_{\trajectoryenergymicro} \rbrace )$ is the probability to observe a microscopic nonautonomous work $\invtemperature^{(1)} \delta \trajectoryworkmicro_{\bm{\protocol}}$, the microscopic time-integrated autonomous work currents $\lbrace \delta \microcurrent^{(\reservoir)}_{\bm{\force}} \rbrace $ and the microscopic time-integrated energy currents $\lbrace \delta \microcurrent^{(\reservoir)}_{\trajectoryenergymicro} \rbrace $ along the forward process in the microscopic state space.
Conversely, $\tilde{\microprobability}( - \invtemperature^{(1)} \delta \trajectoryworkmicro_{\bm{\protocol}} , \lbrace - \delta \microcurrent^{(\reservoir)}_{\bm{\force}} \rbrace , \lbrace - \delta \microcurrent^{(\reservoir)}_{\trajectoryenergymicro} \rbrace )$ is the probability to observe a microscopic nonautonomous work $ - \invtemperature^{(1)} \delta \trajectoryworkmicro_{\bm{\protocol}}$, the microscopic time-integrated autonomous work currents $\lbrace - \delta \microcurrent^{(\reservoir)}_{\bm{\force}} \rbrace $ and the microscopic time-integrated energy currents $ \lbrace - \delta \microcurrent^{(\reservoir)}_{\trajectoryenergymicro} \rbrace $ along the time-reversed backward process in the microscopic state space.

The validity of the last equation can be seen by marginalizing its l.h.s. which gives the l.h.s. of Eq. \eqref{eq:microdft}. Equation \eqref{eq:microdftjoint} can also be derived via the following symmetry of the associated microscopic generating function
\begin{align} \label{eq:microgeneratingfunctionsymmetry} 
\microgeneratingfunction \big( \countingfield_{_{\bm{\protocol}}}, \lbrace \countingfield_{_{\bm{\force}}}^{(\reservoir)} \rbrace , \lbrace \countingfield_{_{\trajectoryenergymicro}}^{(\reservoir)} \rbrace ,t \big) = \tilde{\microgeneratingfunction} \big(1 - \countingfield_{_{\bm{\protocol}}}, \lbrace 1 - \countingfield_{_{\bm{\force}}}^{(\reservoir)} \rbrace , \lbrace 1 - \countingfield_{_{\trajectoryenergymicro}}^{(\reservoir)} \rbrace ,t \big) \; \exp \big[ - \invtemperature^{(1)} \Delta \microfreeenergy^{eq}_1(\bm{\protocol}) \big] ,
\end{align}
as demonstrated in appendix \ref{sec:dftappendix}.

Analogously, we can define the forward and backward process as above also in the mesoscopic state space. In this case, the equilibrum distributions for the forward and, in reversed order for the backward trajectory, read, respectively
\begin{align} \label{eq:initialequilibriumeso} 
\mesoprobability_{\mesostate_{_0}}^{eq}(\bm{\protocol}_0) &= \exp \big\lbrace - \invtemperature^{(1)} [ \mesofreeenergy_{\mesostate_{_0}}(\bm{\protocol}_0) - \mesofreeenergy^{eq}(\bm{\protocol}_0)] \big\rbrace  \\
\mesoprobability_{\mesostate_{_{\size}}}^{eq}(\bm{\protocol}_t) &= \exp \big\lbrace - \invtemperature^{(1)} [ \mesofreeenergy_{\mesostate_{_{\size}}}(\bm{\protocol}_t) - \mesofreeenergy^{eq}(\bm{\protocol}_t)] \rbrace \, .  \label{eq:finalequilibriummeso} 
\end{align}
Crucially, all fluctuating quantities appearing in the microscopic detailed fluctuation theorem \eqref{eq:microdftjoint} are invariant under the dynamically exact coarse-graining \eqref{eq:mesoscopicmasterequation}. Consequently, the symmetry for the microscopic generating function \eqref{eq:microgeneratingfunctionsymmetry} is also exhibited at the mesoscopic level, 
\begin{align} \label{eq:mesogeneratingfunctionsymmetry} 
\mesogeneratingfunction( \bm{\countingfield} ,t) = \tilde{\mesogeneratingfunction}(\tilde{\bm{\countingfield}} ,t) \; \exp \big[ - \invtemperature^{(1)} \Delta \mesofreeenergy^{eq}_1(\bm{\protocol}) \big] , 
\end{align}
where $\Delta \mesofreeenergy^{eq}_1 = \mesofreeenergy^{eq}_1(\bm{\protocol}_t) - \mesofreeenergy^{eq}_1(\bm{\protocol}_0) $. Moreover, for brevity we introduced the notation
\begin{align} \label{eq:countingfieldnotation}
\bm{\countingfield} \equiv \countingfield_{_{\bm{\lambda}}}, \lbrace \countingfield_{_{\bm{\force}}}^{(\reservoir)} \rbrace , \lbrace \countingfield_{\trajectoryenergymeso}^{(\reservoir)} \rbrace, \quad  \tilde{\bm{\countingfield}} \equiv 1- \countingfield_{_{\bm{\lambda}}}, \lbrace 1 - \countingfield_{_{\bm{\force}}}^{(\reservoir)} \rbrace , \lbrace 1 - \countingfield_{\trajectoryenergymeso}^{(\reservoir)} \rbrace , 
\end{align}
and the mesoscopic time-integrated autonomous work currents $ \lbrace \delta \mesocurrent^{(\reservoir)}_{\bm{\force}} \rbrace \equiv \big( \, \invtemperature^{(1)} \delta \trajectoryworkmeso_{\bm{\force}}^{(1)}, \ldots , \invtemperature^{(\reservoirdimension)} \delta \trajectoryworkmeso_{\bm{\force}}^{(\reservoirdimension)} )$ as well as the mesoscopic time-integrated energy currents $ \lbrace \delta \mesocurrent^{(\reservoir)}_{\trajectoryenergymeso} \rbrace ~\equiv~ ( [ \invtemperature^{(1)} - \invtemperature^{(2)} ] \delta \trajectoryenergymeso^{(2)} , \ldots , [\invtemperature^{(1)} - \invtemperature^{(\reservoirdimension)} ]   \delta \trajectoryenergymeso^{(\reservoirdimension)} \, \big)$. Thus, the detailed fluctuation theorem \eqref{eq:microdftjoint} also holds at the mesoscopic level,
\begin{align} \label{eq:mesodftjoint}
\ln \frac{ \mesoprobability \Big( \invtemperature^{(1)} \delta \trajectoryworkmeso_{\bm{\protocol}} \, , \, \lbrace \delta \mesocurrent^{(\reservoir)}_{\bm{\force}} \rbrace \, , \, \lbrace \delta \mesocurrent^{(\reservoir)}_{\trajectoryenergymeso} \rbrace \Big) }{
\tilde{\mesoprobability} \Big( - \invtemperature^{(1)} \delta \trajectoryworkmeso_{\bm{\protocol}} \, , \, - \lbrace \delta \mesocurrent^{(\reservoir)}_{\bm{\force}} \rbrace \, , \, - \lbrace \delta \mesocurrent^{(\reservoir)}_{\trajectoryenergymeso} \rbrace \Big) }  = \invtemperature^{(1)} \big[ \delta \trajectoryworkmeso_{\bm{\protocol}} - \Delta \mesofreeenergy^{eq}_1 \big] + \sum\limits_{\reservoir = 1}^{\reservoirdimension} \big[ \invtemperature^{(\reservoir)} \delta \trajectoryworkmeso_{\bm{\force}}^{(\reservoir)} + \big[ \invtemperature^{(1)} - \invtemperature^{(\reservoir)} \big] \delta \trajectoryenergymeso^{(\reservoir)}\big] ,
\end{align}
where $\mesoprobability \Big( \invtemperature^{(1)} \delta \trajectoryworkmeso_{\bm{\protocol}} \, , \, \lbrace \delta \mesocurrent^{(\reservoir)}_{_{\bm{\force}}} \rbrace \, , \, \lbrace \delta \mesocurrent^{(\reservoir)}_{\trajectoryenergymeso} \rbrace \Big)$ is the probability to observe a mesoscopic nonautonomous work $\invtemperature^{(1)} \delta \trajectoryworkmeso_{\bm{\protocol}}$, the mesoscopic time-integrated autonomous work currents $\lbrace \delta \mesocurrent^{(\reservoir)}_{\bm{\force}} \rbrace $ and the mesoscopic time-integrated energy currents $\lbrace \delta \mesocurrent^{(\reservoir)}_{\trajectoryenergymeso} \rbrace $ along the forward process in the mesoscopic state space. Conversely, $\tilde{\mesoprobability} \Big( - \invtemperature^{(1)} \delta \trajectoryworkmeso_{\bm{\protocol}} \, , \, \lbrace - \delta \mesocurrent^{(\reservoir)}_{\bm{\force}} \rbrace \, , \, \lbrace - \delta \mesocurrent^{(\reservoir)}_{\trajectoryenergymeso} \rbrace \Big)$ is the probability to observe a mesoscopic nonautonomous work $- \invtemperature^{(1)} \delta \trajectoryworkmeso_{\bm{\protocol}}$, the mesoscopic time-integrated autonomous work currents $ \lbrace - \delta \mesocurrent^{(\reservoir)}_{_{\bm{\force}}} \rbrace $ and the mesoscopic time-integrated energy currents $ \lbrace - \delta \mesocurrent^{(\reservoir)}_{\trajectoryenergymeso} \rbrace $ along the time-reversed backward process in the mesoscopic state space.

\begin{figure}[h!]
\begin{center} 

\includegraphics[scale=1]{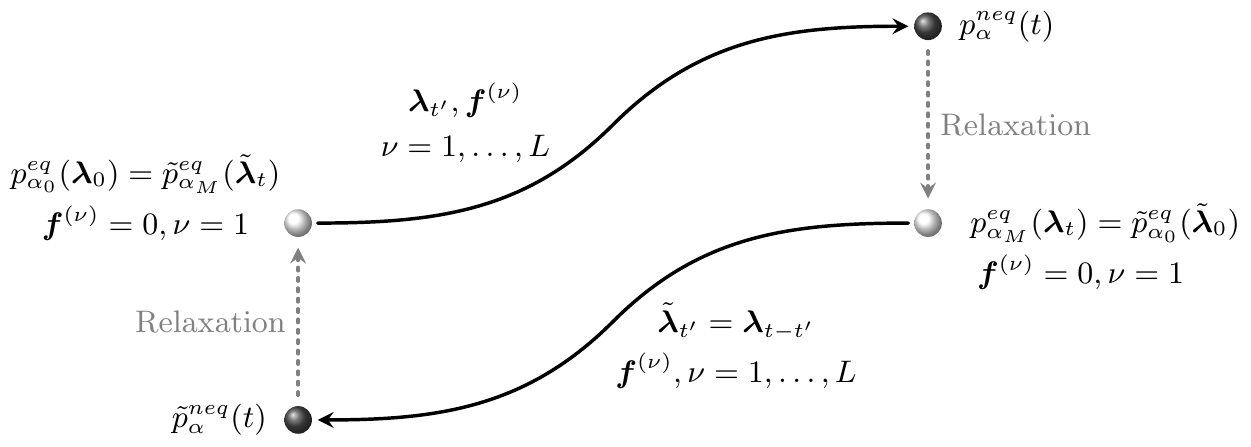}

\caption[Schematic depcition of the backward and forward process]{Schematic representation of the backward and forward process. \label{fig:representationforwardbackwardprocess} }
\end{center}
\end{figure}

Having stated the various detailed fluctuation theorems across scales, we now proceed to show that the latter are relations for the entropy production of the forward processes including the relaxation from the nonequilibrium state at time $t$ towards the final equilibrium state of the forward process \eqref{eq:finalequilibriummicro} which coincides with the initial equilibrium state of the backward process.
First, we note that initial state \eqref{eq:initialequilibriummicro} can be prepared by disconnecting all other heat reservoirs, fixing the protocol at value $\bm{\protocol}_0$ and letting the system relax. At time $t' = 0$, all other heat reservoirs are simultaneously connected to the system and both the nonconservative $\bm{\force}^{(\reservoir)}$ and the nonautonomous driving is switched on. As a result, the system evolves under the driven microscopic Markov process according to the forward protocol $\bm{\protocol}_{t'}, \, t' ~\in ~ [0,t]$ towards a nonequilibrium state $\microprobability_{\microstate}^{neq}(t)$. During that evolution heat $\delta \trajectoryheatmicro_{\bm{\protocol}}^{(\reservoir)}[\trajectory_{(\tau)},t]$ is exchanged between the system and the reservoirs $\reservoir$. There is furthermore autonomous $\delta \trajectoryworkmicro_{\bm{\force}}^{(\reservoir)}[\trajectory_{(\tau)},t]$ work done on or by the system as well as nonautonomous work $\delta \trajectoryworkmicro_{\bm{\protocol}}[\trajectory_{(\tau)},t]$ performed on the system by the external driving to change its energy landscape $\microenergy_{\microstate}(\bm{\protocol}_t')$. At time $t$, all heat reservoirs but the reference one $\reservoir =1$ are disconnected, the driving parameter is kept constant at its final value $\bm{\protocol}_t$ and the nonconservative force $\bm{\force}$ is switched off such that the system relaxes into the equilibrium state \eqref{eq:finalequilibriummicro}. The preparation of the starting and ending distribution of the backward process is analogous. The forward and backward process are illustrated in Fig. \ref{fig:representationforwardbackwardprocess}.

\noindent 


Using Eqs. \eqref{eq:timeintegratedstochasticfirstlaw}--\eqref{eq:stochasticworkdecomposition}, \eqref{eq:initialequilibriummicro} and \eqref{eq:finalequilibriummicro}, the fluctuating entropy production \eqref{eq:stochasticentropyproduction} along the forward process can be rewritten as follows
\begin{align}
\delta \trajectoryepmicro[\trajectory_{(\tau)},t] = \invtemperature^{(1)} \big[ \delta \trajectoryworkmicro_{\bm{\protocol}}[\trajectory_{(\tau)},t] - \Delta \microfreeenergy^{eq}_1 \big] + \sum_{\reservoir = 1}^{\reservoirdimension} \big\lbrace  \invtemperature^{(\reservoir)} \delta \trajectoryworkmicro_{\bm{\force}}^{(\reservoir)}[\trajectory_{(\tau)},t] + \big[ \invtemperature^{(1)} - \invtemperature^{(\reservoir)} \big] \delta \trajectoryenergymicro^{(\reservoir)}[\trajectory_{(\tau)},t] \big\rbrace , \label{eq:stochasticentropyproductionequilibriumboundary}
\end{align}
which is exactly the r.h.s of the detailed fluctuation theorem \eqref{eq:microdft}. Thus, the detailed fluctuation theorem \eqref{eq:microdft} is a symmetry relation for the entropy production not only from initial time $0$ to the time of the final protocol value $t$ but including also the relaxation contribution from time $t$ until the final equilibrium distribution is attained. Though, it is a \emph{finite-time} relation since all fluctuating quantities in the entropy production of the forward process \eqref{eq:stochasticentropyproductionequilibriumboundary} stop evolving at time $t$ and do thus not contribute of the statistics of the following relaxation process.


We want to stress that the existence of the detailed fluctuation theorem for the entropy production across scales \eqref{eq:microdftjoint}, \eqref{eq:mesodftjoint} ensures that the thermodynamics formulated at each of these levels is consistent. We will make use of this result further below when we formulate the fluctuations at the macroscopic level, that is fluctuations that scale exponentially with the system size $\dimension$.

\subsection{Microscopic And Mesoscopic First And Second Law}
\label{sec:micromesofirstsecondlaw}

Before turning to the macroscopic limit, for completeness, we want to formulate the thermodynamics at the ensemble level on microscopic and mesoscopic scales and hereby, because of their importance, focus on the laws of thermodynamics. Using Eq. \eqref{eq:generatingfunctiondifferentiation} and Eqs. \eqref{eq:microscopicgeneratingfunctionenergy} -- \eqref{eq:microscopicgeneratingfunctionwork} or Eqs. \eqref{eq:mesoscopicgeneratingfunctionenergy} -- \eqref{eq:mesoscopicgeneratingfunctionwork}, we arrive at the microscopic or mesoscopic first law of thermodynamics, respectively,
\begin{align} \label{eq:firstlaw} 
\d_t \microenergy(t) = \dot{\microheat}(t) + \dot{\microwork}(t) , \quad \d_t \mesoenergy(t) = \dot{\mesoheat}(t) + \dot{\mesowork}(t) ,
\end{align}
with the average internal energy that is equivalent at microscopic and mesoscopic scale,
\begin{align}  \label{eq:averageenergy} 
\microenergy(t) = \sum\limits_{\microstate} \microenergy_{\microstate}(\bm{\protocol}_t) \; \microprobability_{\microstate}(t) = \sum\limits_{\mesostate } \mesoenergy_{\mesostate}(\bm{\protocol}_t) \; \mesoprobability_{\mesostate}(t) = \mesoenergy(t) .
\end{align} 
and with the equivalent microscopic and mesoscopic heat currents
\begin{align}  \label{eq:firstlawheat}
\dot{\microheat}(t) &= \sum\limits_{\reservoir=1}^{\reservoirdimension} \sum\limits_{\microstate, \microstate'} \left[ \microenergy_{\microstate}(\bm{\protocol}_t) - \microenergy_{\microstate'}(\bm{\protocol}_t) - \bm{\force}^{(\reservoir)}_{\microstate \microstate'} \right] \; \microrates_{\microstate \microstate'}^{(\reservoir)}(\bm{\protocol}_t) \; \microprobability_{\microstate'}(t) \\ 
&= \sum\limits_{\reservoir=1}^{\reservoirdimension} \sum\limits_{\mesostate, \mesostate'} \left[ \mesoenergy_{\mesostate}(\bm{\protocol}_t) - \mesoenergy_{\mesostate'}(\bm{\protocol}_t) - \bm{\force}^{(\reservoir)}_{\mesostate \mesostate'} \right] \; \mesorates_{\mesostate \mesostate'}^{(\reservoir)}(\bm{\protocol}_t) \; \mesoprobability_{\mesostate'}(t) = \dot{\mesoheat}(t) \nonumber 
\end{align}
as well as the equivalent microscopic and mesoscopic work currents
\begin{align}  \label{eq:firstlawwork}
\dot{\microwork}(t) &= \sum_{\microstate} \Big[ \dot{\bm{\protocol}}_t \cdot \left[ \nabla_{\bm{\protocol}_t} \, \mesoenergy_{\microstate}(\bm{\protocol}_t) \right] \,  \microprobability_{\microstate}(t) + \sum\limits_{\reservoir=1}^{\reservoirdimension} \sum\limits_{\microstate'} \bm{\force}^{(\reservoir)}_{\microstate \microstate'} \; \microrates_{\microstate,\microstate'}^{(\reservoir)}(\bm{\protocol}_t) \; \microprobability_{\microstate'}(t) \Big]  \\
&= \sum_{\mesostate} \! \Big[ \dot{\bm{\protocol}}_t \cdot \left[ \nabla_{\bm{\protocol}_t} \, \mesoenergy_{\mesostate}(\bm{\protocol}_t) \right] \mesoprobability_{\mesostate}(t) + \sum\limits_{\reservoir=1}^{\reservoirdimension} \sum\limits_{\mesostate'} \bm{\force}^{(\reservoir)}_{\mesostate \mesostate'} \, \mesorates_{\mesostate \mesostate'}^{(\reservoir)}(\bm{\protocol}_t) \, \mesoprobability_{\mesostate'}(t) \Big] = \dot{\mesowork}(t) \nonumber . 
\end{align} 
Next, with Eq. \eqref{eq:mesogeneratingfunctionentropy}, which only holds for equiprobable microstates inside the mesostates \eqref{eq:stationarymicroprobabilitiesTdep}, we find the equivalence of the average system entropy at microscopic and mesoscopic scale,
\begin{align} \label{eq:averageentropy}
\microentropy(t) = - \sum\limits_{\microstate} \microprobability_{\microstate}(t) \, \ln \microprobability_{\microstate}(t) = \sum\limits_{\mesostate} \left[ \mesoentropy^{int}_{\mesostate} - \ln  \mesoprobability_{\mesostate}(t) \right]  \, \mesoprobability_{\mesostate}(t) = \mesoentropy(t).
\end{align}
Using furthermore Eq. \eqref{eq:mesogeneratingfunctionentropyproduction}, which only holds for equiprobable microstates inside the mesostates \eqref{eq:stationarymicroprobabilitiesTdep}, the microscopic and mesoscopic second law of thermodynamics reads
\begin{align} \label{eq:averageentropyproduction}
\dot{\microep}(t) = \sum\limits_{\reservoir=1}^{\reservoirdimension} \sum\limits_{\microstate,\microstate'} \ln \frac{ \microrates^{(\reservoir)}_{\microstate \microstate'}(\bm{\protocol}_t) \, \microprobability_{\microstate'}(t)}{\microrates^{(\reservoir)}_{\microstate' \microstate}(\bm{\protocol}_t) \, \microprobability_{\microstate}(t)} \, \microrates^{(\reservoir)}_{\microstate \microstate'}(\bm{\protocol}_t) \, \microprobability_{\microstate'}(t) = \sum\limits_{\reservoir=1}^{\reservoirdimension} \sum\limits_{\mesostate,\mesostate'} \ln \frac{ \mesorates^{(\reservoir)}_{\mesostate \mesostate'}(\bm{\protocol}_t) \, \mesoprobability_{\mesostate'}(t)}{\mesorates^{(\reservoir)}_{\mesostate' \mesostate}(\bm{\protocol}_t) \, \mesoprobability_{\mesostate}(t)} \, \mesorates^{(\reservoir)}_{\mesostate \mesostate'}(\bm{\protocol}_t) \, \mesoprobability_{\mesostate'}(t) = \dot{\mesoep}(t) \geq 0 .
\end{align}

\section{Macroscopic Theory }
\label{sec:martinsiggiarose}
\subsection{Macroscopic Fluctuations}

Thus far, we have established two equivalent representations of the stochastic dynamics above, the microscopic and mesoscopic representation. We furthermore identified the conditions under which the thermodynamics at these levels coincide. In this section, the question of how to infer the fluctuations in the macroscopic limit, $\dimension \to \infty$, will be addressed. To shed light on this question, we will employ the Martin-Siggia-Rose formalism \cite{rosepra1973,weber2017rpp} which equivalently represents the Markovian jump process via a path integral. As will be demonstrated in the following, this path-integral formalism allows to establish a fluctuating description valid at macroscopic scales in the large deviation sense \cite{touchette2009pr}, that is for fluctuations that scale exponentially with the number of units $\dimension$.

For better readability, we omit a detailed presentation of the elementary concepts underlying the construction of the path integral and refer to Refs. \cite{ritort2004jsm,suarez1995jcp,lazarescu2019jcp,verley2019} where this formalism has been used in a thermodynamic context. The mesoscopic generating function $\mesogeneratingfunction(\countingfield_{_{\trajectoryobservablemeso}}, t)$ associated with a mesoscopic stochastic observable $\trajectoryobservablemeso[\trajectorymeso_{(\tau)},t] $ within the path integral representation generically reads
\begin{align}
\begin{aligned}
\mesogeneratingfunction(\countingfield_{_{\trajectoryobservablemeso}}, t) &= \!\! \int \!\! \pathintegralmeasure [\mesostate] \! \int \!\! \pathintegralmeasure [\bm{\field}]  \; \exp \! \Bigg\lbrace \! \int\limits_0^t \d t' \Big[ \!-\! \bm{\field}(t') \!\cdot \! \dot{\mesostate}(t') \!+\! \hamiltonian_{\countingfield_{_{\trajectoryobservablemeso}}}[\mesostate(t'),\bm{\field}(t')] \!-\! \countingfield_{_{\trajectoryobservablemeso}} \dot{\bm{\protocol}}_{t'} \! \cdot \! \left[ \nabla_{\bm{\protocol}_{t'}} \! \trajectoryobservablemeso_{\mesostate(t')} \right] \!-\! \countingfield_{_{\trajectoryobservablemeso}} \d_{t'} \trajectoryobservablemeso_{\mesostate(t')} \Big] \Bigg\rbrace \mesoprobability_{\mesostate}(0) \\
&\equiv \!\! \int \!\! \pathintegralmeasure [\mesostate] \! \int \!\! \pathintegralmeasure [\bm{\field}] \; \exp \Big\lbrace  \action_{\countingfield_{_{\trajectoryobservablemeso}}}[\mesostate(t'),\bm{\field}(t')] \Big\rbrace , 
\end{aligned}
\label{eq:mesogeneratingfunctionpathintegral}
\end{align} 
where $\pathintegralmeasure[\bm{X}]$ denotes the path-integral measure for the function $\bm{X}$.
The quantity $\bm{\field}$ is the conjugated field and can be physically interpreted as the instantaneous counting field for variations in the mesostates $\d \mesostate$. Moreover, the biased action functional $ \action_{\countingfield_{_{\trajectoryobservablemeso}}}[\mesostate,\bm{\field}] $ consists of the kinetic term $- \bm{\field} \, \dot{\mesostate}$, the biased Hamiltonian that accounts for the current-like contributions to $\mesogeneratingfunction(\countingfield_{_{\trajectoryobservablemeso}}, t)$,
\begin{align} \label{eq:biasedhamiltonianpathintegral} 
\hamiltonian_{\countingfield_{_{\trajectoryobservablemeso}}}[\mesostate(t'),\bm{\field}(t')] = \sum\limits_{\reservoir = 1}^{\reservoirdimension} \sum\limits_{i,j=1}^{\statenumber} \left\lbrace \exp \big[ \field_i(t') - \field_j(t') \big] \; \exp \big[ - \countingfield_{_{\trajectoryobservablemeso}} \trajectoryobservablemeso_{ij}^{(\reservoir)}(\mesostate(t')) \big] - 1 \right\rbrace \mesorates_{ij}^{(\reservoir)}(\bm{\protocol}_{t'},\mesostate(t')) , 
\end{align}
of a contribution due to the nonautonomous driving 
\begin{align} \label{eq:drivingobservablepathintegral}
- \countingfield_{_{\trajectoryobservablemeso}} \int\limits_{0}^{t} \d t' \, \dot{\bm{\protocol}}_{t'} \cdot \left[ \nabla_{\bm{\protocol}_{t'}} \, \trajectoryobservablemeso_{\mesostate(t')} \right] , 
\end{align}
of a state-like contribution
\begin{align} \label{eq:statelikeobservablepathintegral}
- \countingfield_{_{\trajectoryobservablemeso}} \int\limits_{0}^{t} \d t' \; \d_{t'} \trajectoryobservablemeso_{\mesostate}(t') = - \countingfield_{_{\trajectoryobservablemeso}} 
[ \trajectoryobservablemeso_{\mesostate_{_{\size}}}(t) - \trajectoryobservablemeso_{\mesostate_{_0}}(0)] , 
\end{align}
and of the initial condition $ \ln \mesoprobability_{\mesostate}(0) $. The quantities $ \trajectoryobservablemeso_{ij}^{(\reservoir)}(\mesostate) $ and $\mesorates_{ij}^{(\reservoir)}(\bm{\protocol},\mesostate)$ are the change of the fluctuating mesoscopic observable $\trajectoryobservablemeso$ and the mesoscopic transition rate, respectively, along a jump of the trajectory away from the mesostate $\mesostate$ that, at the unit-state level, corresponds to a transition from state $j$ to $i$ induced by the reservoir $\reservoir$. For vanishing bias, $\countingfield_{_{\trajectoryobservablemeso}} = 0$, Eq. \eqref{eq:mesogeneratingfunctionpathintegral} reduces to the path-integral representation of the path probability in the mesoscopic space.

We now rescale the size-extensive state variables to express them in terms of the size-intensive density $\bm{\meanfieldprobability} \equiv \mesostate/\dimension$. Using the Stirling approximation
\begin{align} \label{eq:stirlingapproximation}
\ln \dimension ! = \dimension \ln \dimension - \dimension + \rest(\ln \dimension) ,
\end{align}
we find with Eqs. \eqref{eq:multiplicityfactorpolynomial} and \eqref{eq:internalentropy} that the size-extensive mesoscopic internal entropy can be rewritten as follows,
\begin{align} \label{eq:scalingmesointernalentropy}
\frac{\mesoentropy^{int}_{\mesostate}}{\dimension} = \sum\limits_{i=1}^{\statenumber} \left[ \meanfieldprobability_i \ln \dimension - \meanfieldprobability_i \right] - \sum\limits_{i=1}^{\statenumber} \left[ \meanfieldprobability_i \ln \dimension_i - \meanfieldprobability_i \right] + \rest\left( \frac{\ln \dimension}{\dimension} \right) = \underbrace{ - \sum\limits_{i=1}^{\statenumber}  \meanfieldprobability_i \ln \meanfieldprobability_i }_{\equiv \meanfieldentropy^{int}_{\bm{\meanfieldprobability}} } + \, \rest\left( \frac{\ln \dimension}{\dimension} \right) , 
\end{align}
where $\rest(\cdot)$ gives the order of magnitude of the error made by the approximation.  Using the last equation and Eq. \eqref{eq:hamiltonian}, the size-extensive part of the mesoscopic free energy \eqref{eq:mesofreeenergy} reads
\begin{align}
\label{eq:scalingmesofreenergy}
\frac{ \mesofreeenergy^{(\reservoir)}_{\mesostate}(\bm{\protocol}_t) }{\dimension}
= \underbrace{ \sum_{i=1}^{\statenumber} \Bigg( \meanfieldprobability_i \epsilon_i(\bm{\protocol}_t) + \frac{\potential_i(\bm{\protocol}_t)}{2} \, \meanfieldprobability_i^2 + \sum_j \potential_{ij}(\bm{\protocol}_t) \meanfieldprobability_i \, \meanfieldprobability_j \Bigg) }_{\equiv \meanfieldenergy_{\bm{\meanfieldprobability}}(\bm{\protocol}_t)} + \frac{1}{\beta^{(\nu)}} \sum_{i=1}^{\statenumber} \meanfieldprobability_i\log{\meanfieldprobability_i} 
+ \rest\left( \frac{\ln \dimension}{\dimension} \right)
\equiv \meanfieldfreeenergy_{\bm{\meanfieldprobability}}^{(\reservoir)}(\bm{\protocol}_t) + \rest\left( \frac{\ln \dimension}{\dimension} \right) .
\end{align}
In order to proceed, we now make the crucial assumption that the functional form of the leading order of the mesoscopic rates in $\dimension$ is invariant under scaling by $1/\dimension$, \idest the size-extensive contributions of the mesoscopic rates are exactly homogeneous,
\begin{align} \label{eq:mesorateshomogeneity} 
\frac{ \mesorates_{ij}^{(\reservoir)}(\bm{\protocol}_t,\mesostate) }{\dimension} = \meanfieldrates_{ij}^{(\reservoir)}(\bm{\protocol}_t,\bm{\meanfieldprobability}) \; \meanfieldprobability_j + \mathbbm{o}(1) , 
\end{align}
where $\mathbbm{o}(1)$ refers to all terms that are subextensive in $\dimension$.
From the homogeneity property of the leading order of the mesoscopic rates and the change in the size-extensive part of the free energy for a transition from unit state $j \to i$,
\begin{align} \label{eq:scalingmesofreenergychange}
\mesofreeenergy^{(\reservoir)}_{\mesostate}(\bm{\protocol}_t) - \mesofreeenergy^{(\reservoir)}_{\mesostate'}(\bm{\protocol}_t) \overset{\substack{ \mesostate' \to \mesostate \\ j \to i }}{\equiv} (\D_{\meanfieldprobability_i} - \D_{\meanfieldprobability_j} ) \meanfieldfreeenergy_{\bm{\meanfieldprobability}}(\bm{\protocol}_t) + \rest\left(1/\dimension \right) , 
\end{align}
follows that the macroscopic, size-intensive transition rates $\meanfieldrates_{ij}^{(\reservoir)}(\bm{\protocol}_t,\bm{\meanfieldprobability}) \; \meanfieldprobability_j$ satisfy, up to non-extensive terms in the mesoscopic free energy, the following local detailed balance condition, \idest 
\begin{align} \label{eq:meanfieldlocaldetailedbalance} 
\begin{aligned}
& \ln \frac{ \mesorates_{ij}^{(\reservoir)}(\bm{\protocol}_t,\mesostate) }{ \mesorates_{ji}^{(\reservoir)}(\bm{\protocol}_t,\mesostate) } = 
\ln \frac{ \meanfieldrates_{ij}^{(\reservoir)}(\bm{\protocol}_t,\bm{\meanfieldprobability}) \; \meanfieldprobability_j }{ \meanfieldrates_{ji}^{(\reservoir)}(\bm{\protocol}_t,\bm{\meanfieldprobability}) \; \meanfieldprobability_i } + \rest\left(1/\dimension\right) = 
-\invtemperature^{(\reservoir)} \Big[ \hspace*{-.45cm} \underbrace{ ( \D_{\meanfieldprobability_i} - \D_{\meanfieldprobability_j} ) \meanfieldfreeenergy_{\bm{\meanfieldprobability}}(\bm{\protocol}_t) }_{ (\D_{\meanfieldprobability_i} - \D_{\meanfieldprobability_j} ) \meanfieldenergy_{\bm{\meanfieldprobability}}(\bm{\protocol}_t) - \frac{1}{\invtemperature^{(\reservoir)}} \ln \frac{ \meanfieldprobability_j }{ \meanfieldprobability_i } 
} \hspace*{-.45cm} - \force_{ij}^{(\reservoir)} \Big] + \rest\left(1/\dimension\right) \\ 
&= \! -\invtemperature^{(\reservoir)} \Big\lbrace \underbrace{ \underbrace{ \stateenergy_i (\protocol_{t}) \!-\! \stateenergy_j(\protocol_{t}) + [ \potential_i(\protocol'_{t}) \, \meanfieldprobability_i \!-\! \potential_j(\protocol'_{t}) \, \meanfieldprobability_j ] + \sum\limits_{k \neq i} \potential_{ik}(\protocol'_{t}) \meanfieldprobability_k(t) \! - \! \sum\limits_{k \neq j} \potential_{jk}(\protocol'_{t}) \meanfieldprobability_k(t) }_{ \equiv \meanfieldenergy_{ij}^{(\reservoir)}\big(\bm{\protocol}_{t},\bm{\meanfieldprobability} \big) } \!-\! \frac{1}{\invtemperature^{(\reservoir)}} \ln \frac{\meanfieldprobability_j}{\meanfieldprobability_i} }_{ \equiv \meanfieldfreeenergy_{ij}^{(\reservoir)}\big(\bm{\protocol}_{t},\bm{\meanfieldprobability} \big) } - \force_{ij}^{(\reservoir)} \Big\rbrace  + \rest\left(1/\dimension \right) \! . 
\end{aligned} 
\end{align}

Next, the scaled mesoscopic entropy production with bounding Gibbs states \eqref{eq:stochasticentropyproductionequilibriumboundary} expressed in terms of the size-extensive fluctuations reads as follows,
\begin{align}
\label{eq:scalingstochasticentropyproductionequilibriumboundary}
\frac{\delta \trajectoryepmeso[\trajectory_{(\tau)},t]}{\dimension} = \invtemperature^{(1)} \big[ \delta \meanfieldwork_{\bm{\protocol}}[\trajectory_{(\tau)},t] - \Delta \meanfieldfreeenergy^{eq}_1 \big] + \sum_{\reservoir = 1}^{\reservoirdimension} \big\lbrace  \invtemperature^{(\reservoir)} \delta \meanfieldwork_{\bm{\force}}^{(\reservoir)}[\trajectory_{(\tau)},t] + \big[ \invtemperature^{(1)} - \invtemperature^{(\reservoir)} \big] \delta \meanfieldenergy^{(\reservoir)}[\trajectory_{(\tau)},t] \big\rbrace + \rest\left( \tfrac{\ln \dimension}{\dimension} \right) , 
\end{align}
where $\delta \meanfieldwork_{\bm{\protocol}} = \lim_{\dimension \to \infty} \delta \trajectoryworkmeso_{\bm{\protocol}} /\dimension $, $\Delta \meanfieldfreeenergy_1^{eq}(\bm{\protocol}) = \lim_{\dimension \to \infty} \mesofreeenergy^{eq}_1(\bm{\protocol}) /\dimension $, $\delta \meanfieldwork_{\bm{\force}}^{(\reservoir)} = \lim_{\dimension \to \infty} \delta \trajectoryworkmeso_{\bm{\force}}^{(\reservoir)} /\dimension $ and $\delta \meanfieldenergy^{(\reservoir)} = \lim_{\dimension \to \infty} \delta \trajectoryenergymeso^{(\reservoir)}/\dimension $
are the size-intensive scaled nonautonomous work current, the change in size-intensive scaled equilibrium free-energy with respect to the reference reservoir $\reservoir=1$, the size-intensive scaled autonomous work current and the size-intensive scaled energy current, respectively, in the macroscopic limit.

Collecting results, inserting the expression for the stochastic entropy production with bounding Gibbs states \eqref{eq:stochasticentropyproductionequilibriumboundary} into the generic path-integral representation for a mesoscopic generating function \eqref{eq:mesogeneratingfunctionpathintegral}, and expressing the latter in terms of the dominant size-extensive terms, we get
\begin{align} \label{eq:mesogeneratingentropyproductionfunctionpathintegraldominant} 
\begin{aligned}
&\mesogeneratingfunction( \bm{\countingfield} ,t) = \int \pathintegralmeasure [\bm{\meanfieldprobability}] \int \pathintegralmeasure [\bm{\field}] \; \mesoprobability_{\bm{\meanfieldprobability}_{_0}}^{eq}(\bm{\protocol}_0) \cdot \exp \Bigg\lbrace \dimension \int\limits_0^t \d t'\Bigg[ \!-\! \countingfield_{_{\bm{\lambda}}} \invtemperature^{(1)} \dot{\bm{\protocol}}_{t'} \cdot \big[ \nabla_{\bm{\protocol}_{t'}} \meanfieldenergy_{\bm{\meanfieldprobability}}(\bm{\protocol}_{t'}) \big] + \sum\limits_{i=1}^{\statenumber} \Big\lbrace - \field_i(t') \dot{\meanfieldprobability}_i(t') \\ 
&+ \! \sum\limits_{\reservoir = 1}^{\reservoirdimension} \sum\limits_{j=1}^{\statenumber} \! \Big[ \! \exp \Big\lbrace \field_i(t') \!-\! \field_j(t') \!-\! \countingfield_{_{\bm{\force}}}^{(\reservoir)} \invtemperature^{(\reservoir)} \force_{ij}^{(\reservoir)} \!-\! \countingfield_{\meanfieldenergy}^{(\reservoir)} [ \invtemperature^{(1)} \!-\! \invtemperature^{(\reservoir)} ] \, \meanfieldenergy_{ij}^{(\reservoir)}\big(\bm{\protocol}_{t'},\bm{\meanfieldprobability}(t') \big) \! \Big\rbrace \!-\! 1 \Big]  \meanfieldrates_{ij}^{(\reservoir)} \big(\bm{\protocol}_{t'},\bm{\meanfieldprobability}(t') \big) \, \meanfieldprobability_j(t') \Big\rbrace \!+\! \mathbbm{o}(1) \Bigg] \! \Bigg\rbrace \\
&\equiv \int \pathintegralmeasure [\bm{\meanfieldprobability}]  \int  \pathintegralmeasure [\bm{\field}] \; \exp \Big\lbrace \dimension \Big[ \action_{\bm{\countingfield}}[\bm{\meanfieldprobability}(t'),\bm{\field}(t')] + \mathbbm{o}(1) \Big] \Big\rbrace  , 
\end{aligned}
\end{align}
with the shorthand notation from Eq. \eqref{eq:countingfieldnotation} and with
$\countingfield$ that denotes a vector of fields $\countingfield_{\meanfieldobservable} $ that counts the scaled observables, $\meanfieldobservable = \lim_{\dimension \to \infty} \mesoobservable / \dimension $. Moreover, we rewrote the mesoscopic initial equilibrium distribution in terms of size-extensive free energies as follows
\begin{align} \label{eq:mesoscopicequilibriumdistributiondominant} 
\mesoprobability^{eq}_{\mesostate}(\bm{\protocol}_0) = \underbrace{ \exp \Big\lbrace \dimension \Big[ \! - \! \invtemperature^{(1)} \big[ \meanfieldfreeenergy^{(1)}_{\mathbf{\meanfieldprobability}}(\bm{\protocol}_0) - \meanfieldfreeenergy^{eq}_1(\bm{\protocol}_0) \big] }_{ \equiv \mesoprobability^{eq}_{\bm{\meanfieldprobability}}(\bm{\protocol}_0) } + \rest\left( \tfrac{\ln \dimension}{\dimension} \right) \Big] \Big\rbrace  .
\end{align}
As demonstrated in appendix \ref{sec:martinsiggiaroseappendix}, the generating function \eqref{eq:mesogeneratingentropyproductionfunctionpathintegraldominant} satisfies, up to non-extensive fluctuations in the system size $\dimension$, the symmetry relation
\begin{align}
\label{eq:mesogeneratingfunctionsymmetrydominant}
\mesogeneratingfunction( \bm{\countingfield} ,t) = \tilde \mesogeneratingfunction( \tilde{\bm{\countingfield}} ,t) \; \exp \big\lbrace \dimension \big[ - \invtemperature^{(1)} \Delta \meanfieldfreeenergy^{eq}_1(\bm{\protocol}) + \mathbbm{o}(1) \big] \big\rbrace .
\end{align}
In the macroscopic limit, $\dimension \to \infty$, there is a single trajectory that carries all the weight of all possible paths contributing to the path integral \eqref{eq:mesogeneratingfunctionpathintegral}. This trajectory maximizes the size-intensive action functional in Eq. \eqref{eq:mesogeneratingentropyproductionfunctionpathintegraldominant}, $\mathrm{max} \; \action_{\bm{\countingfield}}[\bm{\meanfieldprobability},\bm{\field}] = \action_{ \bm{\countingfield} }[\bm{\meanfieldprobability}^*,\bm{\field}^*] $, and its coordinates are therefore determined as follows
\begin{align} \label{eq:actionfunctionalderivative}
\frac{\delta \action_{\bm{\countingfield}}[\bm{\meanfieldprobability},\bm{\field}] }{\delta \bm{\field}} \Big|_{\bm{\meanfieldprobability}^*,\bm{\field}^*} = 0 , \qquad 
\frac{\delta \action_{\bm{\countingfield}}[\bm{\meanfieldprobability},\bm{\field}] }{\delta \bm{\meanfieldprobability}} \Big|_{\bm{\meanfieldprobability}^*,\bm{\field}^*} = 0 , 
\end{align}
where $\bm{\meanfieldprobability}^*(\bm{\countingfield}) \equiv \lim_{\dimension \to \infty} \mesostate^*(\bm{\countingfield})/\dimension$ is the most likely biased and continuous density. To ease notation, we will omit the ``$*$'' in the following.
We consequently obtain via Eq. \eqref{eq:mesogeneratingfunctionpathintegral} the size-scaled cumulant generating function
\begin{align} \label{eq:scaledcumulantgeneratingfunction}
\scgf(\bm{\countingfield},t) = \lim_{\dimension \to \infty} \frac{1}{\dimension} \, \ln \mesogeneratingfunction( \bm{\countingfield} ,t) = \action_{\bm{\countingfield}}[\bm{\meanfieldprobability},\bm{\field}] . 
\end{align}
Crucially, the macroscopic limit taken in the definition of the the scaled cumulant generating function eliminates the non-extensive terms in Eq. \eqref{eq:mesogeneratingfunctionsymmetrydominant}, \idest the scaled cumulant generating function associated with the entropy production \eqref{eq:stochasticentropyproductionequilibriumboundary} satisfies a symmetry that is formally equivalent to the one exhibited by the mesoscopic generating functions \eqref{eq:mesogeneratingfunctionsymmetry}. Explicitly, we have
\begin{align} \label{eq:macrogeneratingfunctionsymmetry}
\scgf(\bm{\countingfield},t) = \tilde{\scgf}(\tilde{\bm{\countingfield}},t) - \invtemperature^{(1)} \Delta \meanfieldfreeenergy^{eq}_1(\bm{\protocol}) ,
\end{align}
which is a symmetry for the macroscopic fluctuations exponentially dominating the mesoscopic dynamics. 
The last equation immediately stipulates the existence of a finite-time detailed fluctuation theorem in the spirit of Eq. \eqref{eq:mesodftjoint} that asymptotically holds in the macroscopic limit,
\begin{align} \label{eq:macrojointdft}
\lim_{\dimension \to \infty} \frac{1}{\dimension} \ln \frac{ \mesoprobability \Big( \invtemperature^{(1)} \delta \trajectoryworkmeso_{\bm{\protocol}} \, , \, \lbrace \delta \mesocurrent^{(\reservoir)}_{\bm{\force}} \rbrace \, , \, \lbrace \delta \mesocurrent^{(\reservoir)}_{\trajectoryenergymeso} \rbrace \Big) }{ \tilde{\mesoprobability} \Big( - \invtemperature^{(1)} \delta \trajectoryworkmeso_{\bm{\protocol}} \, , \, \lbrace - \delta \mesocurrent^{(\reservoir)}_{\bm{\force}} \rbrace \, , \, \lbrace - \delta \mesocurrent^{(\reservoir)}_{\trajectoryenergymeso} \rbrace \Big) } = \invtemperature^{(1)} \big[ \delta \meanfieldwork_{\bm{\protocol}} - \Delta \meanfieldfreeenergy^{eq}_1 \big] + \sum\limits_{\reservoir = 1}^{\reservoirdimension} \big[ \invtemperature^{(\reservoir)} \delta \meanfieldwork_{\bm{\force}}^{(\reservoir)} + \big[ \invtemperature^{(1)} - \invtemperature^{(\reservoir)} \big] \delta \meanfieldenergy^{(\reservoir)}\big] \, .
\end{align}
The existence of the finite-time detailed fluctuation theorem \eqref{eq:macrojointdft} is an important result as it ensures the thermodynamic consistency of the path-integral approach at macroscopic scales, \idest for fluctuations that are extensive in and thus scale exponentially with the system size $\dimension$.

We stress that the finding of the macroscopic symmetry \eqref{eq:macrogeneratingfunctionsymmetry} and the macroscopic detailed fluctuation theorem detailed is nontrivial since it is mathematically not obvious that the symmetry at microscopic \eqref{eq:microgeneratingfunctionsymmetry} and mesoscopic \eqref{eq:mesogeneratingfunctionsymmetry} scales is also asymptotically preserved at macroscopic scales in spite of discarding subextensive contributions to the current statistics. It is however important to recall that these results rely on the assumption that the leading order of the mesoscopic transitions rates in $\dimension$ is homogeneous \eqref{eq:mesorateshomogeneity}.

\subsection{Mean-Field Description}
\subsubsection{Mean-Field Dynamics}

We proceed by formulating the dynamics and thermodynamics in the macroscopic mean-field limit, where the system behaves deterministically. First we note that for an unbiased dynamics, $\bm{\countingfield} = 0$, that the extremal values of the auxiliary field are $\bm{\field} =0$. Thus, the action functional \eqref{eq:mesogeneratingentropyproductionfunctionpathintegraldominant} needs only to be maximized with respect to the density resulting into the following Hamiltonian equations of motion
\begin{align} \label{eq:hamiltonianequationsofmotion}
\frac{\delta \action[\bm{\meanfieldprobability},\bm{\field}] }{\delta \bm{\field}} \Bigg|_{\bm{\field}=0} = 0 \; \Leftrightarrow \; \dot{\bm{\meanfieldprobability}} = \frac{\delta \hamiltonian[\bm{\meanfieldprobability},\bm{\field}] }{\delta \bm{\field}} \Bigg|_{\bm{\field} = 0} . 
\end{align}
The Hamiltonian equations of motion correspond to the mean-field equation governing the deterministic dynamics of the most likely occupation (mean-field) density and read explicitly,
\begin{align}
\D_t \meanfieldprobability_i(t)  = \sum\limits_{j=1}^{\statenumber} \meanfieldrates_{ij}(\bm{\protocol}_t) \, \meanfieldprobability_j(t)  \, , \quad \sum\limits_{i=1}^{\statenumber} \meanfieldprobability_i(t) = 1 \, , \label{eq:meanfieldmasterequation}
\end{align}
with the mean-field transition rate matrix as defined in Eq. 	\eqref{eq:mesorateshomogeneity} and evaluated at the most likely mean-field density \eqref{eq:hamiltonianequationsofmotion},
\begin{align} \label{eq:meanfieldrates} 
\meanfieldrates_{ij}(\bm{\protocol}_t) \equiv \meanfieldrates_{ij}(\bm{\protocol}_t,\bm{\meanfieldprobability}) = \sum\limits_{\reservoir=1}^{\reservoirdimension}  \meanfieldrates_{ij}^{(\reservoir)}(\bm{\protocol}_t,\bm{\meanfieldprobability}) \, , 
\end{align}
that is stochastic, $\sum_i \meanfieldrates_{ij}(\bm{\protocol}_t) = 0$, and whose contributions corresponding to the different heat reservoirs obey a mean-field local detailed balance \eqref{eq:meanfieldlocaldetailedbalance} ensuring thermodynamic consistency at the mean-field level.
We note that because of probability conservation the nonlinear mean-field equation is $\statenumber - 1$ dimensional.

\subsubsection{Mean-Field First And Second Law}
\label{sec:meanfieldfirstsecondlaw}

Analogously to Sec. \ref{sec:micromesofirstsecondlaw}, we now want to formulate the first and second law in the macroscopic mean-field limit. Following a similar procedure as for the derivation of Eq. \eqref{eq:meanfieldmasterequation}, we obtain from Eqs. \eqref{eq:mesogeneratingfunctionpathintegral} and \eqref{eq:biasedhamiltonianpathintegral} for the mean-field energy,
\begin{align}  \label{eq:meanfieldenergy} 
\meanfieldenergy(t) &= \sum\limits_{i=1}^{\statenumber} \meanfieldenergy_i(\bm{\protocol}_t) \, \meanfieldprobability_i(t) , \quad  \meanfieldenergy_i(\bm{\protocol}_t) 
\equiv  \D_{\meanfieldprobability_i} \meanfieldenergy_{\bm{\meanfieldprobability}}(\bm{\protocol}_t)  = \stateenergy_i(\protocol_{t}) + \potential_i(\protocol'_{t}) + \sum\limits_{k \neq i} \potential_{ik}(\protocol'_{t}) \meanfieldprobability_k(t) ,
\end{align} 
whose time-derivative constitutes the first law in the macroscopic limit,
\begin{align} \label{eq:meanfieldfirstlaw}
\d_t \meanfieldenergy(t) = \dot{\meanfieldheat}(t) + \dot{\meanfieldwork}(t) ,
\end{align}
with the mean-field heat and work current,
\begin{align}  \label{eq:meanfieldfirstlawheat}
\dot{\meanfieldheat}(t) &= \sum\limits_{\reservoir=1}^{\reservoirdimension} \sum\limits_{i,j=1}^{\statenumber} \left[ \meanfieldenergy_i(\bm{\protocol}_t) - \meanfieldenergy_j(\bm{\protocol}_t) - \force^{(\reservoir)}_{ij} \right] \, \meanfieldrates_{ij}^{(\reservoir)}(\bm{\protocol}_t) \;  \meanfieldprobability_{j}(t) = \sum\limits_{\reservoir=1}^{\reservoirdimension} \sum\limits_{i,j=1}^{\statenumber} - \frac{1}{\invtemperature^{(\reservoir)}} \, \ln \Bigg[ \frac{ \meanfieldrates_{ij}^{(\reservoir)}(\bm{\protocol}_t) }{ \meanfieldrates_{ji}^{(\reservoir)}(\bm{\protocol}_t) } \, \Bigg] \; \meanfieldrates_{ij}^{(\reservoir)}(\bm{\protocol}_t) \;  \meanfieldprobability_{j}(t) \\ 
\dot{\meanfieldwork}(t) &= \sum_{i=1}^{\statenumber} \bigg\lbrace \dot{\bm{\protocol}}_t \cdot \nabla_{\bm{\protocol}_t} \, \meanfieldenergy_{i}(\bm{\protocol}_t) \; \meanfieldprobability_{i}(t) + \sum\limits_{\reservoir=1}^{\reservoirdimension} \sum\limits_{j=1}^{\statenumber} \Big[ \sum\limits_{k \neq i} \meanfieldprobability_{i}(t) \, \potential_{ik}(\protocol'_{t}) \, \meanfieldrates^{(\reservoir)}_{kj}(\bm{\protocol}_t) + \force^{(\reservoir)}_{ij} \; \meanfieldrates^{(\reservoir)}_{ij}(\bm{\protocol}_t) \Big] \, \meanfieldprobability_{j}(t) \bigg\rbrace  \label{eq:meanfieldfirstlawwork} . 
\end{align}
A closer inspection of Eqs. \eqref{eq:averageentropy} reveals that in the deterministic macroscopic limit the stochastic (Shannon) part of the mesoscopic system entropy vanishes and only the internal entropy of the mesostates \eqref{eq:internalentropy} remains finite.
Thus, we conclude from Eq. \eqref{eq:scalingmesointernalentropy} that the mean-field entropy reads
\begin{align} \label{eq:meanfieldentropy} 
\meanfieldentropy(t) \equiv \meanfieldentropy^{int}_{\bm{\meanfieldprobability}} = -\sum\limits_{i=1}^{\statenumber} \meanfieldprobability_i(t)\, \ln \meanfieldprobability_i(t).
\end{align}
The entropy in deterministic many-body systems therefore originates from the Boltzmann entropies related to the internal structure of the mesostates. Remarkably, the deterministic mean-field entropy takes the form of a Shannon entropy for the mean-field density.

Next, Equations \eqref{eq:averageentropyproduction} and \eqref{eq:meanfieldentropy} imply the second law in the macroscopic limit
\begin{align} \label{eq:meanfieldentropyproduction}
\dot{\meanfieldentropy}_i(t) = \sum\limits_{\reservoir=1}^{\reservoirdimension} \sum\limits_{i,j=1}^{\statenumber} \ln \frac{ \meanfieldrates^{(\reservoir)}_{ij}(\bm{\protocol}_t) \, \meanfieldprobability_{j}(t)}{\meanfieldrates^{(\reservoir)}_{ji}(\bm{\protocol}_t) \, \meanfieldprobability_{i}(t)} \, \meanfieldrates^{(\reservoir)}_{ij}(\bm{\protocol}_t) \, \meanfieldprobability_{j}(t) = \dot{\meanfieldentropy}(t) - \dot{\meanfieldentropy}_e(t) \geq 0 .  
\end{align}
Hence the microscopic and mesoscopic observables in Eqs. \eqref{eq:firstlaw}--\eqref{eq:averageentropyproduction} converge to the corresponding mean-field ones in Eq. \eqref{eq:meanfieldenergy}--\eqref{eq:meanfieldentropyproduction} if the macroscopic limit is taken,
\begin{align}
\lim_{\dimension \to \infty} \frac{1}{\dimension} \dot{\mesoobservable}(t) = \dot{\mathcal{\mesoobservable}}(t), \qquad \mesoobservable = \mesoenergy,\mesoheat,\mesowork,\mesoentropy,\mesoentropy_e,\mesoep , 
\end{align}
where we recall that the mesoscopic representations for $\mesoobservable = \mesoentropy,\mesoep$ are only valid if the microstates inside each mesostate are equiprobable.

This constitutes our main result: For thermodynamically consistent and discrete many-body systems with all-to-all interactions there is an exact coarse-graining \eqref{eq:mesoscopicmasterequation} of the microscopic stochastic dynamics towards a mesoscopic one that is fully characterized by the system occupation. In the macroscopic limit, $\dimension \to \infty$, the stochastic dynamics asymptotically converges to a deterministic and nonlinear macroscopic (mean-field) master equation \eqref{eq:meanfieldmasterequation}. Hence the stochastic dynamics can be equivalently represented across microscopic and mesoscopic scales and asymptotically on macroscopic scales as $\dimension \to \infty$. Furthermore, the thermodynamics can be equivalently formulated at microscopic and mesoscopic scales if the microstates inside each mesostate are equiprobable \eqref{eq:stationarymicroprobabilities}. The thermodynamic consistency at each of the two levels is encoded in the respective detailed fluctuation theorem, see Eqs. \eqref{eq:microdftjoint} and \eqref{eq:mesodftjoint}. Using a path-integral representation of the stochastic (thermo)dynamics à la Martin-Siggia Rose, the fluctuations which scale exponentially with the system size also satisfy a detailed fluctuation theorem \eqref{eq:macrojointdft} and are therefore also thermodynamically consistent.

\section{Example}
\label{sec:example}

To illustrate the methodology developed in the preceding Sec. \ref{sec:martinsiggiarose} we consider a semi-analytically solvable autonomous Ising model which exhibits a nonequilibrium phase transition, thus representing a suitable model to demonstrate the utility of the aforementioned methods. To this end, let us consider $\dimension \to \infty$ spins with flat energy landscapes, $\stateenergy_1 = \stateenergy_2 $, that globally interact via a pair potential $\potential/\dimension$ if they occupy the same spin state $i=1,2$. The system is in contact with two heat reservoirs at different inverse temperatures $\invtemperature^{h}$ and $\invtemperature^{c}$ with $ \invtemperature^{h} < \invtemperature^{c} $. Related models with a similar phenomenology can be found in Refs. \cite{garrido1987jps, blote1990jpa, bauer2018jpa}.

\begin{figure}[h!]
\begin{center}
%
%

\includegraphics[scale=1]{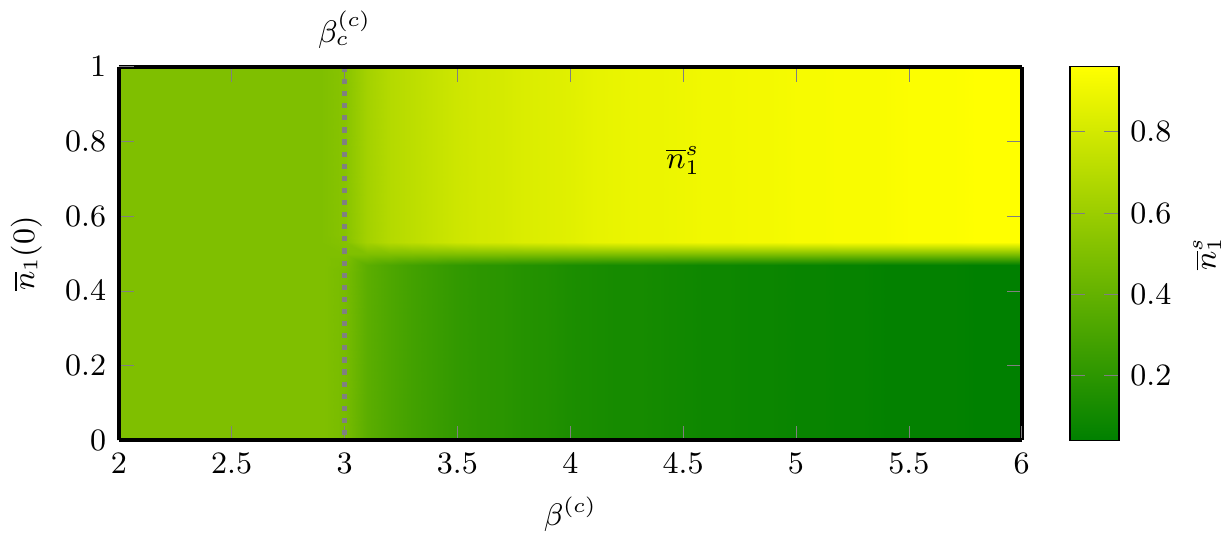}

\caption{Density plot of the stationary solution $\meanfieldprobability_1^s$ for different temperatures $\invtemperature^{(c)}$ and all physical initial conditions $\meanfieldprobability_1(0)$. We choose the following values for the parameters $\arrheniusprefactor = 0.1,\invtemperature^{(h)} = 1, \potential  = -1$ so that $\invtemperature^{(c)}_c = 3$ as indicated by the vertical dotted line.  \label{fig:phasespacexample} }
\end{center}
\end{figure}

According to Eq. \eqref{eq:meanfieldmasterequation}, the mean-field dynamics is governed by the following nonlinear rate equation
\begin{align} \label{eq:meanfieldequationexample}
\D_t \meanfieldprobability_i &= - \Big( \meanfieldrates_{ji}^{(h)} + \meanfieldrates_{ji}^{(c)} \Big) \meanfieldprobability_i + \Big( \meanfieldrates_{ij}^{(h)} + \meanfieldrates_{ij}^{(c)} \Big) \meanfieldprobability_j = - \meanfieldrates_{ji} \, \meanfieldprobability_i + \meanfieldrates_{ij} \, \meanfieldprobability_j  , \quad i,j=1,2 \; ,
\end{align}
with the mean-field transition rates which we assume to be of Arrhenius form
\begin{align}  \label{eq:meanfieldratesexample}
\meanfieldrates_{ij}^{(\reservoir)} = \arrheniusprefactor \, \exp \Big[ -\frac{\invtemperature^{(\reservoir)}}{2} \potential (\meanfieldprobability_i - \meanfieldprobability_j) \Big] , \quad \reservoir = c,h ,
\end{align}
with the constant kinetic prefactor $\arrheniusprefactor$ that sets the time-scale of the Markov jump process. We note that the mean-field dynamics \eqref{eq:meanfieldequationexample} is 	effectively a one-dimensional equation since we have $\meanfieldprobability_2 = 1 - \meanfieldprobability_1$ because the number of spins is conserved. We can immediately read off the stationary solution $\meanfieldprobability_i^s = 1/2, \, i=1,2$ for Eq. \eqref{eq:meanfieldequationexample}. The stability of this symmetric fixed point is encoded in the spectrum of the linearized Jacobian, $ \A_{ij} \equiv [ \D ( \D_t \meanfieldprobability_i) / \D \meanfieldprobability_j ] |_{ \meanfieldprobability_{i,j}=1/2} $, which can be readily determined as follows
\begin{align} \label{eq:linearjacobianspectrumexample}
\eigenvalue_1 = 0, \quad \eigenvalue_2 = - \arrheniusprefactor \big[ 4 + \potential \big( \invtemperature^{(h)} + \invtemperature^{(c)} \big) \big] .
\end{align}
	The zero eigenvalue $\eigenvalue_1$ reflects that the rank of the Jacobian is smaller than its dimension due to the constraint $\sum_i \D_t \meanfieldprobability_i = 0$. More strikingly, the second eigenvalue $\eigenvalue_2$ changes its sign for attractive interactions, $\potential < 0$, at the critical temperatures
\begin{align} \label{eq:criticalpointexample} 
4 + \potential \big( \invtemperature^{(h)}_c + \invtemperature^{(c)}_c \big) = 0 ,
\end{align}
indicative of a supercritical pitchfork bifurcation that destabilizes the symmetric fixed point into two asymmetric fixed points as can be seen in Fig. \ref{fig:phasespacexample}. This density plot depicts the stationary solution $\meanfieldprobability_1^s$ as a function of all physical initial conditions $\meanfieldprobability_1(0)$ and for different cold temperatures $\invtemperature^{(c)}$ while $\invtemperature^{(h)} \equiv 1$ and $\potential = -1$ are kept constant. As can be observed, the symmetric fixed point is stable for $ \invtemperature^{(c)} < \invtemperature^{(c)}_c = 3$. In contrast, for lower temperatures $ \invtemperature^{(c)} > \invtemperature^{(c)}_c = 3$ the symmetric fixed point is unstable and the system dynamics goes to one of the two asymmetric stable fixed points depending on the basin of attraction in which the initial condition lies. These two stable fixed points are related to each other via permutations of their coordinates, in agreement with the invariance of the mean-field Eq. \eqref{eq:meanfieldequationexample} under a permutation operation. The phenomenology observed in Fig. \ref{fig:phasespacexample} can be physically seen as follows. In the high-temperature limit the system behaves entropically, thus occupying the symmetric fixed point. Conversely, in the low-temperature limit the system behaves energetically, thus exhibiting two asymmetric fixed points that converge to the two energy ground states, that is $\meanfieldprobability_1=1, \meanfieldprobability_{2}=0$ and $\meanfieldprobability_1=0, \meanfieldprobability_{2}=1$, as $\invtemperature \to \infty$.
For isothermal systems, Eq. \eqref{eq:criticalpointexample} implies the critical point $ \invtemperature_c = - 2/\potential$. This is in agreement with the $\statenumber$-dependent universal critical temperature, $ \invtemperature_c(\statenumber) = - \statenumber/\potential$ for isothermal and all-to-all interacting $\statenumber$-state clock models derived in Ref. \cite{herpich2019pre}. We add that the isothermal system displays a first-order equilibrium phase transition.

\begin{figure}[h!] 
\begin{center}
%
%
%

\includegraphics[scale=1]{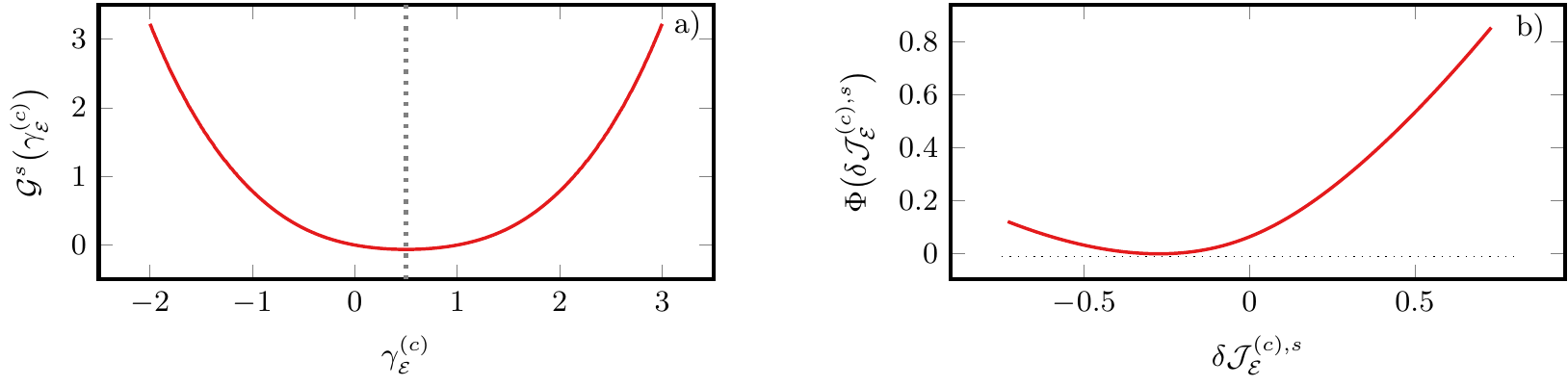}

\caption{The cumulant generating function \eqref{eq:scaledcumulantgeneratingfunctionexample} scaled with time $t$ and size $\dimension$ as a function of the counting field $\countingfield^{(c)}_{_{\meanfieldenergy}}$ in a) and the corresponding rate function $\ratefct\big( \delta \meanfieldcurrent^{(c),s}_{\meanfieldenergy} \big)$ in b). The parameters are chosen as $\invtemperature^{(h)}=3,\invtemperature^{(c)}=5,\potential=-1$ so that for $\countingfield^{(c)}_{_{\meanfieldenergy}} = 0$ the stationary mean-field system is in its energetic phase which has two asymmetric stable fixed points and a symmetric unstable one. \label{fig:scgfsymmetryexample} }
\end{center}
\end{figure}

We now return to the non-isothermal case and consider the fluctuating quantity in Eq. \eqref{eq:stochasticentropyproductionequilibriumboundary} that for the autonomous Ising model simplifies to 
\begin{align}
\delta \trajectoryepmeso[\trajectory_{(\tau)},t] = \sum\limits_{\reservoir = h,c} \big[ \invtemperature^{(h)} - \invtemperature^{(\reservoir)} \big] \delta \trajectoryenergymeso^{(\reservoir)}[\trajectory_{(\tau)},t] \big\rbrace , \quad \delta \trajectoryenergymeso^{(\reservoir)}[\trajectory_{(\tau)},t] = \potential \int\limits_0^t \d t' \sum\limits_{j=1}^{\size} \delta( \reservoir - \reservoir_j) \delta(t'-\tau_j) [\density_j - \density_{j-1}] .
\label{eq:stochasticentropyproductionequilibriumboundaryexample}
\end{align}
According to Eq. \eqref{eq:mesodftjoint}, our model system therefore satisfies a finite-time detailed fluctuation theorem for the time-integrated energy current. Using the path-integral formalism introduced in Sec. \ref{sec:martinsiggiarose}, we however observe that analytical progress is difficult at finite time as it would require to solve the full extremization problem \eqref{eq:actionfunctionalderivative} which is analytically not possible. Instead, we therefore resort to the stationary case which considerable simplifies the problem of finding the dominant trajectory among all paths contributing to the path integral. The biased Hamiltonian \eqref{eq:biasedhamiltonianpathintegral} in the path-integral formulation of the generating function \eqref{eq:mesogeneratingentropyproductionfunctionpathintegraldominant}
associated with the stochastic observable in the last equation reads
\begin{align}
\begin{aligned}
\hamiltonian_{\countingfield^{(c)}_{_{\meanfieldenergy}}}[\bm{\meanfieldprobability},\bm{\field}] &= \meanfieldrates_{21}^{(h)} \, \meanfieldprobability_1 \big[ \exp \big( \field_2 - \field_1 \big) -1 \big] + \meanfieldrates_{21}^{(c)} \, \meanfieldprobability_1 \Big[ \exp \big( \field_2 - \field_1 \big) \exp \big\lbrace \countingfield \potential [ \invtemperature^{(h)} - \invtemperature^{(c)} ] (\meanfieldprobability_2 - \meanfieldprobability_1) \big\rbrace - 1 \Big] \; + \\
&+ \meanfieldrates_{12}^{(h)} \, \meanfieldprobability_2 \big( \exp \big[ -(\field_2 - \field_1) \big] - 1 \big) + \meanfieldrates_{12}^{(c)} \, \meanfieldprobability_2 \Big( \exp \big[ -(\field_2 - \field_1) \big] \exp \big\lbrace - \countingfield \potential [ \invtemperature^{(h)} - \invtemperature^{(c)}  ] (\meanfieldprobability_2 - \meanfieldprobability_1) \big\rbrace - 1 \Big) . 
\end{aligned}
\end{align}
At steady state, the Hamiltonian equations of motion resulting from the extremization of the action functional in Eq. \eqref{eq:actionfunctionalderivative} read
\begin{align} \label{eq:biasedhamiltonianequationexample}
\D_{\meanfieldprobability_i} \hamiltonian_{\countingfield^{(c)}_{_{\meanfieldenergy}}}[\bm{\meanfieldprobability},\bm{\field},\lagrangemultiplier_{\meanfieldprobability},\lagrangemultiplier_{\field}] = 0, \quad \D_{\field_i} \hamiltonian_{\countingfield^{(c)}_{_{\meanfieldenergy}}}[\bm{\meanfieldprobability},\bm{\field},\lagrangemultiplier_{\meanfieldprobability},\lagrangemultiplier_{\field}] = 0,  \quad i=1,2 , 
\end{align}
where we added the Lagrangian multipliers $\lagrangemultiplier_{\meanfieldprobability}$ and $\lagrangemultiplier_{\field}$ to enforce the spin conservation, $\meanfieldprobability_1 + \meanfieldprobability_2 - 1 = 0$ and $\field_1 + \field_2 = 0$. The extremal value for $\field_1$ can be solved analytically,
\begin{align}
\field_1 \!=\! \frac{1}{4} \ln \! \left( \! \frac{\meanfieldprobability_1}{\meanfieldprobability_1-1} \!\cdot \! \frac{ \exp \big[ \invtemperature^{(h)} \potential (2\meanfieldprobability_1-1) \big] + \exp \big\lbrace \frac{\potential}{2} (2\meanfieldprobability_1-1) [\invtemperature^{(c)} \big(2\countingfield^{(c)}_{_{\trajectoryenergymeso}}-1 \big) - \invtemperature^{(h)} \big(1+2\countingfield^{(c)}_{_{\trajectoryenergymeso}}\big)] \big\rbrace   }{ 1 + \exp \big[ \frac{\invtemperature^{(c)} - \invtemperature^{(h)}}{2} \potential (2\meanfieldprobability_1-1) \big(2\countingfield^{(c)}_{_{\trajectoryenergymeso}}-1\big) \big] } \! \right) \!+\! \i \pi(1 + 2 k) , \; k \in \mathbb{Z} , 
\end{align}
and the extremal value $\meanfieldprobability_1$ is subsequently determined numerically. In the $t \to \infty$ limit, the boundary terms in the action functional become negligible so that the time- and size-scaled cumulant generating function is asymptotically equal to the biased Hamiltonian evaluated at the extremal values $\bm{\meanfieldprobability}$ and $\bm{\field}$,
\begin{align} \label{eq:scaledcumulantgeneratingfunctionexample}
\scgf^s\big( \countingfield^{(c)}_{_{\meanfieldenergy}} \big) = \lim_{t \to \infty} \frac{1}{t} \, \lim_{\dimension \to \infty} 
\frac{1}{\dimension}  \, \ln \mesogeneratingfunction \big( \countingfield^{(c)}_{_{\trajectoryenergymeso}} ,t \big) = \hamiltonian^s_{ \countingfield^{(c)}_{_{\meanfieldenergy}}}[\bm{\meanfieldprobability},\bm{\field}] .
\end{align}
The scaled cumulant generating function is plotted in Fig. \ref{fig:scgfsymmetryexample}a). We choose the values $\invtemperature^{(h)}=3,\invtemperature^{(c)}=5,\potential=-1$ corresponding to the phase where the mean-field dynamics exhibits two asymmetric stable and a symmetric unstable fixed point. Similarly, we observe two asymmetric $\countingfield$-dependent fixed points $\meanfieldprobability_1(\countingfield)$ whose coordinates are related to each other via a permutation as well as a symmetric fixed point at $\meanfieldprobability_1 =1/2$. The regime around 1/2 corresponds to the symmetric fixed point and thus to a null observable \eqref{eq:stochasticentropyproductionequilibriumboundaryexample}.
Next, we note that the curve is symmetric with respect to the value $\countingfield=1/2$, thus implying that the scaled cumulant generating function asymptotically satisfies the symmetry relation
\begin{align} \label{eq:scaledcumulantgeneratingfunctionsymmetryexample}
\scgf^s \big(\countingfield^{(c)}_{_{\meanfieldenergy}} \big) = \tilde{\scgf}^s\big(1 - \countingfield^{(c)}_{_{\meanfieldenergy}} \big) , 
\end{align}
which in turn stipulates the existence of a macroscopic steady-state detailed fluctuation theorem for the time-integrated energy current
\begin{align} \label{eq:steadystatedftexample}
\lim_{t \to \infty} \frac{1}{t} \lim_{\dimension \to \infty} \frac{1}{\dimension} \ln 
\frac{ \mesoprobability \big( \delta \mesocurrent^{(c)}_{\trajectoryenergymeso} \big) }{ \tilde{\mesoprobability} \big( - \delta \mesocurrent^{(c)}_{\trajectoryenergymeso} \big) } = \delta \meanfieldcurrent^{(c),s}_{\meanfieldenergy} , \quad \delta \meanfieldcurrent^{(c),s}_{\meanfieldenergy} = [\invtemperature^{(h)} - \invtemperature^{(c)} ] \; \lim_{t \to \infty} \frac{1}{t} \lim_{\dimension \to \infty} \frac{1}{\dimension} \, \delta \trajectoryenergymeso^{(c)} .
\end{align}
The existence of the steady-state fluctuation theorem is by no means obvious, here. In general, the implicit assumption underlying steady-state fluctuation theorems is that the contribution of the boundary terms related to the initial and final state of each trajectory are subextensive in time and thus negligible in the infinite-time limit. There are however situations where this may not be true, \emph{e.g.} in bistable systems for starting distributions of the forward and backward process that are located in the different basins of attraction. Though, in this model the two $\countingfield^{(c)}_{_{\meanfieldenergy}}$-dependent fixed points are related to each other via permutation of their coordinates and the statistics of the corresponding stationary states are thus identical.

\noindent 
Figure \ref{fig:scgfsymmetryexample}b) shows the rate function $\ratefct \big( \delta \meanfieldcurrent^{(c),s}_{\meanfieldenergy} \big) $ associated with the scaled cumulant generating function $\scgf^s\big(\countingfield^{(c)}_{_{\meanfieldenergy}})$ in a). The rate function is defined as \cite{touchette2009pr}
\begin{align} \label{eq:ratefunctiondefinition}
\ratefct\big( \delta \meanfieldcurrent^{(c),s}_{\meanfieldenergy} \big) = - \lim_{t \to \infty} \frac{1}{t} \lim_{\dimension \to \infty} \frac{1}{\dimension} \ln \mesoprobability\big( \delta \mesocurrent^{(c),s}_{\mesoenergy} \big) , 
\end{align}
and is related to its corresponding scaled cumulant generating function via a Legendre-Fenchel transformation
\begin{align} \label{eq:legendrefencheltrafo}
\ratefct\big( \delta \meanfieldcurrent^{(c),s}_{\meanfieldenergy} \big) = \underset{ \delta \meanfieldcurrent^{(c),s}_{\meanfieldenergy}}{\operatorname{sup}} \, [\countingfield^{(c)}_{_{\meanfieldenergy}} \, \delta \meanfieldcurrent^{(c),s}_{\meanfieldenergy} - \scgf^s\big(\countingfield^{(c)}_{_{\meanfieldenergy}}) ] , \qquad \delta \meanfieldcurrent^{(c),s}_{\meanfieldenergy} = \frac{\D \scgf^s\big(\countingfield^{(c)}_{_{\meanfieldenergy}}\big)}{\D \countingfield^{(c)}_{_{\meanfieldenergy}}} , 
\end{align}
where $\operatorname{sup}$ denotes the supremum. As can be seen in Fig. \ref{fig:scgfsymmetryexample}, both the scaled cumulant generating and the rate function are convex functions and the latter has a unique minimum equal to zero.

\begin{figure}[h!] 
\begin{center}
%
%
%
%

\includegraphics[scale=1]{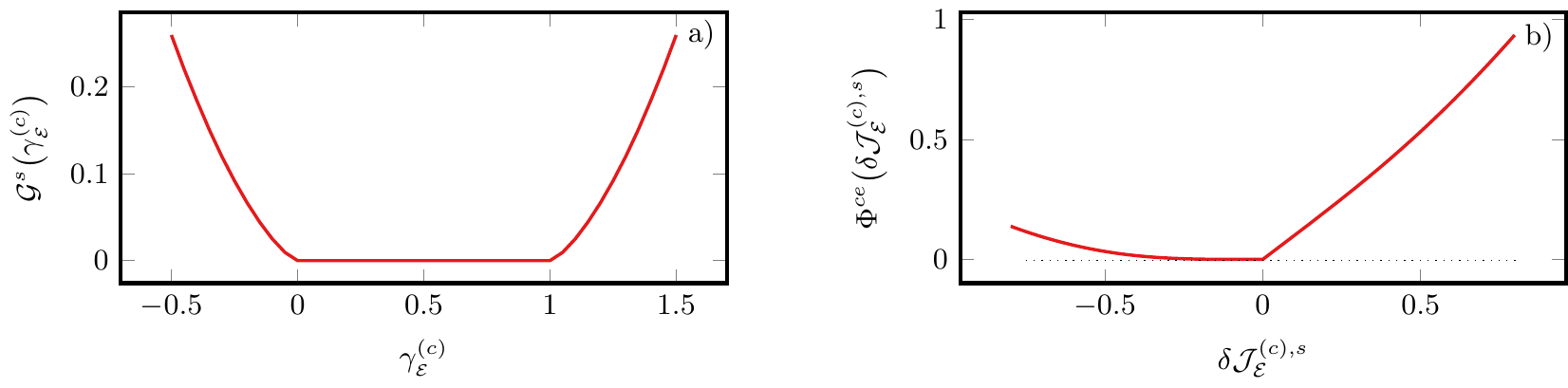}

\caption{The cumulant generating function \eqref{eq:scaledcumulantgeneratingfunctionexample} scaled with time $t$ and size $\dimension$ in a) as a function of the counting field $\countingfield^{(c)}_{_{\meanfieldenergy}}$ and the corresponding convex envelope of the rate function $\ratefct^{ce}\big( \delta \meanfieldcurrent^{(c),s}_{\meanfieldenergy} \big)$ in b). The parameters are chosen as $\invtemperature^{(h)}=1,\invtemperature^{(c)}=3,\potential=-1$ where the unbiased dynamics exhibits a phase transition \eqref{eq:criticalpointexample}. \label{fig:scgfphasetransitionexample} }
\end{center}
\end{figure}

Our thermodynamically consistent framework allows to translate the terminology of nonlinear dynamics, \idest the supercritical pitchfork bifurcation at the critical temperature \eqref{eq:criticalpointexample}, into the language of nonequilibrium statistical mechanics, \idest a nonequilibrium phase transition at the same critical temperature. For this purpose, we prepare the system in its critical state by setting $\invtemperature^{(h)}=1,\invtemperature^{(c)}=3,\potential=-1$. Fig. \ref{fig:scgfphasetransitionexample} depicts in a) the scaled cumulant generating function \eqref{eq:scaledcumulantgeneratingfunctionexample} with the system being in its critical state.
The scaled cumulant generating function exhibits a kink at $\countingfield^{(c)}_{_{\meanfieldenergy}} =0$ indicative of a nonequilibrium phase transition. Owing to the symmetry \eqref{eq:scaledcumulantgeneratingfunctionsymmetryexample}, the scaled cumulant generating function has another kink at $\countingfield^{(c)}_{_{\meanfieldenergy}} =1$.
The non-differentiability of the generating function at $\countingfield^{(c)}_{_{\meanfieldenergy}} =0$ implies that the rate function in Fig.
\ref{fig:scgfphasetransitionexample} b) would be nonconvex over a finite interval. The Legendre-Fenchel transformation \eqref{eq:legendrefencheltrafo} yields not the nononvex rate function but its convex envelope $ \ratefct^{ce}\big( \delta \meanfieldcurrent^{(c),s}_{\meanfieldenergy} \big)$. Here, the part of the convex envelope that replaces the nonconvex regime of the rate function corresponds to the flat part of the curve in the vicinity of the $\delta \meanfieldcurrent^{(c),s}_{\meanfieldenergy} = 0$.
Thus, we find that the time-integrated energy current distribution in Eq. \eqref{eq:ratefunctiondefinition} is bimodal, thus also encoding the nonequilibrium phase transition.

\section{Conclusion}
\label{sec:conclusion}

In this work we demonstrated how to consistently build a stochastic dynamics and thermodynamics description across scales for many-body systems with all-to-all interactions: For this purpose, we considered a system of $\dimension$ all-to-all interacting identical and classical units consisting of $\statenumber$ states. The units undergo transitions due to several heat reservoirs and because of external forces.
The \emph{microscopic} stochastic dynamics characterized by many-body states can be exactly coarse-grained towards a \emph{mesoscopic} one that is determined by the occupation numbers of the different unit states. Here, the all-to-all interactions give rise to equienergetic many-body states which form the mesostates. Importantly, the coarse-graining significantly reduces the complexity of the many-body system as the growth of the state space changes from an exponential to a power-law one.
Employing the formalism of stochastic thermodynamics, it was proven that the stochastic first law of thermodynamics is always invariant under the dynamically exact coarse-graining. Conversely, this only holds true for the stochastic entropy balance if the microstates within each mesostate are equiprobable.

We then considered the macroscopic limit, $\dimension \to \infty$. To consistently determine the macroscopic fluctuations we used the Martin-Siggia-Rose formalism. We showed that the fluctuations that scale exponentially with the system size $\dimension$ are thermodynamically consistent as they obey a detailed fluctuation theorem. Detailed fluctuation theorems of the same form were also derived at the microscopic and mesoscopic level, hence proving thermodynamic consistency across scales. Moreover we proved via the path integral representation of the stochastic dynamics that the mesoscopic master equation asymptotically converges to a nonlinear rate equation. The methodology to determine macroscopic fluctuations was demonstrated via a semi-analytically solvable Ising model in contact with two reservoirs and exhibiting a nonequilibrium phase transition. Our work provides a powerful framework to address the thermodynamics of non-equilibrium phase transitions.

An interesting outcome of this work is that the thermodynamic description of many-body all-to-all interacting systems, when going from a microscopic to an occupation level description, assigns Boltzmann entropies (logarithms of complexion numbers) to each mesostate, despite the fact that the system is driven away from-equilibriun by multiple reservoirs and external forces. Furthermore, in the deterministic macroscopic limit, $\dimension \to \infty$, the ensuing entropy takes the form of a Shannon entropy for the deterministic occupation which exclusively results from these internal mesostate entropies.

Table \ref{tab:summary} provides an overview of the dynamics, fluctuations, local detailed balance (LDB), entropy, entropy production (EP) and detailed fluctuation theorems (DFT) at the microscopic, mesoscopic and macroscopic level.

\begin{table}[h!]
\begin{center}
\begin{tabular}{ | l | c | c | c | c | c | c | } \hline 
   & Dynamics & Fluctuations & LDB & Entropy & EP & DFT \\ \hline 
    Microscopic & \eqref{eq:micromasterequation} & \eqref{eq:generatingfunctionequationofmotion} & \eqref{eq:microlocaldetailedbalance} & \eqref{eq:averageentropy} & \eqref{eq:averageentropyproduction} & \eqref{eq:microdftjoint} \\ \hline 
    Mesoscopic & \eqref{eq:mesoscopicmasterequation} & \eqref{eq:mesoscopicgeneratingfunctionequationofmotion} & \eqref{eq:mesolocaldetailedbalance} & \eqref{eq:averageentropy} & \eqref{eq:averageentropyproduction} & \eqref{eq:mesodftjoint} \\ \hline 
    Macroscopic & \eqref{eq:meanfieldmasterequation} & \eqref{eq:mesogeneratingentropyproductionfunctionpathintegraldominant},\eqref{eq:actionfunctionalderivative} & \eqref{eq:meanfieldlocaldetailedbalance} & \eqref{eq:meanfieldentropy} & \eqref{eq:meanfieldentropyproduction} & \eqref{eq:macrojointdft} \\ \hline 
  \end{tabular} 
\caption{Compilation of the key equations specifying the dynamics, fluctuations, local detailed balance (LDB), entropy, entropy production (EP) and detailed fluctuation theorems (DFT) at the microscopic, mesoscopic and macroscopic level. \label{tab:summary} }  
   \end{center}
\end{table}

We end by placing our findings in the context of the recent works on thermodynamically consistent coarse-graining. Many of these are based on time-scale separation: fast degrees of freedom reach a local stationary state over time-scales much shorter than the slow dynamics and can be adiabatically eliminated. The resulting transition rates of the slow dynamics then satisfy a local detailed balance condition which carries the information about the thermodynamic potentials (energetic and/or entropic) \cite{esposito2012pre, bo2014jsp, herpich2019pre2} or the driving forces \cite{wachtel2018njp, esposito2016pre, esposito2015pre} resulting from the fast dynamics.
Some coarse-grainings do not require time-scale separation and the hidden degrees of freedom have been shown to behave as work sources (pure energy no entropy) on the remaining degrees of freedom, see e.g. Refs. \cite{herpich2019pre2,verley2014njp}.
In the present work, the coarse-graining does not rely on any time-scale separation but results from the all-to-all interactions which do not discriminate energetically between the different microstates leading to the same global occupation. As a result a purely entropic contributions ensues at the occupation level. Models where such coarse grainings appeared can be found in Refs. \cite{herpich2018prx,herpich2019pre,verley2017epl}.

\section*{Acknowledgments}

T. H. warmly thanks Gatien Verley for stimulating exchanges.
We acknowledge support by the National Research Fund, Luxembourg, in the frame of the AFR PhD Grant 2016, No. 11271777 and by the  European Research Council project NanoThermo (ERC-2015-CoG Agreement No. 681456).

\appendix

\section{Derivation of the Equiprobability of Microstates Inside Stationary Mesostates}
\label{sec:spanningtreeappendix}

For finite systems, the stationary microscopic probabilities can be determined via the spanning-tree formula \cite{schnakenberg1976rmp}.
A spanning tree, $\tree(\graph)$ of a graph $\graph$ consists only of edges that are also edges of $\graph$ and contains all vertices (microstates $\microstate$) of $\graph$. Further, a spanning tree $\tree'(\graph)$ is connected and contains no circuits. We write $ \tree^{(\mu)}_{\microstate}(\graph) $ for the $\mu$th spanning tree rooted in $\microstate$, that is a tree with branches that are directed towards the vertex $\microstate$.
The spanning-tree formula reads
\begin{align} \label{eq:spanningtreeformula}
\microprobability^{s}_{\microstate}(\bm{\protocol}) = 
\frac{\sum\limits_{\mu} \tree^{(\mu)}_{\microstate}(\graph) }{\sum\limits_{\microstate} \sum\limits_{\mu} \tree^{(\mu)}_{\microstate}(\graph) } 
= \frac{ \sum\limits_{\tree_{\microstate}(\graph)} \; \prod\limits_{ \substack{ \microstate', \microstate'' \text{ such that } \\ \text{  current is directed to } \microstate }} \microrates_{\microstate'\microstate''}(\bm{\protocol}) }{ \sum\limits_{\microstate}\sum\limits_{\tree_{\microstate}(\graph)} \prod\limits_{ \substack{ \microstate', \microstate'' \text{ such that } \\ \text{  current is directed to } \microstate } } \microrates_{\microstate'\microstate''}(\bm{\protocol}) } \, . 
\end{align} 
First we note that the transition rates do not depend on the microstates $\microstate$ and $\microstate'$ belonging to the same pair of mesostates $(\mesostate,\mesostate')$, \idest $\microrates_{\microstate_{_{\mesostate}} \microstate'_{_{\mesostate'}}}(\bm{\protocol}) = \text{const } \; \forall \, \microstate_{_{\mesostate}}, \microstate'_{_{\mesostate'}} $.
Secondly, the number of possible transitions for any microstate belonging to a given mesostate is a constant such that the number of spanning trees rooted in $\microstate_{_{\mesostate}}$ is constant for all $\microstate_{_{\mesostate}}$. It therefore holds that all microstates belonging to the same mesostate are equally probable,
\begin{align}
\microprobability^{s}_{\microstate_{_{\mesostate}}}(\bm{\protocol}) &= 
\frac{ \sum\limits_{\tree_{\microstate_{\mesostate}}(\graph)} \; \prod\limits_{ \substack{ \microstate'_{\mesostate'},\microstate''_{\mesostate''} \text{ such that } \\ \text{ current is directed to } \microstate_{_{\mesostate}} }}  \microrates_{\microstate'_{\mesostate'},\microstate''_{\mesostate''}} (\bm{\protocol}) }
{ \sum\limits_{\mesostate}  \sum\limits_{ \microstate_{_{\mesostate}} }
 \sum\limits_{\tree_{\microstate_{\mesostate}}(\graph)} 
\; \prod\limits_{ \substack{ \microstate'_{\mesostate'},\microstate''_{\mesostate''} \text{ such that } \\ \text{ current is directed to } \microstate_{_{\mesostate}} }} \microrates_{\microstate'_{\mesostate'},\microstate''_{\mesostate''}}(\bm{\protocol})
} = \mathrm{const} \;\, \forall \, \microstate_{_{\mesostate}} ,
\end{align}
as claimed in Eq. \eqref{eq:stationarymicroprobabilities}.

\section{Derivation of the Detailed Fluctuation Theorem: Time-Evolution Operator}
\label{sec:dftappendix}

In this section we prove the detailed fluctuation theorem \eqref{eq:microdftjoint} following a procedure detailed in Ref. \cite{rao2018njp}. We denote by $\microprobability_{\microstate} \big(\delta \trajectoryworkmicro_{\bm{\protocol}} , \lbrace \delta \microcurrent^{(\reservoir)}_{\bm{\force}} \rbrace , \lbrace \delta \microcurrent^{(\reservoir)}_{\trajectoryenergymicro} \rbrace, t \big)$ the joint probability to observe a work contribution $ \invtemperature^{(1)} \delta \trajectoryworkmicro_{\bm{\protocol}}$ defined in Eq. \eqref{eq:stochasticwork} and time-integrated autonomous work and energy currents $\lbrace \delta \microcurrent^{(\reservoir)}_{\bm{\force}} \rbrace $ and $\lbrace \delta \microcurrent^{(\reservoir)}_{\trajectoryenergymicro} \rbrace $ defined in Eqs. \eqref{eq:stochasticworkdecomposition} and \eqref{eq:stochasticenergydecomposition}, respectively, along a trajectory that is in state $\microstate$ at time $t$. In the following, we note arrays with bold characters whose entries in case of the generating function $\bm{\microgeneratingfunction}$ and the associated probability $\bm{\microprobability}$ correspond to different microstates $\microstate$. According to Eqs. \eqref{eq:stochasticentropyproduction} and \eqref{eq:generatingfunctiondefinitionrewritten}, the microscopic generating function associated with $\delta \trajectoryworkmicro_{\bm{\protocol}}$, $\lbrace \delta \microcurrent^{(\reservoir)}_{\bm{\force}} \rbrace $ and $ \lbrace \delta \microcurrent^{(\reservoir)}_{\trajectoryenergymicro} \rbrace$ is given by
\begin{align} \label{eq:generatingfunctiondefinitionappendix} 
\begin{aligned}
\bm{\microgeneratingfunction}(\i \countingfield_{_{\bm{\protocol}}}, \lbrace \i\countingfield_{_{\bm{\force}}}^{(\reservoir)} \rbrace , \lbrace \i \countingfield_{_{\trajectoryenergymicro}}^{(\reservoir)} \rbrace ,t) &= \int\limits_{-\infty}^{\infty} \prod_{ \reservoir } \d \Big( \invtemperature^{(1)} \delta \trajectoryworkmicro_{\bm{\protocol}} \Big) \d \Big( \delta  \microcurrent^{(\reservoir)}_{\bm{\force}} \Big) \d \Big( \delta \microcurrent^{(\reservoir)}_{\trajectoryenergymicro} \Big) \; \cdot \\ 
&\cdot \exp \Big\lbrace - \i \countingfield_{_{\bm{\protocol}}} \invtemperature^{(1)} \delta \trajectoryworkmicro_{\bm{\protocol}} - \i \countingfield_{_{\bm{\force}}  }^{(\reservoir)} \; \delta \microcurrent^{(\reservoir)}_{\bm{\force}} - \i \countingfield_{_{\trajectoryenergymicro}}^{(\reservoir)} \; \delta \microcurrent^{(\reservoir)}_{\trajectoryenergymicro} \Big\rbrace  \, \bm{\microprobability} \big(\delta \trajectoryworkmicro_{\bm{\protocol}} , \lbrace \delta \microcurrent^{(\reservoir)}_{\bm{\force}} \rbrace , \lbrace \delta \microcurrent^{(\reservoir)}_{\trajectoryenergymicro} \rbrace , t \big) , 
\end{aligned}
\end{align} 
and its time evolution is governed by the following biased stochastic dynamics
\begin{align}  \label{eq:biaseddynamicsappendix}
\d_t \, \bm{\microgeneratingfunction}(\countingfield_{_{\bm{\protocol}}}, \lbrace \countingfield_{_{\bm{\force}}}^{(\reservoir)} \rbrace , \lbrace \countingfield_{_{\trajectoryenergymicro}}^{(\reservoir)} \rbrace ,t) &= \bm{\microrates}(\countingfield_{_{\bm{\protocol}}}, \lbrace \countingfield_{_{\bm{\force}}}^{(\reservoir)} \rbrace , \lbrace \countingfield_{_{\trajectoryenergymicro}}^{(\reservoir)} \rbrace, \bm{\protocol}_t) \cdot \bm{\microgeneratingfunction}(\countingfield_{_{\bm{\protocol}}}, \lbrace \countingfield_{_{\bm{\force}}}^{(\reservoir)} \rbrace , \lbrace \countingfield_{_{\trajectoryenergymicro}}^{(\reservoir)} \rbrace ,t), 
\end{align}
with the biased generator
\begin{align} \label{eq:biasedgeneratorsappendix} 
\microrates_{\microstate \microstate'}(\countingfield_{_{\bm{\protocol}}}, \lbrace \countingfield_{_{\bm{\force}}}^{(\reservoir)} \rbrace , \lbrace \countingfield_{_{\trajectoryenergymicro}}^{(\reservoir)} \rbrace, \bm{\protocol}_t) = - \countingfield_{_{\bm{\protocol}}} \, \invtemperature^{(1)}  \big[ \dot{\bm{\protocol}}_t \cdot \nabla_{\bm{\protocol}_t} \, \trajectoryenergymicro_{\microstate}(\bm{\protocol}_t)\big] \delta_{\microstate,\microstate'} + \microrates_{\microstate \microstate'}(\bm{\protocol}_t) \prod\limits_{\reservoir} \exp \Big[ - \countingfield_{\microcurrent^{(\reservoir)}_{\bm{\force}} } \; \microcurrent^{(\reservoir)}_{_{\bm{\force}_{\microstate \microstate'}}} - \, \countingfield_{\microcurrent^{(\reservoir)}_{\trajectoryenergymicro}} \; \microcurrent^{(\reservoir)}_{_{\microstate \microstate'}} \Big] ,
\end{align}
where $ \microcurrent^{(\reservoir)}_{_{\bm{\force}_{\microstate \microstate'}}} = \invtemperature^{(\reservoir)} \bm{\force}^{(\reservoir)}_{\microstate,\microstate'} $ and $ \microcurrent^{(\reservoir)}_{\trajectoryenergymicro_{\microstate \microstate'}} = [ \invtemperature^{(1)} - \invtemperature^{(\reservoir)} ]  [ \microenergy_{\microstate}(\bm{\protocol}_t) - \microenergy_{\microstate'}(\bm{\protocol}_t) ] $ denote the change in the autonomous work and energy current during a transition $\microstate' \overset{(\reservoir)}{\to} \microstate$, respectively.
We easily verify with Eq. \eqref{eq:microlocaldetailedbalance} that the biased generator satisfies the following symmetry 
\begin{align} \label{eq:biasedgeneratorsymmetrysappendix} 
\bm{\microrates}^{\top}(\countingfield_{_{\bm{\protocol}}}, \lbrace \countingfield_{_{\bm{\force}}}^{(\reservoir)} \rbrace , \lbrace \countingfield_{_{\trajectoryenergymicro}}^{(\reservoir)} \rbrace, \bm{\protocol}_t)  = \bm{A}^{-1}(\bm{\protocol}_t) \cdot \bm{\microrates}(\countingfield_{_{\bm{\protocol}}}, \lbrace 1 - \countingfield_{_{\bm{\force}}}^{(\reservoir)} \rbrace , \lbrace 1 - \countingfield_{_{\trajectoryenergymicro}}^{(\reservoir)} \rbrace, \bm{\protocol}_t) \cdot \bm{A}(\bm{\protocol}_t) ,
\end{align}
with the matrix
\begin{align}
A_{\microstate \microstate'}(\bm{\protocol}_t) = \exp \Big[ -\invtemperature^{(1)} \microenergy_{\microstate}(\bm{\protocol}_t) \Big] \, \delta_{\microstate \microstate'} .
\end{align}
In this notation, the initial Gibbs states \eqref{eq:initialequilibriummicro} can be written as
\begin{align}
\bm{\microgeneratingfunction}(\countingfield_{_{\bm{\protocol}}}, \lbrace \countingfield_{_{\bm{\force}}}^{(\reservoir)} \rbrace , \lbrace \countingfield_{_{\trajectoryenergymicro}}^{(\reservoir)} \rbrace , 0) = \mathbf{\microprobability}^{eq}(\bm{\protocol}_0) =  \bm{A}(\bm{\protocol}_0) \cdot \bm{1} \, \exp \Big[ \invtemperature^{(1)} \microfreeenergy^{eq}_1(\bm{\protocol}_0) \Big] , 
\end{align}
where $\bm{1}$ refers to a vector whose entries are all unity.

Since it will be useful to proceed later, we now prove a preliminary result. To this end, we consider a generic biased dynamics as in Eq. \eqref{eq:biaseddynamicsappendix}
\begin{align} 
\D_t \, \bm{\microgeneratingfunction}(\countingfield,t) &= \bm{\microrates}(\countingfield, \bm{\protocol}_t) \cdot \bm{\microgeneratingfunction}(\countingfield,t) , 
\end{align}
which has the formal solution
\begin{align} 
\bm{\microgeneratingfunction}(\countingfield,t) &= \bm{U}(\countingfield, t) \cdot \bm{\microprobability}(0) , 
\end{align}
with the time-evolution operator
\begin{align} 
\bm{U}(\countingfield, t) = T_{+} \; \exp \Bigg[ \int\limits_0^{t} \d t' \, \bm{\microrates}(\countingfield, \bm{\protocol}_{t'}) \Bigg] , 
\end{align}
where $T_{+}$ is the time-ordering operator. We define a transformed time-evolution operator
\begin{align}
\hat{\bm{U}}(\countingfield, t) = \bm{B}^{-1}(\bm{\protocol}_t)\cdot \bm{U}(\countingfield, t) \cdot \bm{B}(\bm{\protocol}_0)
\end{align}
with a generic but invertible operator $\bm{B}$ and find for its evolution equation
\begin{align} 
\D_t \hat{\bm{U}}(\countingfield, t) = \big\lbrace [ \d_t \bm{B}^{-1}(\bm{\protocol}_t) ] \cdot \bm{B}(\bm{\protocol}_t) + \bm{B}^{-1}(\bm{\protocol}_t) \cdot \bm{\microrates}(\countingfield, \bm{\protocol}_t) \cdot \bm{B}(\bm{\protocol}_t) \big\rbrace \, \hat{\bm{U}}(\countingfield, t) \equiv \hat{\bm{\microrates}}(\countingfield, \bm{\protocol}_t) \cdot \hat{\bm{U}}(\countingfield, t) , 
\end{align}
which implies for the transformed time-evolution operator
\begin{align} \label{eq:timeevolutionoperatortransformedappendix}
\hat{\bm{U}}(\countingfield, t) = T_{+} \; \exp \Bigg[ \int\limits_0^{t} \d t' \, \hat{\bm{\microrates}}(\countingfield, \bm{\protocol}_{t'}) \Bigg] . 
\end{align}
Combining the last three equations, we arrive at the preliminary result
\begin{align} \label{eq:timeevolutionoperatortransformedresultappendix}
\bm{B}^{-1}(\bm{\protocol}_t) \cdot \bm{U}(\countingfield, t) \cdot \bm{B}(\bm{\protocol}_0) = T_{+} \; \exp \Bigg[ \int\limits_0^{t} \d t' \,  \Big\lbrace [ \d_{t'} \bm{B}^{-1}(\bm{\protocol}_{t'}) ] \cdot \bm{B}(\bm{\protocol}_{t'}) + \bm{B}^{-1}(\bm{\protocol}_{t'}) \cdot \bm{\microrates}(\countingfield, \bm{\protocol}_{t'}) \cdot \bm{B}(\bm{\protocol}_{t'}) \Big\rbrace \Bigg] . 
\end{align}

We now return to the specific biased stochastic dynamics considered in Eq. \eqref{eq:biaseddynamicsappendix} and obtain for the generating function 
\begin{align}  
\begin{aligned}
\microgeneratingfunction(\countingfield_{_{\bm{\protocol}}}, \lbrace \countingfield_{_{\bm{\force}}}^{(\reservoir)} \rbrace , \lbrace \countingfield_{_{\trajectoryenergymicro}}^{(\reservoir)} \rbrace ,t) &= \bm{1} \cdot \bm{U}(\countingfield_{_{\bm{\protocol}}}, \lbrace \countingfield_{_{\bm{\force}}}^{(\reservoir)} \rbrace , \lbrace \countingfield_{_{\trajectoryenergymicro}}^{(\reservoir)} \rbrace ,t) \cdot \bm{A}(\bm{\protocol}_0) \cdot \bm{1} \, \exp \Big[ \invtemperature^{(1)} \microfreeenergy^{eq}_1(\bm{\protocol}_0) \Big] \\ 
&= \! \exp \!\Big[ \invtemperature^{(1)} \microfreeenergy^{eq}_1(\bm{\protocol}_t) \big] \bm{1} \!\cdot \! \bm{A}(\bm{\protocol}_t) \!\cdot \! \bm{A}^{-1}(\bm{\protocol}_t) \!\cdot \! \bm{U}(\countingfield_{_{\bm{\protocol}}}, \lbrace \countingfield_{_{\bm{\force}}}^{(\reservoir)} \rbrace , \lbrace \countingfield_{_{\trajectoryenergymicro}}^{(\reservoir)} \rbrace ,t) \! \cdot \! \bm{A}(\bm{\protocol}_0) \cdot \! \bm{1} \! \, \exp \Big[ \!-\! \invtemperature^{(1)} \Delta \microfreeenergy^{eq}_1(\bm{\protocol}) \Big]  \label{eq:generatingfunctionrelationoneappendix} , 
\end{aligned}
\end{align}
where $ \bm{U}(\countingfield_{_{\bm{\protocol}}}, \lbrace \countingfield_{_{\bm{\force}}}^{(\reservoir)} \rbrace , \lbrace \countingfield_{_{\trajectoryenergymicro}}^{(\reservoir)} \rbrace ,t) $ is the time-evolution operator for that biased stochastic dynamics. Owing to the assumption of initial equilibrium distributions for the forward and backward process, $ \microprobability^{eq}(\bm{\protocol}_0) $ and $ \microprobability^{eq}(\bm{\protocol}_t) $, we have 
\begin{align}
\mathbf{\microprobability}^{eq}(\bm{\protocol}_t) = \bm{1} \cdot \bm{A}(\bm{\protocol}_t) \, \exp \Big[ \invtemperature^{(1)} \microfreeenergy^{eq}_1(\bm{\protocol}_t) \Big] .
\end{align}
Substituting the last and Eq. \eqref{eq:timeevolutionoperatortransformedresultappendix} into Eq. \eqref{eq:generatingfunctionrelationoneappendix} gives
\begin{align}
\begin{aligned}
& \microgeneratingfunction(\countingfield_{_{\bm{\protocol}}}, \lbrace \countingfield_{_{\bm{\force}}}^{(\reservoir)} \rbrace , \lbrace \countingfield_{_{\trajectoryenergymicro}}^{(\reservoir)} \rbrace ,t) \, = \\ 
&= \!\! \mathbf{\microprobability}^{eq}(\bm{\protocol}_t) \!\cdot \! T_{+} \! \exp \! \Bigg[ \! \int\limits_0^{t} \! \d t' \! \big\lbrace [ \dot{\bm{\protocol}}_{t'} \!\cdot\! \! \nabla_{\bm{\protocol}_{t'}} \bm{A}^{-1}(\bm{\protocol}_{t'}) ] \!\cdot \!\! \bm{A}(\bm{\protocol}_{t'}) \!\!+\!\! \bm{A}^{-1}(\bm{\protocol}_{t'}) \!\cdot \! \bm{\microrates}(\countingfield_{_{\bm{\protocol}}}, \lbrace  \countingfield_{_{\bm{\force}}}^{(\reservoir)} \rbrace , \lbrace \countingfield_{_{\trajectoryenergymicro}}^{(\reservoir)} \rbrace, \bm{\protocol}_{t'}) \!\cdot \! \bm{A}(\bm{\protocol}_{t'}) \big\rbrace \! \Bigg] \!\! \cdot \! \bm{1} \exp \!\! \big[ \! -\! \invtemperature^{(1)} \Delta \microfreeenergy^{eq}_1(\bm{\protocol}) \big] \\
&= \mathbf{\microprobability}^{eq}(\bm{\protocol}_t) \cdot T_{+} \; \exp \Bigg[ \int\limits_0^{t} \d t' \, \bm{A}^{-1}(\bm{\protocol}_{t'}) \cdot \bm{\microrates}(\countingfield_{_{\bm{\protocol}}} - 1, \lbrace \countingfield_{_{\bm{\force}}}^{(\reservoir)} \rbrace , \lbrace \countingfield_{_{\trajectoryenergymicro}}^{(\reservoir)} \rbrace,  \bm{\protocol}_{t'}) \cdot \bm{A}(\bm{\protocol}_{t'}) \Bigg] \cdot \bm{1} \; \exp \big[ - \invtemperature^{(1)} \Delta \microfreeenergy^{eq}_1(\bm{\protocol}) \big] \\ 
&= \mathbf{\microprobability}^{eq}(\bm{\protocol}_t) \cdot T_{+} \; \exp \Bigg[ \int\limits_0^{t} \d t' \, \bm{\microrates}^{\top}(\countingfield_{_{\bm{\protocol}}} - 1, \lbrace 1 - \countingfield_{_{\bm{\force}}}^{(\reservoir)} \rbrace , \lbrace 1 - \countingfield_{_{\trajectoryenergymicro}}^{(\reservoir)} \rbrace,  \bm{\protocol}_{t'}) \Bigg] \cdot \bm{1} \; \exp \big[ - \invtemperature^{(1)} \Delta \microfreeenergy^{eq}_1(\bm{\protocol}) \big] 
\label{eq:generatingfunctionrelationtwoappendix}  , 
\end{aligned}
\end{align}
where we used $ [ \dot{\bm{\protocol}}_{t'} \cdot \nabla_{\bm{\protocol}_{t'}} \bm{A}^{-1}(\bm{\protocol}_{t'}) ] \cdot \bm{A}(\bm{\protocol}_{t'}) = \invtemperature^{(1)} \, \mathrm{diag} \lbrace \dot{\bm{\protocol}}_{t'} \cdot \nabla_{\bm{\protocol}_{t'}} \mesoenergy_{\mesostate}(\bm{\protocol}_t) \rbrace $ in the second and Eq. \eqref{eq:biasedgeneratorsymmetrysappendix} in the last equality.
Next, we transform the time from $t' $ to $\tilde{t'} = t - t' $ corresponding to a time-reserved process. As a result of this transformation, the time-ordering operator becomes an anti-time-ordering one $T_{-}$ and the diagonal entries of the biased generator \eqref{eq:biasedgeneratorsappendix} become 
\begin{align}
\microrates_{\microstate \microstate}(\countingfield_{_{\bm{\protocol}}}, \lbrace \countingfield_{_{\bm{\force}}}^{(\reservoir)} \rbrace , \lbrace \countingfield_{_{\trajectoryenergymicro}}^{(\reservoir)} \rbrace, \bm{\protocol}_{t - \tilde{t'}}) =  \countingfield_{_{\bm{\protocol}}} \, \dot{\bm{\protocol}}_{t - \tilde{t'}} \cdot \nabla_{\bm{\protocol}_{t - \tilde{t'}}} \, \trajectoryenergymicro_{\microstate}(\bm{\protocol}_{t - \tilde{t'}}) + \microrates_{\microstate \microstate}(\bm{\protocol}_{t - \tilde{t'}}) 
= - \countingfield_{_{\bm{\protocol}}} \, \dot{\bm{\protocol}}_{t - \tilde{t'}} \cdot \nabla_{\bm{\protocol}_{ \tilde{t'}}} \, \trajectoryenergymicro_{\microstate}(\bm{\protocol}_{t - \tilde{t'}}) + \microrates_{\microstate \microstate}(\bm{\protocol}_{t - \tilde{t'}}) .
\end{align}
Thus, we conclude that
\begin{align}
\microrates_{\microstate \microstate'}(\countingfield_{_{\bm{\protocol}}}, \lbrace \countingfield_{_{\bm{\force}}}^{(\reservoir)} \rbrace , \lbrace \countingfield_{_{\trajectoryenergymicro}}^{(\reservoir)} \rbrace, \bm{\protocol}_{t - \tilde{t'}}) = \microrates_{\microstate \microstate'}(- \countingfield_{_{\bm{\protocol}}}, \lbrace \countingfield_{_{\bm{\force}}}^{(\reservoir)} \rbrace , \lbrace \countingfield_{_{\trajectoryenergymicro}}^{(\reservoir)} \rbrace, \bm{\protocol}_{t - \tilde{t'}}) \equiv 
\tilde{\microrates}_{\microstate \microstate'}( - \countingfield_{_{\bm{\protocol}}}, \lbrace \countingfield_{_{\bm{\force}}}^{(\reservoir)} \rbrace , \lbrace \countingfield_{_{\trajectoryenergymicro}}^{(\reservoir)} \rbrace, \tilde{\bm{\protocol}}_{\tilde{t}'}) , 
\end{align}
where we introduced the biased generator of the time-reversed stochastic dynamics $\tilde{\microrates}_{\microstate \microstate'}(\countingfield_{_{\bm{\protocol}}}, \lbrace \countingfield_{_{\bm{\force}}}^{(\reservoir)} \rbrace , \lbrace \countingfield_{_{\trajectoryenergymicro}}^{(\reservoir)} \rbrace, \tilde{\bm{\protocol}}_{t'})$ that is naturally a function of the time-reversed protocol, $\tilde{\bm{\protocol}}_{t'} = \bm{\protocol}_{t-t'}, \, t' ~\in ~ [0,t]$. Consequently, Eq. \eqref{eq:generatingfunctionrelationtwoappendix} becomes
\begin{align}
\begin{aligned}
\microgeneratingfunction(\countingfield_{\trajectoryworkmicro_{\tilde{\bm{\protocol}}}}, \lbrace \countingfield_{_{\bm{\force}}}^{(\reservoir)} \rbrace , \lbrace \countingfield_{_{\trajectoryenergymicro}}^{(\reservoir)} \rbrace ,t) &= \tilde{\mathbf{\microprobability}}^{eq}(\tilde{\bm{\protocol}}_0) \cdot T_{-} \; \exp \Bigg[ \int\limits_0^{t} \d \tilde{t'} \, \tilde{\bm{\microrates}}^{\top}( 1 - \countingfield_{_{\bm{\protocol}}} , \lbrace 1 - \countingfield_{_{\bm{\force}}}^{(\reservoir)} \rbrace , \lbrace 1 - \countingfield_{_{\trajectoryenergymicro}}^{(\reservoir)} \rbrace,  \tilde{\bm{\protocol}}_{t'}) \Bigg] \cdot \bm{1} \; \exp \big[ - \invtemperature^{(1)} \Delta \microfreeenergy^{eq}_1(\bm{\protocol}) \big] \\
&= \bm{1} \cdot T_{+} \; \exp \Bigg[ \int\limits_0^{t} \d \tilde{t'} \, \tilde{\bm{\microrates}}( 1 - \countingfield_{_{\bm{\protocol}}} , \lbrace 1 - \countingfield_{_{\bm{\force}}}^{(\reservoir)} \rbrace , \lbrace 1 - \countingfield_{_{\trajectoryenergymicro}}^{(\reservoir)} \rbrace,  \tilde{\bm{\protocol}}_{t'}) \Bigg] \cdot \tilde{\mathbf{\microprobability}}^{eq}(\tilde{\bm{\protocol}}_0) \; \exp \big[ - \invtemperature^{(1)} \Delta \microfreeenergy^{eq}_1(\bm{\protocol}) \big]  
\label{eq:generatingfunctionrelationthreeappendix}  .
\end{aligned}
\end{align}
In the last equality we applied a global transposition and used the relationship
\begin{align}
T_{+} \Big( \prod\limits_i \bm{C}(\bm{\protocol}_{t_i})^{\top} \Big) = \Big( T_{-} \prod\limits_i \bm{C}(\bm{\protocol}_{t_i}) \Big)^{\top} , 
\end{align}
that is valid for a generic operator $\bm{C}$. Inserting Eq. \eqref{eq:timeevolutionoperatortransformedappendix} into Eq. \eqref{eq:generatingfunctionrelationthreeappendix} yields
\begin{align}
\begin{aligned}
\microgeneratingfunction(\countingfield_{_{\bm{\protocol}}}, \lbrace \countingfield_{_{\bm{\force}}}^{(\reservoir)} \rbrace , \lbrace \countingfield_{_{\trajectoryenergymicro}}^{(\reservoir)} \rbrace ,t) &= \bm{1} \cdot \tilde{\bm{U}}(1 - \countingfield_{_{\bm{\protocol}}}, \lbrace 1 - \countingfield_{_{\bm{\force}}}^{(\reservoir)} \rbrace , \lbrace 1 - \countingfield_{_{\trajectoryenergymicro}}^{(\reservoir)} \rbrace ,t) \cdot \tilde{\mathbf{\microprobability}}^{eq}(\tilde{\bm{\protocol}}_t) \; \exp \big[ - \invtemperature^{(1)} \Delta \microfreeenergy^{eq}_1(\bm{\protocol}) \big]  
\label{eq:generatingfunctionrelationfourappendix}  ,
\end{aligned}
\end{align}
from which we conclude the following symmetry
\begin{align}
\microgeneratingfunction(\countingfield_{_{\bm{\protocol}}}, \lbrace \countingfield_{_{\bm{\force}}}^{(\reservoir)} \rbrace , \lbrace \countingfield_{_{\trajectoryenergymicro}}^{(\reservoir)} \rbrace ,t) = \tilde{\microgeneratingfunction}(1 - \countingfield_{_{\bm{\protocol}}}, \lbrace 1 - \countingfield_{_{\bm{\force}}}^{(\reservoir)} \rbrace , \lbrace 1 - \countingfield_{_{\trajectoryenergymicro}}^{(\reservoir)} \rbrace ,t) \; \exp \big[ - \invtemperature^{(1)} \Delta \microfreeenergy^{eq}_1(\bm{\protocol}) \big] , 
\label{eq:generatingfunctionrelationfourappendix} 
\end{align}
which via inverse Fourier transformation of the definition \eqref{eq:generatingfunctiondefinitionappendix} stipulates the detailed fluctuation theorem
\begin{align} \label{eq:microdftjointappendix}
\ln \frac{ \microprobability \Big( \invtemperature^{(1)} \delta  \trajectoryworkmicro_{\bm{\protocol}} , \lbrace \delta \microcurrent^{(\reservoir)}_{\bm{\force}} \rbrace , \lbrace \delta \microcurrent^{(\reservoir)}_{\trajectoryenergymicro} \rbrace \Big) }{
\tilde{\microprobability} \Big( - \invtemperature^{(1)} \delta \trajectoryworkmicro_{\bm{\protocol}} \, , - \lbrace \delta \microcurrent^{(\reservoir)}_{\bm{\force}} \rbrace , - \lbrace \delta \microcurrent^{(\reservoir)}_{\trajectoryenergymicro} \rbrace \Big) } = \invtemperature^{(1)} \big[ \delta \trajectoryworkmicro_{\bm{\protocol}} - \Delta \microfreeenergy^{eq}_1 \big] + \sum\limits_{\reservoir = 1}^{\reservoirdimension} \big[ \invtemperature^{(\reservoir)} \delta \trajectoryworkmicro_{\bm{\force}}^{(\reservoir)} + \big[ \invtemperature^{(1)} - \invtemperature^{(\reservoir)} \big] \delta \trajectoryenergymicro^{(\reservoir)} \big] ,
\end{align}
as stated in Eq. \eqref{eq:microdftjoint}.

\section{Derivation of the Detailed Fluctuation Theorem: Path-Integral Formalism}
\label{sec:martinsiggiaroseappendix}

We finally present the proof of the asymptotic symmetry in Eq. \eqref{eq:macrogeneratingfunctionsymmetry}.
The path-integral representation of the dominant contribution to the generating function associated with the size-extensive fluctuations of the mesoscopic entropy production with bounding Gibbs states \eqref{eq:mesogeneratingentropyproductionfunctionpathintegraldominant} reads
\begin{align} \label{eq:mesogeneratingfunctionentropyproductionpathintegralappendix} 
\begin{aligned}
&\mesogeneratingfunction( \bm{\countingfield} ,t) = \int \pathintegralmeasure [\bm{\meanfieldprobability}] \int \pathintegralmeasure [\bm{\field}] \; \mesoprobability_{\bm{\meanfieldprobability}_{_0}}^{eq}(\bm{\protocol}_0) \cdot \exp \Bigg\lbrace \dimension \int\limits_0^t \d t'\Bigg[ \!-\! \countingfield_{_{\bm{\lambda}}} \invtemperature^{(1)} \dot{\bm{\protocol}}_{t'} \cdot \big[ \nabla_{\bm{\protocol}_{t'}} \meanfieldenergy_{\bm{\meanfieldprobability}}(\bm{\protocol}_{t'}) \big] + \sum\limits_{i=1}^{\statenumber} \Big\lbrace - \field_i(t') \dot{\meanfieldprobability}_i(t') \\ 
&+ \! \sum\limits_{\reservoir = 1}^{\reservoirdimension} \sum\limits_{j=1}^{\statenumber} \! \Big[ \! \exp \Big\lbrace \field_i(t') \!-\! \field_j(t') \!-\! \countingfield_{_{\bm{\force}}}^{(\reservoir)} \invtemperature^{(\reservoir)} \force_{ij}^{(\reservoir)} \!-\! \countingfield_{\meanfieldenergy}^{(\reservoir)} [ \invtemperature^{(1)} \!-\! \invtemperature^{(\reservoir)} ] \, \meanfieldenergy_{ij}^{(\reservoir)}\big(\bm{\protocol}_{t'},\bm{\meanfieldprobability}(t') \big) \! \Big\rbrace \!-\! 1 \Big]  \meanfieldrates_{ij}^{(\reservoir)} \big(\bm{\protocol}_{t'},\bm{\meanfieldprobability}(t') \big) \, \meanfieldprobability_j(t') \Big\rbrace \Bigg] \Bigg\rbrace , 
\end{aligned}
\end{align}
where, for better readability, the subextensive terms in the exponential, $ \mathbbm{o}(1) $, are omitted in the following.

The crucial step of the derivation is to define physically consistent transformation rules to time-reverse the biased stochastic dynamics. Time-reversal transformations of unbiased Langevin dynamics have been investigated in Ref. \cite{aron2010jsm}. For the generating function in question \eqref{eq:mesogeneratingfunctionentropyproductionpathintegralappendix}, we define the time-reversed biased stochastic dynamics as follows
\begin{align} \label{eq:timereversaltransformationappendix}
\begin{aligned}
&\tilde{t}' = t - t', \quad \tilde{\bm{\meanfieldprobability}} = \bm{\meanfieldprobability}, \quad \tilde{\bm{\protocol}}_{t'} = \bm{\protocol}_{t-t'}, \\
& \tilde{\bm{\field}} = - \bm{\field} +  \invtemperature^{(1)} \, \nabla_{\bm{\meanfieldprobability}} \, \meanfieldfreeenergy_{\bm{\meanfieldprobability}}^{(\reservoir)}(\bm{\protocol}_t) = - \bm{\field} + \invtemperature^{(1)} \,  \nabla_{\bm{\meanfieldprobability}} \, \meanfieldenergy_{\bm{\meanfieldprobability}}(\bm{\protocol}_t) -  \nabla_{\bm{\meanfieldprobability}} \, \meanfieldentropy^{int}_{\bm{\meanfieldprobability}} , \quad \tilde{\bm{\countingfield}} = 1 - \bm{\countingfield} ,
\end{aligned}
\end{align}
while reusing the shorthand notation from Eq. \eqref{eq:countingfieldnotation}.
The definitions of the time-reversed physical quantities in the first line are trivial. Less obvious is the transformation rule of the auxiliary field $\bm{\field}$. This transformation rule amounts to inverting the directions of the edges corresponding to a reversion of the Markov dynamics: The change of the sign in front of $\bm{\field}$ can be seen by noting that the latter is a counting field for variations in the state variables $\d \bm{\meanfieldprobability}$. Moreover, the affinity along an edge is inverted by the free energy shift.

We proceed by demonstrating that the above transformation, up to a non-fluctuating quantity, indeed leaves the generating function invariant. For better readability, we will split the action functional \eqref{eq:mesogeneratingfunctionentropyproductionpathintegralappendix} into two parts and investigate how they transform under the time reversal in Eq. \eqref{eq:timereversaltransformationappendix}. First, the invariance of the biased Hamiltonian under this time-reversal transformation can be seen as follows,
{\begin{align} 
\begin{aligned}
&\tilde{\hamiltonian}_{\tilde{\bm{\countingfield}}}[\bm{\meanfieldprobability}(t'),\bm{\field}(t')] = \\ 
&= \sum\limits_{\reservoir = 1}^{\reservoirdimension} \sum\limits_{i,j=1}^{\statenumber} \Big[ \exp \Big\lbrace \field_j(t') - \field_i(t') + \invtemperature^{(1)} ( \D_{\meanfieldprobability_i} - \D_{\meanfieldprobability_j} )  \meanfieldfreeenergy_{\bm{\meanfieldprobability}}^{(1)}(\bm{\protocol}_t') + \big( \countingfield_{_{\bm{\force}}}^{(\reservoir)} - 1 \big) \invtemperature^{(\reservoir)} \force_{ij}^{(\reservoir)} \; + \\
&+ \big( \countingfield_{\trajectoryenergymeso}^{(\reservoir)} - 1 \big) [ \invtemperature^{(1)} - \invtemperature^{(\reservoir)} ] ( \D_{\meanfieldprobability_i} - \D_{\meanfieldprobability_j} ) \meanfieldenergy_{\bm{\meanfieldprobability}}^{(1)}(\bm{\protocol}_t')  \Big\rbrace - 1 \Big] \meanfieldrates_{ij}^{(\reservoir)} (\bm{\protocol}_{t'},\bm{\meanfieldprobability}(t')) \, \meanfieldprobability_j(t') \\ 
&= \sum\limits_{\reservoir = 1}^{\reservoirdimension} \sum\limits_{i,j=1}^{\statenumber}  \Big[ \exp \Big\lbrace \field_j(t') - \field_i(t') - \invtemperature^{(1)} ( \D_{\meanfieldprobability_j} - \D_{\meanfieldprobability_i} ) \meanfieldfreeenergy_{\bm{\meanfieldprobability}}^{(1)}(\bm{\protocol}_t') + \big( 1 - \countingfield_{_{\bm{\force}}}^{(\reservoir)} \big) \invtemperature^{(\reservoir)} \force_{ji}^{(\reservoir)} \; + \\
&+ \big( 1 - \countingfield_{\trajectoryenergymeso}^{(\reservoir)} \big) [\invtemperature^{(1)} - \invtemperature^{(\reservoir)} ] ( \D_{\meanfieldprobability_j} - \D_{\meanfieldprobability_i} ) \meanfieldenergy_{\bm{\meanfieldprobability}}^{(\reservoir)}(\bm{\protocol}_t') \Big\rbrace - 1 \Big] \meanfieldrates_{ij}^{(\reservoir)} (\bm{\protocol}_{t'},\bm{\meanfieldprobability}(t')) \, \meanfieldprobability_j(t') \\ 
&= \sum\limits_{\reservoir = 1}^{\reservoirdimension} \sum\limits_{i,j=1}^{\statenumber}  \Big[ \exp \Big\lbrace \field_j(t') - \field_i(t') - \invtemperature^{(\reservoir)} \big[  ( \D_{\meanfieldprobability_j} - \D_{\meanfieldprobability_i} ) \meanfieldfreeenergy_{\bm{\meanfieldprobability}}^{(\reservoir)}(\bm{\protocol}_t') - \force_{ji}^{(\reservoir)} \big] - \countingfield_{_{\bm{\force}}}^{(\reservoir)} \invtemperature^{(\reservoir)} \force_{ji}^{(\reservoir)} \; - \\ &- \countingfield_{\trajectoryenergymeso}^{(\reservoir)} [ \invtemperature^{(1)} - \invtemperature^{(\reservoir)} ] ( \D_{\meanfieldprobability_j} - \D_{\meanfieldprobability_i} ) \meanfieldenergy_{\bm{\meanfieldprobability}}^{(\reservoir)}(\bm{\protocol}_t') \Big\rbrace - 1 \Big] \meanfieldrates_{ij}^{(\reservoir)} (\bm{\protocol}_{t'},\bm{\meanfieldprobability}(t')) \, \meanfieldprobability_j(t') \\ 
&= \sum\limits_{\reservoir = 1}^{\reservoirdimension} \sum\limits_{i,j=1}^{\statenumber}  \Big[ \exp \big[ \field_j(t') - \field_i(t') \big] \, \frac{\meanfieldrates_{ji}^{(\reservoir)} (\bm{\protocol}_{t'},\bm{\meanfieldprobability}(t')) \, \meanfieldprobability_i(t') }{\meanfieldrates_{ij}^{(\reservoir)} (\bm{\protocol}_{t'},\bm{\meanfieldprobability}(t')) \, \meanfieldprobability_j(t') }  \, \cdot \\
&\cdot \exp \Big\lbrace - \countingfield_{_{\bm{\force}}}^{(\reservoir)} \invtemperature^{(\reservoir)} \force_{ji}^{(\reservoir)} - \countingfield_{\trajectoryenergymeso}^{(\reservoir)} [ \invtemperature^{(1)} - \invtemperature^{(\reservoir)} ] ( \D_{\meanfieldprobability_j} - \D_{\meanfieldprobability_i} ) \meanfieldenergy_{\bm{\meanfieldprobability}}^{(\reservoir)}(\bm{\protocol}_t') \Big\rbrace - 1 \Big] \meanfieldrates_{ij}^{(\reservoir)} (\bm{\protocol}_{t'},\bm{\meanfieldprobability}(t')) \, \meanfieldprobability_j(t') \\
&= \sum\limits_{\reservoir = 1}^{\reservoirdimension} \sum\limits_{i,j=1}^{\statenumber}  \Big[ \exp \Big\lbrace \field_j(t') - \field_i(t') - \countingfield_{_{\bm{\force}}}^{(\reservoir)} \invtemperature^{(\reservoir)} \force_{ji}^{(\reservoir)} - \countingfield_{\trajectoryenergymeso}^{(\reservoir)} [ \invtemperature^{(1)} - \invtemperature^{(\reservoir)} ] ( \D_{\meanfieldprobability_j} - \D_{\meanfieldprobability_i} )  \meanfieldenergy_{\bm{\meanfieldprobability}}^{(\reservoir)}(\bm{\protocol}_t') \Big\rbrace - 1 \Big] \meanfieldrates_{ji}^{(\reservoir)} (\bm{\protocol}_{t'},\bm{\meanfieldprobability}(t')) \, \meanfieldprobability_i(t') \\
&= \hamiltonian_{\bm{\countingfield}}[\bm{\meanfieldprobability}(t'),\bm{\field}(t')] .
\end{aligned}
\end{align} }

Furthermore, we find for the sum of the kinetic and non-autonomous driving terms together with the initial condition under time-reversal,
\begin{align} 
&\int\limits_0^t \d t' \, \Big\lbrace \big[  \, \invtemperature^{(1)} \, \nabla_{\bm{\meanfieldprobability}} \, \meanfieldfreeenergy_{\bm{\meanfieldprobability}}^{(1)}(\bm{\protocol}_t') - \bm{\field}(t') \big] \cdot \dot{\bm{\meanfieldprobability}}(t') +  (1 - \countingfield_{_{\bm{\lambda}}}) \, \invtemperature^{(1)} \dot{\bm{\protocol}}_{t'} \cdot \left[ \nabla_{\bm{\protocol}_{t'}} \meanfieldenergy_{\bm{\meanfieldprobability}}(\bm{\protocol}_t') \right] \Big\rbrace  + \ln \mesoprobability_{\bm{\meanfieldprobability}_t}^{eq}(\bm{\protocol}_t) \, = \nonumber \\
= &\int\limits_0^t \d t' \Big\lbrace  \invtemperature^{(1)} \big( \d_{t'} \meanfieldfreeenergy_{\bm{\meanfieldprobability}}^{(1)}(\bm{\protocol}_t') \!-\! \dot{\bm{\protocol}}_{t'} \! \cdot \! \big[ \nabla_{\bm{\protocol}_{t'}} \meanfieldfreeenergy_{\bm{\meanfieldprobability}}^{(1)}(\bm{\protocol}_t') \big] \!-\! \bm{\field}(t') \! \cdot \! \dot{\bm{\meanfieldprobability}}(t') \!+\! (1 \!-\! \countingfield_{_{\bm{\lambda}}}) \, \invtemperature^{(1)} \dot{\bm{\protocol}}_{t'} \!\cdot \! \left[ \nabla_{\bm{\protocol}_{t'}} \meanfieldenergy_{\bm{\meanfieldprobability}}(\bm{\protocol}_t') \right] \Big\rbrace \!+\! \ln  \mesoprobability_{\bm{\meanfieldprobability}_t}^{eq}(\bm{\protocol}_t) \nonumber \\ 
= & - \int\limits_0^t \d t' \, \Big\lbrace \bm{\field}(t') \cdot \dot{\bm{\meanfieldprobability}}(t') + \countingfield_{_{\bm{\lambda}}} \, \invtemperature^{(1)} \dot{\bm{\protocol}}_{t'} \cdot \left[ \nabla_{\bm{\protocol}_{t'}} \meanfieldenergy_{\bm{\meanfieldprobability}}(\bm{\protocol}_t') \right] \Big\rbrace + \invtemperature^{(1)} \big[ \meanfieldfreeenergy_{\bm{\meanfieldprobability}}^{(1)}(\bm{\protocol}_t) - \meanfieldfreeenergy_{\bm{\meanfieldprobability}}^{(1)}(\bm{\protocol}_0) \big] + \ln \mesoprobability_{\bm{\meanfieldprobability}_t}^{eq}(\bm{\protocol}_t) \nonumber \\
= & - \int\limits_0^t \d t'\, \Big\lbrace \bm{\field}(t') \cdot \dot{\bm{\meanfieldprobability}}(t') + \countingfield_{_{\bm{\lambda}}} \, \invtemperature^{(1)} \dot{\bm{\protocol}}_{t'} \cdot \left[ \nabla_{\bm{\protocol}_{t'}} \meanfieldenergy_{\bm{\meanfieldprobability}}(\bm{\protocol}_t') \right] \Big\rbrace
+ \ln \frac{\mesoprobability_{\bm{\meanfieldprobability}_0}^{eq}(\bm{\protocol}_0)}{\mesoprobability_{\bm{\meanfieldprobability}_t}^{eq}(\bm{\protocol}_t)} + \invtemperature^{(1)} \big[ \meanfieldfreeenergy^{eq}(\bm{\protocol}_t) - \meanfieldfreeenergy^{eq}(\bm{\protocol}_0) \big] + \ln \mesoprobability_{\bm{\meanfieldprobability}_t}^{eq}(\bm{\protocol}_t) \nonumber \\
= & - \int\limits_0^t \d t'\, \Big\lbrace \bm{\field}(t') \cdot \dot{\bm{\meanfieldprobability}}(t') + \countingfield_{_{\bm{\lambda}}} \, \invtemperature^{(1)} \dot{\bm{\protocol}}_{t'} \cdot \left[ \nabla_{\bm{\protocol}_{t'}} \meanfieldenergy_{\bm{\meanfieldprobability}}(\bm{\protocol}_t') \right] \Big\rbrace
+ \ln \mesoprobability_{\bm{\meanfieldprobability}_0}^{eq}(\bm{\protocol}_0) + \invtemperature^{(1)} \big[ \meanfieldfreeenergy^{eq}(\bm{\protocol}_t) - \meanfieldfreeenergy^{eq}(\bm{\protocol}_0) \big] . 
\end{align} 

Collecting results, we thus find that the size-intensive action functional is invariant under the time reversal \eqref{eq:timereversaltransformationappendix} up to a non-fluctuating term corresponding to the change in the size-intensive part of the equilibrium free energy, \idest 
\begin{align}
\action_{ \bm{\countingfield}}[\bm{\meanfieldprobability},\bm{\field}] = \tilde{\action}_{ \tilde{\bm{\countingfield}} }[\bm{\meanfieldprobability},\bm{\field}] -  \invtemperature^{(1)} \Delta \meanfieldfreeenergy^{eq}_1(\bm{\protocol}) .
\end{align}
In the macroscopic limit, the scaled cumulant generating function is equal to the extremal action functional, cf. Eq. \eqref{eq:scaledcumulantgeneratingfunction}. Moreover, the action functional contains the initial condition of the trajectories so that its extremization does not give rise to additional boundary terms,
\begin{align}
\delta \action_{\bm{\countingfield} }[\bm{\meanfieldprobability},\bm{\field}] = \delta \tilde{\action}_{ \tilde{\bm{\countingfield}} }[\bm{\meanfieldprobability},\bm{\field}] . 
\end{align}
Hence the invariance of the action functional is preserved in the macroscopic limit that in turn stipulates the following symmetry for the scaled-cumulant generating function,
\begin{align}
\scgf(\bm{\countingfield},t) = \tilde{\scgf}(\tilde{\bm{\countingfield}},t) - \invtemperature^{(1)} \Delta \meanfieldfreeenergy^{eq}_1(\bm{\protocol}) , 
\end{align}
which is exactly Eq. \eqref{eq:macrogeneratingfunctionsymmetry}.

\bibliography{bibliography}

\end{document}